\RequirePackage[l2tabu, orthodox]{nag}
\documentclass[journal, 10pt, onecolumn, letterpaper, oneside]{IEEEtran}

\usepackage{ifpdf}

\usepackage[noadjust]{cite}

\ifCLASSINFOpdf
  \usepackage[pdftex]{graphicx}
  \else
  \usepackage[dvips]{graphicx}
\fi

\usepackage[cmex10]{amsmath}
\usepackage[caption=false, font=footnotesize]{subfig}

\usepackage{array}

\ifCLASSOPTIONcompsoc
  \usepackage[caption=false,font=normalsize,labelfont=sf,textfont=sf]{subfig}
\else
  \usepackage[caption=false,font=footnotesize]{subfig}
\fi

\usepackage{fixltx2e}

\usepackage{url}

\usepackage[utf8]{inputenc} 
\usepackage[T1]{fontenc}
\usepackage{ifthen}

\usepackage{mathtools}
\usepackage{amssymb}
\usepackage[only, varoast, boxast]{stmaryrd}
\usepackage{relsize}
\usepackage{bbm}
\usepackage{xfrac}
\usepackage{amsthm}
\theoremstyle{plain}
\newtheorem{definition}{Definition}
\newtheorem{lemma}{Lemma}

\newtheorem{theorem}{Theorem}
\newtheorem{corollary}{Corollary}
\newtheorem{remark}{Remark}

\newtheorem{example}{Example}
\newtheorem{proposition}{Proposition}
\newtheorem{claim}{Claim}
\usepackage[varg, cmbraces]{newtxmath}

\usepackage{xcolor}
\definecolor{burgundy}{rgb}{0.545098,0,0}
\definecolor{navyblue}{rgb}{0.0, 0.0, 0.5}
\definecolor{leafgreen}{rgb}{0.290196, 0.470588, 0.0}
\definecolor{bluegreen}{rgb}{0, 0.470588, 0.415686}
\definecolor{zuhl}{rgb}{0.1875, 0.26171875, 0.46484375}
\definecolor{orange}{rgb}{1, 0.6470588235, 0}
\definecolor{red}{rgb}{1, 0, 0}

\usepackage{overpic}
\usepackage{pict2e}
\usepackage{booktabs}

\newcommand{\bvec}[1]{\boldsymbol{#1}}

\newcommand{\lcm}{\operatorname{lcm}}

\newcommand{\tabref}[1]{Table~\ref{#1}}
\newcommand{\figref}[1]{Fig.~\ref{#1}}

\newcommand{\lemref}[1]{Lemma~\ref{#1}}
\newcommand{\propref}[1]{Proposition~\ref{#1}}
\newcommand{\thref}[1]{Theorem~\ref{#1}}
\newcommand{\defref}[1]{Definition~\ref{#1}}
\newcommand{\corref}[1]{Corollary~\ref{#1}}
\newcommand{\sectref}[1]{Section~\ref{#1}}
\newcommand{\remref}[1]{Remark~\ref{#1}}
\newcommand{\exref}[1]{Example~\ref{#1}}
\newcommand{\appref}[1]{Appendix~\ref{#1}}
\newcommand{\clref}[1]{Claim~\ref{#1}}

\allowdisplaybreaks[3]

\usepackage{array}
\usepackage[linesnumbered, algoruled, boxed, lined, vlined]{algorithm2e}
\SetKwRepeat{Do}{do}{while}
\SetKw{KwBreak}{break}
\SetKw{KwAnd}{and}

\usepackage{soulutf8}
\setstcolor{red}

\usepackage{tikz}
\usetikzlibrary{positioning}
\usetikzlibrary{arrows.meta}
\usetikzlibrary{fit}
\usepackage{tikz-3dplot}
\usetikzlibrary{positioning}
\usetikzlibrary{decorations.pathreplacing}
\usetikzlibrary{decorations.markings}
\usetikzlibrary{positioning}
\usetikzlibrary{patterns}
\usetikzlibrary{shapes,arrows}
\tikzstyle{vecArrow} = [thick, decoration={markings,mark=at position
   1 with {\arrow[semithick]{open triangle 90}}},
   double distance=3pt, shorten >= 4pt,
   preaction = {decorate},
   postaction = {draw,line width=1.4pt, white,shorten >= 4.5pt}]
\tikzstyle{innerWhite} = [semithick, white,line width=1.4pt, shorten >= 4.5pt]
\tikzset{
    partial ellipse/.style args={#1:#2:#3}{
        insert path={+ (#1:#3) arc (#1:#2:#3)}
    }
}
\tikzset{%
  highlight/.style = {rectangle, rounded corners, fill = green!20, draw, fill opacity = 0.2, thick, inner sep = 0pt}
}

\usepackage{hyperref}
\hypersetup{
    colorlinks,
    linkcolor={red!60!black},
    citecolor={blue!60!black},
    urlcolor={blue!50!black}
}
\usepackage{microtype}

\begin{document}

\title{Modular Arithmetic Erasure Channels \\ and Their Multilevel Channel Polarization}

\IEEEoverridecommandlockouts

\author{%
\IEEEauthorblockN{%
Yuta~Sakai,~\IEEEmembership{Member,~IEEE,}
Ken-ichi~Iwata,~\IEEEmembership{Member,~IEEE,}
and Hiroshi~Fujisaki,~\IEEEmembership{Member,~IEEE,}%
\thanks{This work was supported in part by JSPS KAKENHI Grant Numbers 26420352, 17K06422, 17J11247, and 18K11465.
This article was presented in part at the 2016 IEEE Information Theory Workshop \cite{itw2016} \emph{(Corresponding author: Yuta Sakai)} and at the 2018 IEEE International Symposium on Information Theory \cite{isit2018} \emph{(Corresponding author: Yuta Sakai)}.}%
\thanks{Y.~Sakai was with Graduate School of Engineering, University of Fukui, Japan. He is currently with the Department of Electrical and Computer Engineering, National University of Singapore, Singapore, Email: \url{eleyuta@nus.edu.sg}.}%
\thanks{K.~Iwata is with Graduate School of Engineering,
University of Fukui,
Japan,
Email: \url{k-iwata@u-fukui.ac.jp}.}%
\thanks{H.~Fujisaki is with Graduate School of Natural Science and Technology,
Kanazawa University,
Japan,
Email: \url{fujisaki@ec.t.kanazawa-u.ac.jp}.}
}%
}

\maketitle

\begin{abstract}
This study proposes \emph{modular arithmetic erasure channels} (MAECs), a novel class of erasure-like channels with an input alphabet that need not be binary.
This class contains the binary erasure channel (BEC) and some other known erasure-like channels as special cases.
For MAECs, we provide recursive formulas of Ar{\i}kan-like polar transform to simulate channel polarization.
In other words, we show that the synthetic channels of MAECs are equivalent to other MAECs.
This is a generalization of well-known recursive formulas of the polar transform for BECs.
Using our recursive formulas, we also show that a recursive application of the polar transform for MAECs results in \emph{multilevel channel polarization,} which is an asymptotic phenomenon that is characteristic of non-binary polar codes.
Specifically, we establish a method to calculate the limiting proportions of the partially noiseless and noisy channels that are generated as a result of multilevel channel polarization for MAECs.
In the particular case of MAECs, this calculation method solves an open problem posed by Nasser (2017) in the study of non-binary polar codes.
\end{abstract}


\IEEEpeerreviewmaketitle

\section{Introduction}

Ar{\i}kan \cite{arikan_it2009} proposed binary polar codes as a class of channel codes that provably achieves the symmetric capacity of a binary-input discrete memoryless channel (DMC), admits a deterministic construction, and has low encoding/decoding complexities.
A key operation employed in polar codes is the polar transform.
This transform results in almost noiseless and useless synthetic channels as the number of polarization steps increases.
This phenomenon is called channel polarization, and the limiting proportions of noiseless and useless synthetic channels coincide with $I(W)$ and $1 - I(W)$, respectively, where $I(W)$ stands for the symmetric capacity of the given binary-input DMC $W$.

In the study of non-binary polar codes, there are two types of channel polarization: two-level channel polarization \cite{sasoglu_isit2012, mori_tanaka_it2014} and multilevel channel polarization \cite{abbe_telatar_it2012, park_barg_it2013, sahebi_pradhan_it2013, nasser_telatar_it2016, nasser_it2016_ergodic1, nasser_it2017_ergodic2, nasser_it2017_fourier, nasser_PhD, nasser_isit2019_level, nasser_isit2019_topology}.
In the context of two-level channel polarization, the synthetic channels converge to either \emph{noiseless} or \emph{useless} channels.
In contrast, in the context of multilevel channel polarization, the synthetic channels converge to several types of \emph{partially noiseless} channels.
It was independently shown in \cite{sasoglu_isit2012, abbe_telatar_it2012, park_barg_it2013, sahebi_pradhan_it2013, mori_tanaka_it2014, nasser_telatar_it2016, nasser_it2017_fourier, nasser_it2016_ergodic1, nasser_it2017_ergodic2, nasser_PhD} that two-level and multilevel channel polarization can achieve the symmetric capacity $I(W)$ of the DMC $W$.
However, it is difficult to characterize the limiting proportions of the partially noiseless synthetic channels in the context of multilevel channel polarization (see \cite[Section~9.2.1]{nasser_PhD}).
In this study, we term these limiting proportions as the \emph{asymptotic distribution} of multilevel channel polarization.

To construct and analyze polar codes, we have to calculate channel parameters, e.g., the symmetric capacity, the Bhattacharyya parameter, etc., of the synthetic channels induced by the polar transform.
However, the computational complexities of these channel parameters grow doubly-exponentially in the number of polar transforms.
In the binary-input case, Tal and Vardy \cite{tal_vardy_it2013} solved this issue by applying approximation algorithms for the synthetic channels at each polar transform.
Such an approximation method was recently extended from the binary to non-binary settings by Gulch, Ye, and Barg \cite{gulcu_ye_barg_it2018}.
On the other hand, it is well-known that for binary erasure channels (BECs), one can avoid the use of any approximation arguments.
Obviously, the asymptotic distribution of a BEC can be simply characterized by its erasure probability.
Therefore, BECs are excellent toy problems in the study of binary polar codes.
In non-binary polar codes, similar easily-analyzable channel models have been proposed by Park and Barg \cite[Section~III]{park_barg_it2013} and Sahebi and Pradhan \cite[Figs.~3 and~4]{sahebi_pradhan_it2013}, and the recursive formulas of the polar transform were given therein.%
\footnote{Note that the recursive formula \cite[Equation~(4)]{sahebi_pradhan_it2013} for the minus transform is valid, but the recursive formula \cite[Equation~(3)]{sahebi_pradhan_it2013} for the plus transform is incorrect.
\thref{th:recursive_V} of \sectref{sect:ease} corrects this error (see \exref{ex:recursive_6} of \sectref{sect:bec}).}

\subsection{Main Contributions}
\label{sect:contribution}

\begin{figure*}[!t]
\centering
\begin{overpic}[width = 0.9\hsize, clip]{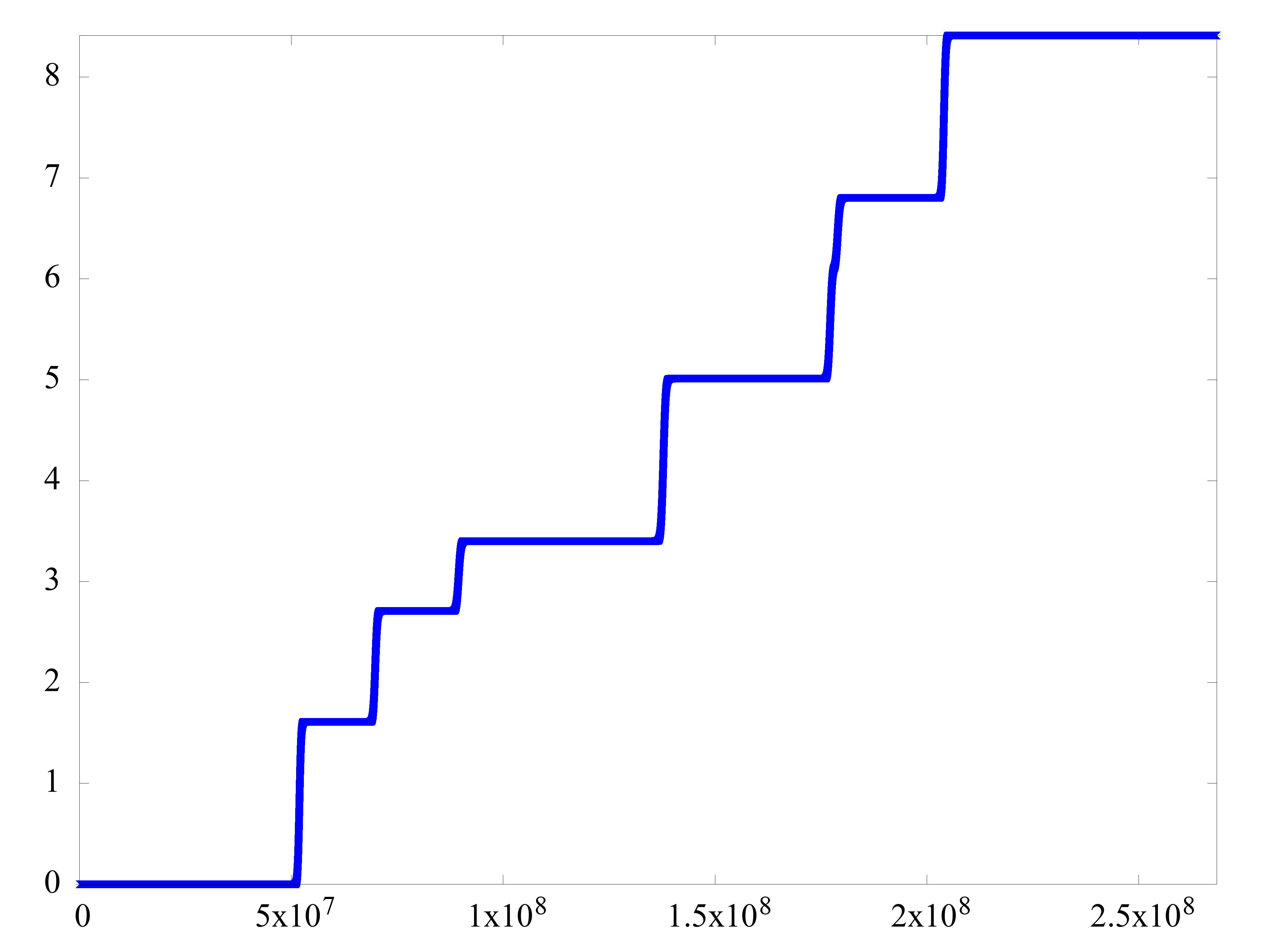}
\put(-2, 70){\footnotesize [nats]}
\put(-1, 25){\rotatebox{90}{symmetric capacity $I( V_{\bvec{\varepsilon}}^{\bvec{s}} )$}}
\put(28, -1){indices of $\bvec{s}$ (sorted in increasing order of $I( V_{\bvec{\varepsilon}}^{\bvec{s}} )$)}
\put(8, 35){$\mu_{1}^{(\infty)} = \sfrac{29}{150} \approx 0.193333$}
\put(15, 33){\vector(0, -1){26}}
\put(27, 7.5){$\mu_{5}^{(\infty)} = \sfrac{1}{15} \approx 0.066667$}
\put(28, 10.5){\vector(-1, 4){1.7}}
\put(34, 15){$\mu_{15}^{(\infty)} = \sfrac{11}{150} \approx 0.073333$}
\put(35, 18){\vector(-1, 4){2}}
\put(22, 45){$\mu_{30}^{(\infty)} = \sfrac{9}{50} = 0.18$}
\put(35, 43){\vector(1, -1){9}}
\put(60, 28){$\mu_{150}^{(\infty)} = \sfrac{11}{75} \approx 0.146667$}
\put(61, 31){\vector(0, 1){13}}
\put(70, 48.5){$\mu_{450}^{(\infty)} = \sfrac{1}{150} \approx 0.006667$}
\put(71, 51){\vector(-2, 1){4.75}}
\put(45, 67){$\mu_{900}^{(\infty)} = \sfrac{7}{75} \approx 0.093333$}
\put(63, 66){\vector(1, -1){5.5}}
\put(77, 60){$\mu_{4500}^{(\infty)} = \sfrac{6}{25} = 0.24$}
\put(88, 62){\vector(0, 1){9}}
\end{overpic}
\caption{Plot of the symmetric capacities $\{ I(V_{\bvec{\varepsilon}}^{\bvec{s}}) \mid \bvec{s} \in \{ -, + \}^{n} \}$ under $n = 28$-steps of the polar transform.
The initial MAEC $V_{\bvec{\varepsilon}}$ is given in \exref{ex:mu_d} of \sectref{sect:composite} (see also \tabref{table:mu_d} of \sectref{sect:composite}).
Note that the logarithm is taken to be the natural logarithm so the units of the symmetric capacity is nats.
The proportion of synthetic channels $V_{\bvec{\varepsilon}}^{\bvec{s}}$ satisfying $I(V_{\bvec{\varepsilon}}^{\bvec{s}}) \approx \log d$ and $I(V_{\bvec{\varepsilon}}^{\bvec{s}}[ \ker \varphi_{d} ]) \approx \log d$ is roughly equal to $\mu_{d}^{(\infty)}$ for each $d|q$ (cf.\ \corref{cor:multilevel} of \sectref{sect:composite}), where $\mu_{d}^{(\infty)}$ is defined in \eqref{def:mu_d_infty} of \sectref{sect:asymptotic_distribution_MAEC}.
For example, the proportion of synthetic channels $V_{\bvec{\varepsilon}}^{\bvec{s}}$ satisfying $I(V_{\bvec{\varepsilon}}^{\bvec{s}}) \approx \log 30$ and $I(V_{\bvec{\varepsilon}}^{\bvec{s}}[ \ker \varphi_{30} ]) \approx \log 30$ is roughly equal to $\mu_{30}^{(\infty)} = 0.18$.}
\label{fig:mu_d}
\end{figure*}

The main contributions of this study can be broadly divided into the following two parts:
Firstly, we propose a novel channel model called \emph{modular arithmetic erasure channels} (MAECs), which can be naturally specialized to the following erasure-like channels: BECs, a na\"{i}ve definition of $q$-ary erasure channels ($q$-ECs) (see, e.g., \cite[p.~589]{mackay_2003}), $q$-ary input ordered erasure channels (OECs) proposed by Park and Barg \cite[p.~2285]{park_barg_isit2011}, and Sahebi and Pradhan's senary-input channels \cite[Fig.~4: Channel~2]{sahebi_pradhan_it2013}.
Then, we show that analyzing the polarization properties for MAECs is a straightforward endeavor under our framework.
Similar to the polar transform for BECs, we show that the synthetic channels generated from an MAEC are again equivalent to other MAECs with certain transition probabilities.
Secondly, we characterize the asymptotic distribution of multilevel channel polarization for MAECs.
Specifically, we establish an algorithm for calculating the asymptotic distribution for a given MAEC.
Figure~\ref{fig:mu_d} illustrates the symmetric capacities of the synthetic channels induced by the polar transform for an MAEC with input alphabet size $q = 4500$;
this channel results in multilevel channel polarization.
This figure is plotted using our proposed recursive formulas of the polar transform for MAECs and its asymptotic distribution is calculated by our proposed algorithm.
This result solves an open problem in the study of multilevel channel polarization (cf.\ \cite[Section~9.2.1]{nasser_PhD}) in the particular case of MAECs.

\subsection{Paper Organization}

The rest of this paper is organized as follows:
\sectref{sect:preliminaries} introduces basic notations and definitions for this study.
Specifically, modular arithmetic is introduced in \sectref{sect:modular-arithmetic}, DMCs and their channel parameters are defined in \sectref{sect:dmc}, the Ar{\i}kan-like polar transform is defined in \sectref{sect:polar-transform}, and a notion of the channel equivalence is given in \sectref{sect:output_equiv}.
\sectref{sect:ease} introduces MAECs, and characterizes the ease of analyzing polar transform for MAECs.
The definition of MAECs is given in \defref{def:V}, and the recursive formulas of the polar transform for MAECs are stated in \thref{th:recursive_V}.
In \sectref{sect:bec}, some reductions of MAECs to known erasure-like channels are introduced.
\sectref{sect:two-notions} briefly compares the notions of two-level and multilevel channel polarization, these are revisited in Sections~\ref{sect:strong} and~\ref{sect:multilevel}, respectively.
An open problem in the study of multilevel channel polarization is described in \sectref{sect:multilevel}.
Some numerical simulations for MAECs are provided in \sectref{sect:sim}.
\sectref{sect:asymptotic_distribution_MAEC} discusses our solution to the asymptotic distribution of multilevel channel polarization for MAECs.
The main statement is given in \corref{cor:multilevel}.
In \sectref{sect:primepower}, we characterize the asymptotic distribution in the simplest case when the input alphabet size is a prime power.
In \sectref{sect:composite}, we consider the general case when the input alphabet size is not necessarily a prime power.
We then give Algorithm~\ref{alg:main} for calculating the asymptotic distribution.
A formal statement of the asymptotic distribution is given in \sectref{sect:asymptotic_distribution}.
Finally, \sectref{sect:conclusion} concludes this study.

\section{Preliminaries and Problem Presentations}
\label{sect:preliminaries}

\subsection{Basic Notations in Elementary Number Theory}
\label{sect:modular-arithmetic}

Firstly, we introduce standard notations in elementary number theory.
Let $\mathbb{Z}$ be the set of integers, and $\mathbb{N}$ the set of positive integers.
Given two positive integers $a, b \in \mathbb{N}$, define the following three sets:
\begin{align}
a \mathbb{Z}
& \coloneqq
\{ a z \mid z \in \mathbb{Z} \}
= \{ \dots, -2 a, - a, 0, a, 2 a, \dots \} ,
\\
b + a \mathbb{Z}
& \coloneqq
\{ b + z \mid z \in a \mathbb{Z} \}
= \{ \dots, b - 2 a, b - a, b, b + a, b + 2 a, \dots \} ,
\\
\frac{ \mathbb{Z} }{ a\mathbb{Z} }
& \coloneqq
\{ z + a \mathbb{Z} \mid z \in \mathbb{Z} \}
= \{ a \mathbb{Z}, 1 + a \mathbb{Z}, \dots, (a-1) + a \mathbb{Z} \} .
\end{align}
For two positive integers $a, b \in \mathbb{N}$, let $a|b$ be a shorthand for ``$a$ divides $b$,'' which means that there exists a positive integer $c \in \mathbb{N}$ satisfying $a c = b$.
If we define the sum set%
\footnote{The term \emph{sum set} or \emph{sumset} is used in additive combinatorics \cite{tao_vu_2006}.}
$\mathcal{S} + \mathcal{T} \coloneqq \{ s + t \mid s \in \mathcal{S} \ \mathrm{and} \ t \in \mathcal{T} \}$ for given two subsets $\mathcal{S}, \mathcal{T} \subset \mathbb{Z}$, then it is clear that $a \mathbb{Z} + b \mathbb{Z} = a \mathbb{Z}$ whenever $a|b$.
These definitions naturally introduce the congruence relation on the integers modulo $q$ (see \eqref{def:congruence} of \appref{app:recursive_V} for details).
Given $q \in \mathbb{N}$, the multiplication $\cdot$ on $\mathbb{Z}/q\mathbb{Z}$ is defined as $(a + q\mathbb{Z}) \cdot (b + q\mathbb{Z}) = a b + q\mathbb{Z}$.

\subsection{DMCs and Channel Parameters}
\label{sect:dmc}

We now define DMCs as follows:
The input alphabet is given by $\mathbb{Z}/q\mathbb{Z}$ for some integer $q \ge 2$.
The output alphabet $\mathcal{Y}$ is a nonempty and countable set.
The transition probability from an input symbol $x \in \mathbb{Z}/q\mathbb{Z}$ to an output symbol $y \in \mathcal{Y}$ is denoted by $W(y \mid x)$.
Let $W : \mathbb{Z}/q\mathbb{Z} \to \mathcal{Y}$, or simply $W$, be a shorthand for such a DMC.
The \emph{$\alpha$-symmetric capacity} of $W$, which is the $\alpha$-mutual information \cite{ho_verdu_isit2015, verdu_ita2015} between the input and output of $W$ under a uniform input distribution on $\mathbb{Z}/q\mathbb{Z}$, is defined by
\begin{align}
I_{\alpha}( W )
& \coloneqq
\begin{dcases}
\min_{y \in \mathcal{Y}} \left( \log \frac{ q }{ |\{ x \in \mathbb{Z}/q\mathbb{Z} \mid W(y \mid x) > 0 \}| } \right)
& \mathrm{if} \ \alpha = 0 ,
\\
I( W )
& \mathrm{if} \ \alpha = 1 ,
\\
\log \left( \sum_{y \in \mathcal{Y}} \max_{x \in \mathbb{Z}/q\mathbb{Z}} W(y \mid x) \right)
& \mathrm{if} \ \alpha = \infty ,
\\
\frac{ \alpha }{ \alpha - 1 } \log \left( \sum_{y \in \mathcal{Y}} \left( \sum_{x \in \mathbb{Z}/q\mathbb{Z}} \frac{ 1 }{ q } W(y \mid x)^{\alpha} \right)^{1/\alpha} \right)
& \mathrm{otherwise}
\end{dcases}
\label{def:alpha}
\end{align}
for each order $\alpha \in [0, \infty]$, where the symmetric capacity $I(W)$ is defined by
\begin{align}
I(W)
\coloneqq
\sum_{y \in \mathcal{Y}} \sum_{x \in \mathbb{Z}/q\mathbb{Z}} \frac{ 1 }{ q } W(y \mid x) \log \frac{ W(y \mid x) }{ \sum_{x^{\prime} \in \mathbb{Z}/q\mathbb{Z}} (1/q) W(y \mid x^{\prime}) } .
\end{align}
Unless stated otherwise, assume throughout this paper that the base of logarithms is $q$.
Several relations between the $\alpha$-symmetric capacity $I_{\alpha}( W )$ and other channel parameters are summarized in the following remark.

\begin{remark}[connections between the $\alpha$-symmetric capacity and the other channel parameters]
\label{remark:connections_alpha_mutual}
The following identities hold:
\begin{align}
I_{\alpha}( W )
& =
\frac{ \alpha }{ 1 - \alpha } E_{0}\left( \frac{ 1 - \alpha }{ \alpha } , W \right)
\end{align}
for $\alpha \in (0, 1) \cup (1, \infty)$ and
\begin{align}
I_{1/2}( W )
& \: =
E_{0}( 1, W )
=
\log \left( \frac{ q }{ 1 + (q-1) Z( W ) } \right) ,
\\
I_{\infty}( W )
& \: =
(\log q) + \log( 1 - P_{\mathrm{e}}( W ) ) ,
\\
P_{\mathrm{e}}( W )
& \coloneqq
1 - \sum_{y \in \mathcal{Y}} \frac{ 1 }{ q } \max_{x \in \mathbb{Z}/q\mathbb{Z}} W(y \mid x)
\end{align}
denotes the average probability of maximum likelihood decoding error of uncoded communication via a channel $W$,
\begin{align}
Z( W )
\coloneqq
\frac{ 1 }{ q (q - 1) } \sum_{\substack{ x, x^{\prime} \in \mathbb{Z}/q\mathbb{Z} : \\ x \neq x^{\prime} }} \sum_{y \in \mathcal{Y}} \sqrt{ W(y \mid x) \, W(y \mid x^{\prime}) }
\end{align}
denotes the average Bhattacharyya distance of a channel $W$ \cite{sasoglu_telatar_arikan_itw2009}, and
\begin{align}
E_{0}(\rho, W)
\coloneqq
- \log \left( \sum_{y \in \mathcal{Y}} \left( \sum_{x \in \mathbb{Z}/q\mathbb{Z}} \frac{ 1 }{ q } W(y \mid x)^{1/(1+\rho)} \right)^{1+\rho} \right)
\end{align}
denotes Gallager's reliability function $E_{0}$ of a channel $W$ under a uniform input distribution for $\rho \in (-1, \infty)$ \cite[Equation~(5.6.14)]{gallager_1968}.
\end{remark}

\subsection{Ar{\i}kan-like Polar Transform over Modular Arithmetic}
\label{sect:polar-transform}

Let $\gamma \in \mathbb{Z}/q\mathbb{Z}$ be a unit of the ring, i.e., it has a multiplicative inverse element $\gamma^{-1} \in \mathbb{Z}/q\mathbb{Z}$ satisfying $\gamma \cdot \gamma^{-1} = \gamma^{-1} \cdot \gamma = 1 + q \mathbb{Z}$.
Given two DMCs $W_{1} : \mathbb{Z}/q\mathbb{Z} \to \mathcal{Y}_{1}$ and $W_{2} : \mathbb{Z}/q\mathbb{Z} \to \mathcal{Y}_{2}$, the polar transform creates two synthetic channels:
the worse channel $W_{1} \boxast W_{2} : \mathbb{Z}/q\mathbb{Z} \to \mathcal{Y}_{1} \times \mathcal{Y}_{2}$ defined by
\begin{align}
(W_{1} \boxast W_{2}) (y_{1}, y_{2} \mid u_{1})
\coloneqq
\sum_{u_{2}^{\prime} \in \mathbb{Z}/q\mathbb{Z}} \frac{ 1 }{ q } W_{1}(y_{1} \mid u_{1} + \gamma \cdot u_{2}^{\prime}) \, W_{2}(y_{2} \mid u_{2}^{\prime}) ,
\label{def:minus}
\end{align}
and the better channel $W_{1} \varoast W_{2} : \mathbb{Z}/q\mathbb{Z} \to \mathcal{Y}_{1} \times \mathcal{Y}_{2} \times \mathbb{Z}/q\mathbb{Z}$ defined by
\begin{align}
(W_{1} \varoast W_{2}) (y_{1}, y_{2}, u_{1} \mid u_{2})
\coloneqq
\frac{ 1 }{ q } W_{1}(y_{1} \mid u_{1} + \gamma \cdot u_{2}) \, W_{2}(y_{2} \mid u_{2}) .
\label{def:plus}
\end{align}
These polar transforms with a unit $\gamma \in \mathbb{Z}/q\mathbb{Z}$ are inspired by the study of entropy weighted sums (see \cite{abbe_li_madiman_2017}).
Since this polar transform is an analogue of the polar transform with a $2 \times 2$ kernel, in this paper, we call these polar transforms Ar{\i}kan-like polar transform.
Note that when $\gamma = 1 + q \mathbb{Z}$, one can think of our polar transform as being defined over a cyclic group $(\mathbb{Z}/q\mathbb{Z}, +)$.

Ar{\i}kan-like polar transforms with distinct initial channels $W_{1} \neq W_{2}$ have been studied in the study of polar codes for non-stationary memoryless channels \cite{alsan_telatar_it2016, mahdavifar_isit2017}.
When both $W_{1}$ and $W_{2}$ are identical to a given channel $W : \mathbb{Z}/q\mathbb{Z} \to \mathcal{Y}$, the polar transform stated in \eqref{def:minus} and \eqref{def:plus} can be specialized to standard polar transform for a stationary DMC $W$.
We then simply write
\begin{align}
W^{-}
& \coloneqq
W \boxast W ,
\\
W^{+}
& \coloneqq
W \varoast W .
\end{align}
After applying the polar transform $n$ times, the synthetic channel $W^{\bvec{s}} : \mathbb{Z}/q\mathbb{Z} \to \mathcal{Y}^{2^{n}} \times (\mathbb{Z}/q\mathbb{Z})^{w(\bvec{s})}$ is given by
\begin{align}
W^{\bvec{s}}
& \coloneqq
( \cdots (W^{s_{1}})^{s_{2}} \cdots )^{s_{n}}
\label{def:n-step}
\end{align}
for each $\bvec{s} = s_{1}s_{2} \cdots s_{n} \in \{ -, + \}^{n}$, where the function%
\footnote{%
The set $\mathbb{N}_{0} \coloneqq \mathbb{N} \cup \{ 0 \}$ consists of all nonnegative integers.
}
$w : \{ -, + \}^{\ast} \to \mathbb{N}_{0}$ is recursively defined by%
\footnote{%
For example, we observe that $w(+, -, +) = 2 \, w(+, -) + 1 = 2 \cdot 2 \, w(+) + 1 = 2 \cdot 2 \cdot 1 + 1 = 5$.
As $w( \cdot )$ seems binary expansions by replacing $(-, +)$ with $(0, 1)$, it is clear that $w : \{ -, + \}^{n} \to \{ 0, 1, \dots, 2^{n}-1 \}$ is bijective.
}
\begin{align}
w( s_{1}, \dots, s_{n} )
\coloneqq
\begin{cases}
2 \, w( s_{1}, \dots, s_{n-1} )
& \text{if} \ n \ge 1 \ \mathrm{and} \ s_{n} = - ,
\\
2 \, w( s_{1}, \dots, s_{n-1} ) + 1
& \text{if} \ n \ge 1 \ \mathrm{and} \ s_{n} = + ,
\\
0
& \mathrm{otherwise} ,
\end{cases}
\label{def:weight}
\end{align}
and
$\{ -, + \}^{\ast} \coloneqq \{ \varnothing, -, +, --, -+, +-, ++, \dots \}$ denotes the set of $\{ -, + \}$-valued finite-length sequences containing the empty sequence $\varnothing$.
Note that the output alphabet size $|\mathcal{Y}^{2^{n}} \times (\mathbb{Z}/q\mathbb{Z})^{w(\bvec{s})}|$ of the synthetic channel $W^{\bvec{s}}$ grows doubly-exponentially in $n$.
The difficulties in analyzing the performance of polar codes are mainly due to this issue as the computational complexities for calculating the channel parameters depends on the size of the output alphabet $|\mathcal{Y}^{2^{n}} \times (\mathbb{Z}/q\mathbb{Z})^{w(\bvec{s})}|$; see \sectref{sect:dmc}.

\subsection{Output Degradedness and Equivalence of Channels}
\label{sect:output_equiv}

We now introduce an equivalence relation between two channels having the same input alphabet $\mathcal{X}$ as follows:

\begin{definition}[stochastic degradedness and equivalence]
\label{def:output_equiv}
A channel $W : \mathcal{X} \to \mathcal{Y}$ is said to be \emph{degraded} with respect to another channel $\tilde{W} : \mathcal{X} \to \mathcal{Z}$ if there exists an intermediate channel $Q : \mathcal{Z} \to \mathcal{Y}$ satisfying
\begin{align}
W(y \mid x)
& =
\sum_{z \in \mathcal{Z}} \tilde{W}(z \mid x) \, Q(y \mid z)
\end{align}
for every $(x, y) \in \mathcal{X} \times \mathcal{Y}$.
We denote this degradedness relation as $W \preceq \tilde{W}$.
In particular, we say that $W$ and $\tilde{W}$ are \emph{equivalent} if $W \preceq \tilde{W}$ and $\tilde{W} \preceq W$.
We denote this equivalence as $W \equiv \tilde{W}$.
\end{definition}

\begin{remark}
To rigorously deal with the convergence of synthetic channels, Nasser \cite{nasser_isit2019_topology} introduced an equivalent class of DMCs via this equivalence relation.
A different notion of an equivalence relation has been discussed by Mori and Tanaka \cite[Section~IV]{mori_tanaka_it2014} and Gulcu, Ye, and Barg \cite[Definition~3]{gulcu_ye_barg_it2018} in the context of non-binary polar source and channel coding, respectively.
\end{remark}

The following lemma implies that the above equivalence relation preserves the $\alpha$-symmetric capacity.

\begin{lemma}
\label{lem:degraded_mutual}
For any $\alpha \in [0, \infty]$, it holds that
\begin{align}
W \preceq \tilde{W}
\quad \Longrightarrow \quad
I_{\alpha}( W )
\le
I_{\alpha}( \tilde{W} ) .
\label{eq:degraded_alpha_mutual}
\end{align}
Consequently, for any $\alpha \in [0, \infty]$, it holds that
\begin{align}
W \equiv \tilde{W}
\quad \Longrightarrow \quad
I_{\alpha}( W )
=
I_{\alpha}( \tilde{W} ) .
\label{eq:equiv_alpha_mutual}
\end{align}
\end{lemma}

\begin{IEEEproof}[Proof of \lemref{lem:degraded_mutual}]
Equation~\eqref{eq:degraded_alpha_mutual} is a direct consequence of the data-processing lemma%
\footnote{Note that the data-processing lemma \cite[Theorem~5]{polyanskiy_verdu_allerton2010} is usually stated in terms of the conditional independence between two random variables given a third one; such a notion is stronger than the stochastic degradedness assumed in \defref{def:output_equiv}.} for the $\alpha$-mutual information 
(see \cite[Theorem~5]{polyanskiy_verdu_allerton2010}).
\end{IEEEproof}

\lemref{lem:degraded_mutual} is a minor extension of \cite[Lemma~3]{tal_vardy_it2013} because the $\alpha$-symmetric capacity $I_{\alpha}( W )$ can be specialized to the symmetric capacity $I( W )$, the average Bhattacharyya distance $Z( W )$, and the probability of error $P_{\mathrm{e}}( W )$; see \remref{remark:connections_alpha_mutual}.
The following lemma shows that channel degradedness is preserved under the polar transform.

\begin{lemma}
\label{lem:invariant_degradedness}
Given four channels $W_{1} : \mathbb{Z}/q\mathbb{Z} \to \mathcal{Y}_{1}$, $\tilde{W}_{1} : \mathbb{Z}/q\mathbb{Z} \to \mathcal{Z}_{1}$, $W_{2} : \mathbb{Z}/q\mathbb{Z} \to \mathcal{Y}_{2}$, and $\tilde{W}_{2} : \mathbb{Z}/q\mathbb{Z} \to \mathcal{Z}_{2}$, it holds that
\begin{align}
W_{1} \preceq \tilde{W}_{1}
\ \mathrm{and} \
W_{2} \preceq \tilde{W}_{2}
\quad \Longrightarrow \quad
W_{1} \boxast W_{2} \preceq \tilde{W}_{1} \boxast \tilde{W}_{2}
\ \mathrm{and} \
W_{1} \varoast W_{2} \preceq \tilde{W}_{1} \varoast \tilde{W}_{2} .
\label{eq:invariant_degradedness}
\end{align}
Consequently, it holds that
\begin{align}
W_{1} \equiv \tilde{W}_{1}
\ \mathrm{and} \
W_{2} \equiv \tilde{W}_{2}
\quad \Longrightarrow \quad
W_{1} \boxast W_{2} \equiv \tilde{W}_{1} \boxast \tilde{W}_{2}
\ \mathrm{and} \
W_{1} \varoast W_{2} \equiv \tilde{W}_{1} \varoast \tilde{W}_{2} .
\label{eq:invariant_equivalence}
\end{align}
\end{lemma}

\begin{IEEEproof}[Proof of \lemref{lem:invariant_degradedness}]
See \appref{app:invariant_degradedness}.
\end{IEEEproof}

Note that \lemref{lem:invariant_degradedness} is a straightforward extension of \cite[Lemma~4.7]{korada_thesis} and \cite[Lemma~5]{tal_vardy_it2013}.

\section{Modular Arithmetic Erasure Channels and Their Polar Transform}
\label{sect:ease}

In the following, we propose a general type of erasure-like channels with input alphabet $\mathbb{Z}/q\mathbb{Z}$.

\begin{definition}[Modular arithmetic erasure channels (MAECs)]
\label{def:V}
Given a probability vector $\bvec{\varepsilon} = ( \varepsilon_{d} )_{d|q}$,%
\footnote{A nonnegative real vector is called a \emph{probability vector} if the sum of elements is unity.}
the MAEC $V_{\bvec{\varepsilon}} : \mathbb{Z}/q\mathbb{Z} \to \mathcal{Y}$ is defined by 
\begin{align}
V_{\bvec{\varepsilon}}(y \mid x)
\coloneqq
\begin{dcases}
\varepsilon_{d}
& \text{\emph{if $y = x + d \mathbb{Z}$ for some divisor $d$ of $q$}} ,
\\
0
& \mathrm{otherwise}
\end{dcases}
\label{eq:V}
\end{align}
for each $(x, y) \in \mathcal{X} \times \mathcal{Y}$, where the output alphabet is given by
\begin{align}
\mathcal{Y}
& =
\bigcup_{d|q} \frac{ \mathbb{Z} }{ d\mathbb{Z} }
=
\left\{ z + d \mathbb{Z} \ \middle|
\begin{array}{l}
z \in \mathbb{Z} , \\
\text{\emph{$d$ runs over all positive divisors of $q$}}
\end{array}\!\!
\right\} .
\label{def:alphabetY}
\end{align}
We denote this channel model as $\mathrm{MAEC}_{q}( \bvec{\varepsilon} )$.
\end{definition}

\begin{figure}[!t]
\centering
\begin{tikzpicture}
\node (x) at (0, 0) {$\mathbb{Z}/q\mathbb{Z} \ni X$};
\node (plus) at (3, 0) {$\bigoplus$};
\node (y) at (6, 0) {$Y = X + Z$};
\node (z) at (3, 2) {$Z$};
\draw [thick, -latex] (x) -- (plus);
\draw [thick, -latex] (z) -- (plus);
\draw [thick, -latex] (plus) -- (y);
\end{tikzpicture}
\caption{An interpretation of $\mathrm{MAEC}_{q}( \bvec{\varepsilon} )$ defined in \defref{def:V}.
The noise symbol $Z$ follows the probability law $\mathbb{P}\{ Z = d \mathbb{Z} \} = \varepsilon_{d}$ for each $d|q$, where $Z$ takes values in the set $\{ d \mathbb{Z} \mid \text{$d$ runs over all positive divisors of $q$} \}$.}
\label{fig:additive}
\end{figure}

The MAEC can be thought of as being similar to a channel with additive noise.
To wit, the input symbol is modeled by a random variable (r.v.) $X$ taking values in $\mathbb{Z}/q\mathbb{Z}$, and the noise symbol is modeled by a r.v.\ $Z$ taking values in $\{ d \mathbb{Z} \mid \text{$d$ divides $q$} \}$ with the probability law $\mathbb{P}\{ Z = d \mathbb{Z} \} = \varepsilon_{d}$ for each $d|q$.
Then, the output symbol is modeled by the r.v.\ $Y = X + Z$.
In this case, it can be verified that the conditional probability distribution $P_{Y|X}$ of $Y$ given $X$ is equal to the transition probability distribution $V_{\bvec{\varepsilon}}$ given in \eqref{eq:V}.
This observation implies that the input symbol $X$ is erased according to the modular arithmetic rule.
See \figref{fig:additive} for this interpretation.

It can be easily verified that every MAEC is Gallager-symmetric \cite[p.~94]{gallager_1968} (see also \cite[Definition~4]{ieeeit2018-01}), i.e., its channel capacity coincides with the symmetric capacity $I( V_{\bvec{\varepsilon}} )$ (cf. \cite[Theorem~4.5.2]{gallager_1968}).
In addition, note that $\mathrm{MAEC}_{q}( \bvec{\varepsilon} )$ is determined by the pair of an input alphabet size $q$ and a probability vector $\bvec{\varepsilon} = ( \varepsilon_{d} )_{d|q}$, so is its $\alpha$-symmetric capacity $I_{\alpha}( V_{\bvec{\varepsilon}} )$.
The following proposition provides formulas for the $\alpha$-symmetric capacity of an MAEC.

\begin{proposition}
\label{prop:I(V)}
For any probability vector $\bvec{\varepsilon} = ( \varepsilon_{d} )_{d|q}$, it holds that
\begin{align}
I_{\alpha}( V_{\bvec{\varepsilon}} )
& =
\begin{dcases}
\min_{d|q : \varepsilon_{d} > 0} \Big( \log d \Big)
& \mathrm{if} \ \alpha = 0 ,
\\
\sum_{d|q} (\log d) \, \varepsilon_{d}
& \mathrm{if} \ \alpha = 1 ,
\\
\log \left( \sum_{d|q} d \, \varepsilon_{d} \right)
& \mathrm{if} \ \alpha = \infty ,
\\
\frac{ \alpha }{ \alpha - 1 } \log \left( \sum_{d|q} d^{(\alpha-1)/\alpha} \, \varepsilon_{d} \right)
& \mathrm{otherwise}
\end{dcases}
\end{align}
for each $\alpha \in [0, \infty]$.
\end{proposition}

\begin{IEEEproof}[Proof of \propref{prop:I(V)}]
See \appref{app:proof_prop:I(V)}.
\end{IEEEproof}

\begin{remark}
\label{rem:I(V)}
By \remref{remark:connections_alpha_mutual} and \propref{prop:I(V)}, after some algebra, we observe that
\begin{align}
Z( V_{\bvec{\varepsilon}} )
& =
\frac{ 1 }{ q - 1 } \left( \sum_{d|q} \left( \frac{ q }{ d } \right) \, \varepsilon_{d} - 1 \right) ,
\\
P_{\mathrm{e}}( V_{\bvec{\varepsilon}} )
& =
1 - \sum_{d|q} \left( \frac{ d }{ q } \right) \, \varepsilon_{d} .
\end{align}
\end{remark}

The following theorem is our main result establishing recursive formulas of the polar transform for MAECs.

\begin{theorem}
\label{th:recursive_V}
Let $q \ge 2$ be an integer, $\gamma \in \mathbb{Z}/q\mathbb{Z}$ a unit of the ring, and $\bvec{\varepsilon} = ( \varepsilon_{d} )_{d|q}$ and $\bvec{\varepsilon}^{\prime} = ( \varepsilon_{d}^{\prime} )_{d|q}$ two probability vectors.
Then, it holds that
\begin{align}
\mathrm{MAEC}_{q}( \bvec{\varepsilon} ) \boxast \mathrm{MAEC}_{q}( \bvec{\varepsilon}^{\prime} )
& \equiv
\mathrm{MAEC}_{q}( \bvec{\varepsilon} \boxast \bvec{\varepsilon}^{\prime} ) ,
\label{eq:maec_minus_recursive} \\
\mathrm{MAEC}_{q}( \bvec{\varepsilon} ) \varoast \mathrm{MAEC}_{q}( \bvec{\varepsilon}^{\prime} )
& \equiv
\mathrm{MAEC}_{q}( \bvec{\varepsilon} \varoast \bvec{\varepsilon}^{\prime} ) ,
\label{eq:maec_plus_recursive} 
\end{align}
where two probability vectors $\bvec{\varepsilon} \boxast \bvec{\varepsilon}^{\prime} \coloneqq ( \varepsilon_{d}^{\boxast} )_{d|q}$ and $\bvec{\varepsilon} \varoast \bvec{\varepsilon}^{\prime} \coloneqq ( \varepsilon_{d}^{\varoast} )_{d|q}$ are given by
\begin{align}
\varepsilon_{d}^{\boxast}
& =
\varepsilon_{d}^{\boxast}( \bvec{\varepsilon}, \bvec{\varepsilon}^{\prime} )
\coloneqq
\sum_{\substack{ d_{1}|q, d_{2}|q : \\ \gcd(d_{1}, d_{2}) = d }} \varepsilon_{d_{1}} \, \varepsilon_{d_{2}}^{\prime} ,
\label{def:eps_minus}
\\
\varepsilon_{d}^{\varoast}
& =
\varepsilon_{d}^{\varoast}( \bvec{\varepsilon}, \bvec{\varepsilon}^{\prime} )
\coloneqq
\sum_{\substack{ d_{1}|q, d_{2}|q : \\ \lcm(d_{1}, d_{2}) = d }} \varepsilon_{d_{1}} \, \varepsilon_{d_{2}}^{\prime} ,
\label{def:eps_plus}
\end{align}
respectively, for each $d|q$.
\end{theorem}

\begin{IEEEproof}[Proof of \thref{th:recursive_V}]
See \appref{app:recursive_V}.
\end{IEEEproof}

It is worth mentioning that while the polar transform of a DMC depends on the unit $\gamma \in \mathbb{Z}/q\mathbb{Z}$ in general (see \cite{abbe_li_madiman_2017}), the statement of \thref{th:recursive_V} is independent of the choice of the unit $\gamma \in \mathbb{Z}/q\mathbb{Z}$.

\begin{remark}
An interesting observation from \thref{th:recursive_V} is that the recursive formulas stated in \eqref{def:eps_minus} and \eqref{def:eps_plus} are derived from the Chinese remainder theorem (see \appref{app:recursive_V} for details).
Namely, \thref{th:recursive_V} characterizes an algebraic structure of the polar transform over the ring $\mathbb{Z}/q\mathbb{Z}$.
More specifically, one of the key technical tools used in the proof of \thref{th:recursive_V} is the second isomorphism theorem of a group when the polar transform is defined by the group operation (see \cite{isit2019}).
\end{remark}

Combining \lemref{lem:invariant_degradedness} and \thref{th:recursive_V}, we readily obtain the following corollary.

\begin{corollary}
\label{cor:recursive_V}
Let $q \ge 2$ be an integer, $\gamma \in \mathbb{Z}/q\mathbb{Z}$ a unit of the ring, and $\bvec{\varepsilon} = ( \varepsilon_{d} )_{d|q}$ a probability vector.
Then, it holds that
\begin{align}
\mathrm{MAEC}_{q}( \bvec{\varepsilon} )^{\bvec{s}}
\equiv
\mathrm{MAEC}_{q}( \bvec{\varepsilon}^{\bvec{s}} )
\end{align}
for every $\bvec{s} \in \{ -, + \}^{\ast}$, where the probability vector $\bvec{\varepsilon}^{\bvec{s}} = ( \varepsilon_{d}^{\bvec{s}} )_{d|q}$ is recursively given by
\begin{align}
\left\{
\begin{array}{l}
\varepsilon_{d}^{\bvec{s}-}
=  \displaystyle
\sum_{\substack{ d_{1}|q, d_{2}|q : \\ \gcd(d_{1}, d_{2}) = d }} \varepsilon_{d_{1}}^{\bvec{s}} \varepsilon_{d_{2}}^{\bvec{s}} ,
\\[10pt]
\varepsilon_{d}^{\bvec{s}+}
= \displaystyle
\sum_{\substack{ d_{1}|q, d_{2}|q : \\ \lcm(d_{1}, d_{2}) = d }} \varepsilon_{d_{1}}^{\bvec{s}} \varepsilon_{d_{2}}^{\bvec{s}}
\end{array}
\right.
\label{def:eps_s}
\end{align}
for each $d|q$.
\end{corollary}

\subsection{Specializations to Binary Erasure Channels and Other Erasure-Like Channels}
\label{sect:bec}

This subsection considers the reduction of MAECs to known erasure-like channels.
Given an erasure probability $0 \le \varepsilon \le 1$, the BEC $W_{\mathrm{BEC}( \varepsilon )} : \mathbb{Z}/2\mathbb{Z} \to \mathcal{Y}$ can be defined by
\begin{align}
W_{\mathrm{BEC}(\varepsilon)}(y \mid x)
\coloneqq
\begin{cases}
1 - \varepsilon
& \mathrm{if} \ y = x ,
\\
\varepsilon
& \mathrm{if} \ y = \mathbb{Z} ,
\\
0
& \mathrm{otherwise}
\end{cases}
\label{def:bec}
\end{align}
for each $(x, y) \in \mathbb{Z}/2\mathbb{Z} \times \mathcal{Y}$, where the output alphabet is given as $\mathcal{Y} = (\mathbb{Z}/2\mathbb{Z}) \cup (\mathbb{Z}/\mathbb{Z}) = \{ \mathbb{Z}, 2\mathbb{Z}, 1 + 2\mathbb{Z} \}$.
This BEC is indeed equivalent to $\mathrm{MAEC}_{q}( \bvec{\varepsilon} )$ with $q = 2$ and $\bvec{\varepsilon} = (\varepsilon_{1}, \varepsilon_{2}) = (\varepsilon, 1 - \varepsilon)$.
Note that the erasure symbol of this BEC corresponds to $\mathbb{Z}$.
For the sake of brevity, we denote this channel model as $\mathrm{BEC}( \varepsilon )$.

Now, consider the polar transform as stated in \eqref{def:minus} and \eqref{def:plus} with $\gamma = 1 + 2\mathbb{Z}$.
As summarized in the following proposition, it is well-known that both synthetic channels $\mathrm{BEC}( \varepsilon ) \boxast \mathrm{BEC}( \varepsilon^{\prime} )$ and $\mathrm{BEC}( \varepsilon ) \varoast \mathrm{BEC}( \varepsilon^{\prime} )$ are equivalent to BECs with modified erasure probabilities.

\begin{proposition}[{\cite[Proposition~6]{arikan_it2009}; see also \cite[Corollary~1]{parizi_telatar_isit2013}}]
\label{prop:bec}
For any $0 \le \varepsilon, \varepsilon^{\prime} \le 1$, it holds that
\begin{align}
\mathrm{BEC}( \varepsilon ) \boxast \mathrm{BEC}( \varepsilon^{\prime} )
& \equiv
\mathrm{BEC}( \varepsilon + \varepsilon^{\prime} - \varepsilon \varepsilon^{\prime} ) ,
\label{eq:bec_minus} \\
\mathrm{BEC}( \varepsilon ) \varoast \mathrm{BEC}( \varepsilon^{\prime} )
& \equiv
\mathrm{BEC}( \varepsilon \varepsilon^{\prime} ) .
\label{eq:bec_plus}
\end{align}
\end{proposition}

It is clear that \thref{th:recursive_V} can be specialized to \propref{prop:bec}.
Analogously, \corref{cor:recursive_V} can be specialized to the following corollary.

\begin{corollary}
\label{cor:bec}
For each $\bvec{s} \in \{ -, + \}^{\ast}$ and each $0 \le \varepsilon \le 1$, it holds that
\begin{align}
\mathrm{BEC}( \varepsilon )^{\bvec{s}}
\equiv
\mathrm{BEC}( \varepsilon^{\bvec{s}} ) ,
\end{align}
where the erasure probability $0 \le \varepsilon^{\bvec{s}} \le 1$ can be recursively calculated by
\begin{align}
\left\{
\begin{array}{l}
\varepsilon^{\bvec{s}-}
=
2 \varepsilon^{\bvec{s}} - ( \varepsilon^{\bvec{s}} )^{2} ,
\\
\varepsilon^{\bvec{s}+}
=
( \varepsilon^{\bvec{s}} )^{2} .
\end{array}
\right.
\label{eq:recursive_bec}
\end{align}
\end{corollary}

By \corref{cor:bec}, to analyze the polar transform of a stationary BEC, it suffices to propagate its erasure probability by using the recursive formulas in \eqref{eq:recursive_bec} and to analyze the propagated erasure probabilities.
This is a well-known fact in the study of binary polar codes.
Moreover, we can verify from \corref{cor:bec} that for any fixed $0 < \delta < 1$,
\begin{align}
\lim_{n \to \infty} \frac{ 1 }{ 2^{n} } \Big| \Big\{ \bvec{s} \in \{ -, + \}^{n} \ \Big| \ I( W_{\mathrm{BEC}(\varepsilon)}^{\bvec{s}} ) > 1 - \delta \Big\} \Big|
& =
1 - \varepsilon ,
\label{eq:noiseless_bec} \\
\lim_{n \to \infty} \frac{ 1 }{ 2^{n} } \Big| \Big\{ \bvec{s} \in \{ -, + \}^{n} \ \Big| \ I( W_{\mathrm{BEC}(\varepsilon)}^{\bvec{s}} ) < \delta \Big\} \Big|
& =
\varepsilon .
\label{eq:useless_bec} 
\end{align}
These relations imply that the asymptotic distribution of two-level channel polarization for a BEC can be simply characterized by the initial erasure probability $\varepsilon$.

The following three examples introduce reductions of MAECs to other erasure-like channels.

\begin{example}[{$q$-ary erasure channels ($q$-ECs), see, e.g., \cite[p.~589]{mackay_2003}}]
Let $q \ge 2$ be an arbitrary integer.
Suppose that the probability vector $\bvec{\varepsilon} = ( \varepsilon_{d} )_{d|q}$ satisfies $\varepsilon_{1} + \varepsilon_{q} = 1$, i.e., $\varepsilon_{d} = 0$ for every $d|q$ in which $1 < d < q$.
Then,
\begin{align}
V_{\bvec{\varepsilon}}(y \mid x)
=
\begin{cases}
\varepsilon_{q}
& \mathrm{if} \ y = x ,
\\
\varepsilon_{1}
& \mathrm{if} \ y = \mathbb{Z} ,
\\
0
& \mathrm{otherwise}
\end{cases}
\end{align}
for each $(x, y) \in \mathbb{Z}/q\mathbb{Z} \times \mathcal{Y}$.
In this case, the output alphabet $\mathcal{Y}$ and the probability vector $\bvec{\varepsilon} = ( \varepsilon_{d} )_{d|q}$ can be simplified as $\mathcal{Y}^{\prime} = (\mathbb{Z}/q\mathbb{Z}) \cup (\mathbb{Z}/\mathbb{Z})$ and $\bvec{\varepsilon}^{\prime} = (\varepsilon_{1}, \varepsilon_{q})$, respectively.
The erasure symbol corresponds to $\mathbb{Z}$ as in the BEC.
\end{example}

\begin{example}[{ordered erasure channels (OECs) \cite[p.~2285]{park_barg_isit2011}}]
\label{ex:oec}
Let $q = p^{r}$ be a prime power.
Note that each divisor $d|q$ can be written by $d = p^{t}$ for some $0 \le t \le r$.
Given a probability vector $\bvec{\varepsilon} = ( \varepsilon_{1}, \varepsilon_{p}, \varepsilon_{p^{2}}, \dots, \varepsilon_{p^{r-1}}, \varepsilon_{p^{r}} )$, it holds that
\begin{align}
V_{\bvec{\varepsilon}}(y \mid x)
& =
\begin{cases}
\varepsilon_{p^{r}}
& \mathrm{if} \ y = x ,
\\
\varepsilon_{p^{r-1}}
& \mathrm{if} \ y = x + p^{r-1} \mathbb{Z} ,
\\
\vdots
& \vdots
\\
\varepsilon_{p}
& \mathrm{if} \ y = x + p \mathbb{Z} ,
\\
\varepsilon_{1}
& \mathrm{if} \ y = \mathbb{Z} ,
\\
0
& \mathrm{otherwise}
\end{cases}
\end{align}
for each $(x, y) \in \mathbb{Z}/q\mathbb{Z} \times \mathcal{Y}$.
Note that if $q = 4$, then this channel model is also equivalent to Sahebi and Pradhan's quaternary-input erasure-like channel \cite[Fig.~3: Channel~1]{sahebi_pradhan_it2013}.
\end{example}

\begin{example}[{Sahebi and Pradhan's senary-input erasure-like channel \cite[Fig.~4: Channel~2]{sahebi_pradhan_it2013}}]
\label{ex:sahebi_pradhan}
Consider the case in which $q = 6$.
Then, the output alphabet is given by $\mathcal{Y} = (\mathbb{Z}/\mathbb{Z}) \cup (\mathbb{Z}/2\mathbb{Z}) \cup (\mathbb{Z}/3\mathbb{Z}) \cup (\mathbb{Z}/6\mathbb{Z}) = \{ \mathbb{Z}, 2 \mathbb{Z}, 1 + 2 \mathbb{Z}, 3 \mathbb{Z}, 1 + 3 \mathbb{Z}, 2 + 3 \mathbb{Z}, 6 \mathbb{Z}, 1 + 6 \mathbb{Z}, 2 + 6 \mathbb{Z}, 3 + 6 \mathbb{Z}, 4 + 6 \mathbb{Z}, 5 + 6 \mathbb{Z} \}$, and the transition probability is given by
\begin{align}
V_{\bvec{\varepsilon}}(y \mid x)
=
\begin{cases}
\varepsilon_{6}
& \mathrm{if} \ y = x ,
\\
\varepsilon_{3}
& \mathrm{if} \ y = x + 3 \mathbb{Z} ,
\\
\varepsilon_{2}
& \mathrm{if} \ y = x + 2 \mathbb{Z} ,
\\
\varepsilon_{1}
& \mathrm{if} \ y = \mathbb{Z} ,
\\
0
& \mathrm{otherwise}
\end{cases}
\end{align}
for each $(x, y) \in \mathbb{Z}/6\mathbb{Z} \times \mathcal{Y}$.
\end{example}

The following example gives a special case of \thref{th:recursive_V} for the channel model given in \exref{ex:sahebi_pradhan}.

\begin{example}
\label{ex:recursive_6}
The minus channel $\mathrm{MAEC}_{6}( \varepsilon_{1}, \varepsilon_{2}, \varepsilon_{3}, \varepsilon_{6} )^{-}$ is equivalent to $\mathrm{MAEC}_{6}( \varepsilon_{1}^{-}, \varepsilon_{2}^{-}, \varepsilon_{3}^{-}, \varepsilon_{6}^{-} )$, where
\begin{align}
\left\{
\begin{array}{l}
\varepsilon_{6}^{-}
=
1 - \varepsilon_{1}^{-} - \varepsilon_{2}^{-} - \varepsilon_{3}^{-} ,
\\[5pt]
\varepsilon_{3}^{-}
=
2 \varepsilon_{3} - (\varepsilon_{3}^{2} + 2 \varepsilon_{1} \varepsilon_{3} + 2 \varepsilon_{2} \varepsilon_{3}) ,
\\[5pt]
\varepsilon_{2}^{-}
=
2 \varepsilon_{2} - (\varepsilon_{2}^{2} + 2 \varepsilon_{1} \varepsilon_{2} + 2 \varepsilon_{2} \varepsilon_{3}) ,
\\[5pt]
\varepsilon_{1}^{-}
=
2 \varepsilon_{1} + 2 \varepsilon_{2} \varepsilon_{3} - \varepsilon_{1}^{2} ,
\end{array}
\right.
\label{eq:recursive_6_minus}
\end{align}
the plus channel $\mathrm{MAEC}_{6}( \varepsilon_{1}, \varepsilon_{2}, \varepsilon_{3}, \varepsilon_{6} )^{+}$ is equivalent to $\mathrm{MAEC}_{6}( \varepsilon_{1}^{+}, \varepsilon_{2}^{+}, \varepsilon_{3}^{+}, \varepsilon_{6}^{+} )$, where
\begin{align}
\left\{
\begin{array}{l}
\varepsilon_{6}^{+}
=
1 - \varepsilon_{1}^{+} - \varepsilon_{2}^{+} - \varepsilon_{3}^{+} ,
\\[5pt]
\varepsilon_{3}^{+}
=
\varepsilon_{3}^{2} + 2 \varepsilon_{1} \varepsilon_{3} ,
\\[5pt]
\varepsilon_{2}^{+}
=
\varepsilon_{2}^{2} + 2 \varepsilon_{1} \varepsilon_{2} ,
\\[5pt]
\varepsilon_{1}^{+}
=
\varepsilon_{1}^{2} .
\end{array}
\right.
\label{eq:recursive_6_plus}
\end{align}
Note that \eqref{eq:recursive_6_minus} coincides with Sahebi and Pradhan's recursive formula \cite[Equation~(4)]{sahebi_pradhan_it2013} for the minus transform.
\end{example}

\section{Two Types of Channel Polarization}
\label{sect:two-notions}

We review two-level and multilevel channel polarization in the context of non-binary polar coding in Sections~\ref{sect:strong} and~\ref{sect:multilevel}, respectively.
These subsections can be omitted if readers are aware of these differences.
Some numerical simulations of multilevel channel polarization for MAECs are given in \sectref{sect:sim}.

\subsection{Two-level Channel Polarization}
\label{sect:strong}

When the input alphabet size $q$ is a prime number, \c{S}a\c{s}o\u{g}lu, Telatar, and Ar{\i}kan \cite{sasoglu_telatar_arikan_itw2009} showed that for any $q$-ary input DMC $W: \mathbb{Z}/q\mathbb{Z} \to \mathcal{Y}$ and any fixed $0 < \delta < 1$, both identities
\begin{align}
\lim_{n \to \infty} \frac{ 1 }{ 2^{n} } \Big| \Big\{ \bvec{s} \in \{ -, + \}^{n} \ \Big| \ I( W^{\bvec{s}} ) > 1 - \delta \Big\} \Big|
& =
I( W ) ,
\label{eq:strong1} \\
\lim_{n \to \infty} \frac{ 1 }{ 2^{n} } \Big| \Big\{ \bvec{s} \in \{ -, + \}^{n} \ \Big| \ I( W^{\bvec{s}} ) < \delta \Big\} \Big|
& =
1 - I( W )
\label{eq:strong2}
\end{align}
hold under the polar transform stated in \eqref{def:n-step} with $\gamma = 1 + q \mathbb{Z}$.
The left-hand sides of \eqref{eq:strong1} and \eqref{eq:strong2} are the limiting proportions of \emph{almost noiseless} and \emph{almost useless} synthetic channels, respectively.
Moreover, Equations \eqref{eq:strong1} and \eqref{eq:strong2} imply that the limiting proportion of \emph{mediocre} synthetic channels is zero, i.e.,
\begin{align}
\lim_{n \to \infty} \frac{ 1 }{ 2^{n} } \Big| \Big\{ \bvec{s} \in \{ -, + \}^{n} \ \Big| \ \delta \le I( W^{\bvec{s}} ) \le 1 - \delta \Big\} \Big|
= 0
\label{eq:strong}
\end{align}
for every fixed $0 < \delta < 1$.
In this paper, we call the phenomenon exhibited in \eqref{eq:strong} as \emph{two-level channel polarization}.
When the input alphabet size $q \ge 2$ is not necessarily a prime number, two-level polarization was investigated by \c{S}a\c{s}o\u{g}lu \cite{sasoglu_isit2012} and Mori and Tanaka \cite{mori_tanaka_it2014}.

\subsection{Multilevel Channel Polarization}
\label{sect:multilevel}

In contrast to \sectref{sect:strong}, when the input alphabet size $q$ is a composite number, there are polar transforms in which the two-level channel polarization stated in \eqref{eq:strong} does not hold in general (cf. \cite[Example~1]{sasoglu_isit2012}).
In this case, another polarization phenomenon called \emph{multilevel channel polarization} occurs.
Following \cite[Section~VI]{nasser_telatar_it2016}, we now introduce a more precise notion of multilevel channel polarization as follows:
Let $G$ be a finite group, and $N \lhd G$ a shorthand for a normal subgroup $N$ of a group $G$.
Given a DMC $W : G \to \mathcal{Y}$ and a normal subgroup $N \lhd G$, the \emph{homomorphism channel} $W[N] : G/N \to \mathcal{Y}$ is defined by
\begin{align}
W[N](y \mid a N)
\coloneqq
\frac{ 1 }{ |N| } \sum_{x \in a N} W(y \mid x) ,
\label{def:homomorphism}
\end{align}
where the quotient group of $G$ by $N \lhd G$ is denoted by $G/N$.
Then, Nasser and Telatar \cite[Theorem~6]{nasser_telatar_it2016} showed that%
\footnote{In \cite[Theorem~6]{nasser_telatar_it2016}, the rate of polarization for Bhattacharyya parameters is also shown.}
\begin{align}
\sum_{N \lhd G} \lim_{n \to \infty} \frac{ 1 }{ 2^{n} } \left| \left\{ \bvec{s} \in \{ -, + \}^{n} \, \middle| \!
\begin{array}{l}
| I(W^{\bvec{s}}) - \log [G : N] | < \delta ,
\\[5pt]
| I(W^{\bvec{s}}[N]) - \log [G : N] | < \delta
\end{array}\!\!\!
\right\} \right|
=
1
\label{eq:multilevel}
\end{align}
for fixed $\delta > 0$ small enough,%
\footnote{For example, it suffices to take small $\delta$ so that $\delta < \log |G| - \log (|G| - 1)$.}
where the synthetic channel $W^{\bvec{s}}$ is generated by a certain polar transform defined on a group $G$, and $[G : N] = |G/N|$ denotes the index of a normal subgroup $N$ in a group $G$.
Thus, the limiting proportions of mediocre (partially noiseless) synthetic channels are allowed to be positive in the context of multilevel channel polarization.

\begin{remark}
Notions of multilevel channel polarization have been independently introduced by several researchers \cite{park_barg_it2013, abbe_telatar_it2012, sahebi_pradhan_it2013, nasser_it2017_fourier, nasser_it2016_ergodic1, nasser_it2017_ergodic2, nasser_telatar_it2016, nasser_PhD, nasser_isit2019_level, nasser_isit2019_topology} in different forms.
In particular, formulations of multilevel channel polarization are more complicated if the polar transform is defined on a quasigroup \cite{nasser_telatar_it2016} or a weaker algebraic structure \cite{nasser_it2016_ergodic1, nasser_it2017_ergodic2, nasser_PhD}.
\end{remark}

We now consider each term in the sum of \eqref{eq:multilevel}.
It is clear that the left-hand sides of \eqref{eq:strong1} and \eqref{eq:strong2} coincide with the terms in the sum with the trivial subgroup $N = \{ e \}$ and with the whole group $N = G$, respectively, where $e$ stands for the identity element of $G$.
Thus, two-level channel polarization \eqref{eq:strong} is a special case of \eqref{eq:multilevel}.
Other terms in the sum refer to the limiting proportions of partially noiseless synthetic channels $W^{\bvec{s}}$.
Roughly speaking, the first condition
\begin{align}
\big| I(W^{\bvec{s}}[N]) - \log [G : N] \big| < \delta
\quad (\mathrm{for} \ \delta > 0 \ \mathrm{small} \ \mathrm{enough})
\label{eq:partially_noiseless}
\end{align}
implies that the homomorphism channel $W^{\bvec{s}}[N]$ is almost noiseless, and the second condition
\begin{align}
\big| I(W^{\bvec{s}}) - \log [G : N] \big| < \delta
\quad (\mathrm{for} \ \delta > 0 \ \mathrm{small} \ \mathrm{enough})
\end{align}
implies that the original synthetic channel $W^{\bvec{s}}$ has almost the same symmetric capacity as $W^{\bvec{s}}[N]$.
These observations give us intuition as to why polar codes can achieve the symmetric capacity with multilevel channel polarization.

While the limiting proportions stated in the left-hand sides of \eqref{eq:strong1} and \eqref{eq:strong2} are fully and simply characterized by the symmetric capacity $I(W)$, the exact characterization of each term in the sum of \eqref{eq:multilevel} remains an open problem (see \cite[Section~9.2.1]{nasser_PhD}).
Recently, Nasser \cite{nasser_isit2019_level} showed that a term in the sum of \eqref{eq:multilevel} is positive only if $N$ is a characteristic subgroup, provided that $G$ is abelian and $W$ is \emph{automorphic-symmetric.}
While every DMC $W : \mathbb{Z}/q\mathbb{Z} \to \mathcal{Y}$ is automorphic-symmetric (see \cite[Example~3]{nasser_isit2019_level}), since every subgroup of the cyclic group $(\mathbb{Z}/q\mathbb{Z}, +)$ is characteristic, the results in \cite{nasser_isit2019_level} are insufficient to characterize the polarization levels induced by the polar transform stated in \eqref{def:minus} and \eqref{def:plus}.

\subsection{Simulations of Multilevel Channel Polarization for Modular Arithmetic Erasure Channels}
\label{sect:sim}

\begin{figure*}[!t]
\centering
\subfloat[input alphabet size $q = 6$ (case 1) \label{subfig:maec_6_case1}]{
\begin{overpic}[width = 0.45\hsize, clip]{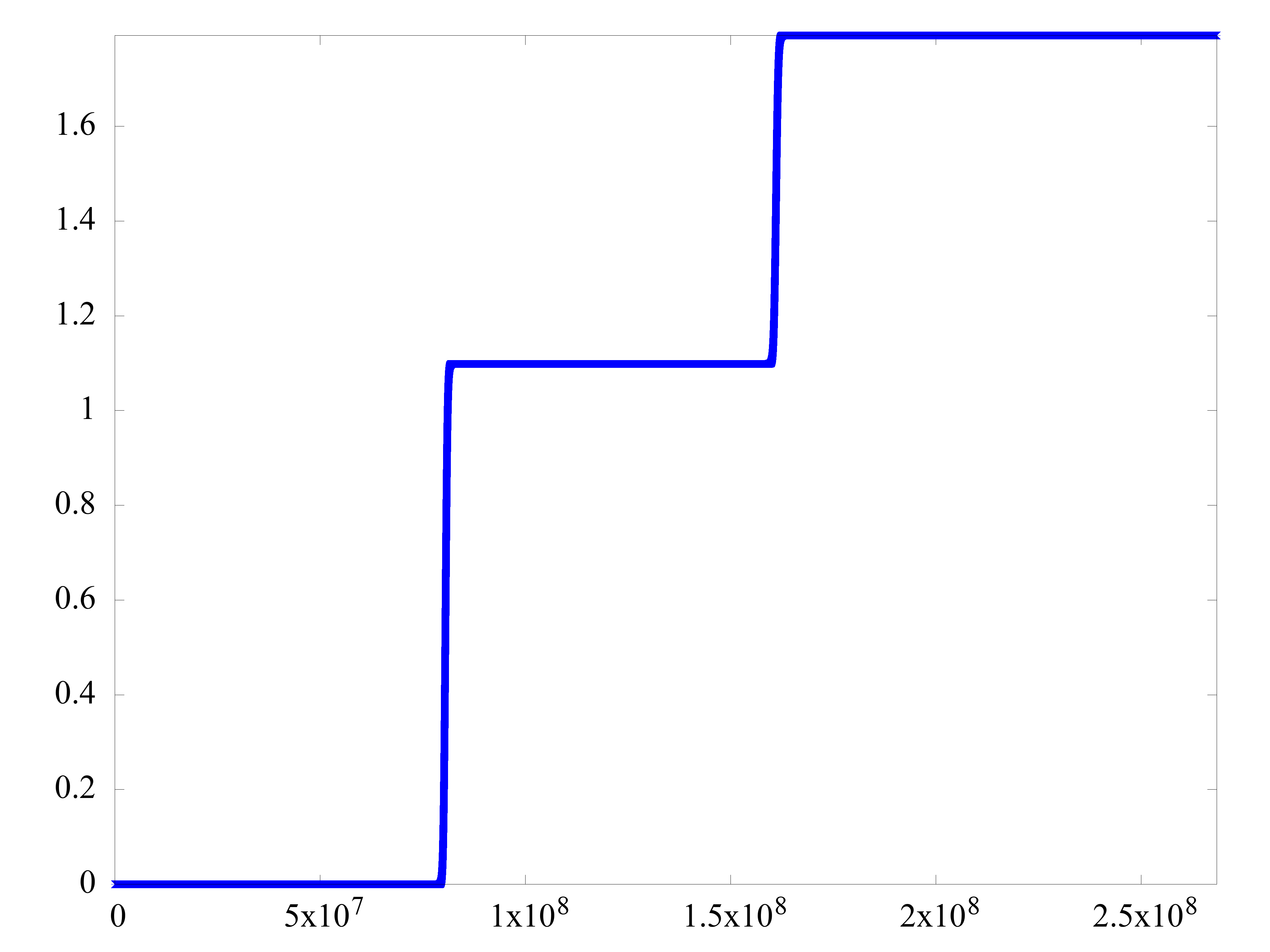}
\put(13, -1){\footnotesize indices of $\bvec{s}$ (sorted in increasing order of $I( V_{\bvec{\varepsilon}}^{\bvec{s}} )$)}
\put(-1, 16){\rotatebox{90}{\footnotesize symmetric capacity $I( V_{\bvec{\varepsilon}}^{\bvec{s}} )$}}
\put(0, 70){\footnotesize [nats]}
\end{overpic}
}
\hfill
\subfloat[input alphabet size $q = 6$ (case 2) \label{subfig:maec_6_case2}]{
\begin{overpic}[width = 0.45\hsize, clip]{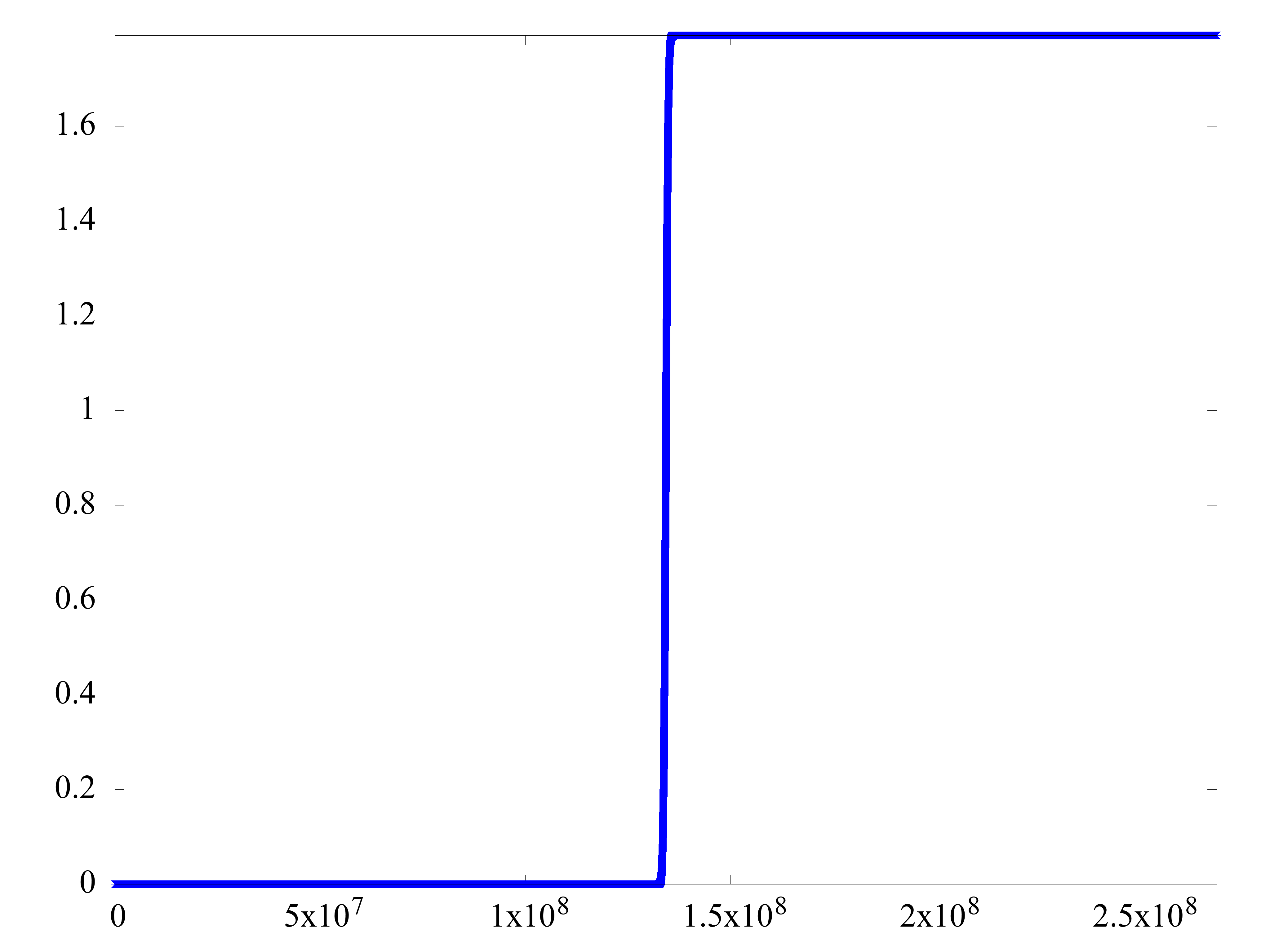}
\put(13, -1){\footnotesize indices of $\bvec{s}$ (sorted in increasing order of $I( V_{\bvec{\varepsilon}}^{\bvec{s}} )$)}
\put(-1, 16){\rotatebox{90}{\footnotesize symmetric capacity $I( V_{\bvec{\varepsilon}}^{\bvec{s}} )$}}
\put(0, 70){\footnotesize [nats]}
\end{overpic}
}
\\
\subfloat[input alphabet size $q = 45$ \label{subfig:maec_45}]{
\begin{overpic}[width = 0.45\hsize, clip]{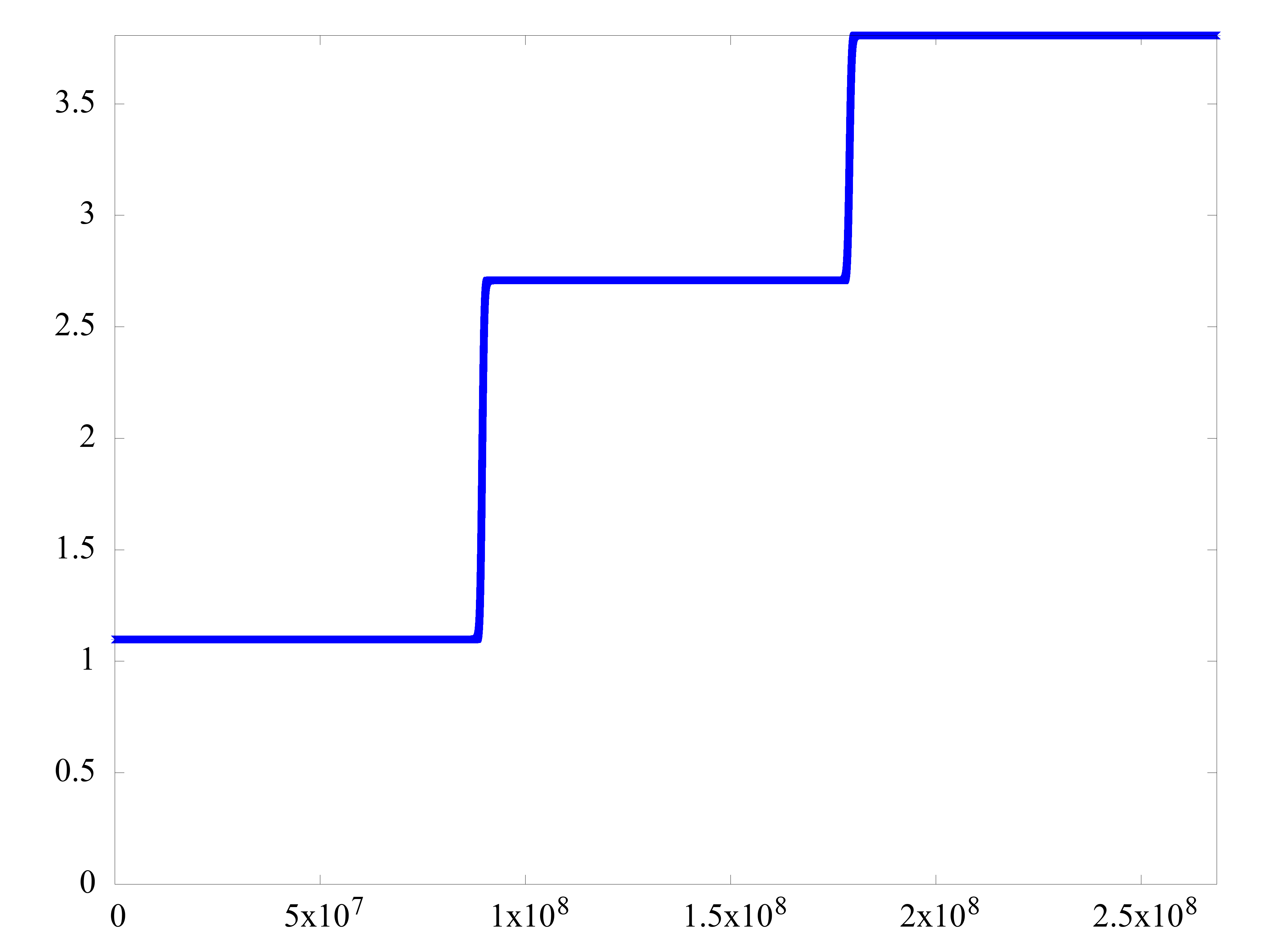}
\put(13, -1){\footnotesize indices of $\bvec{s}$ (sorted in increasing order of $I( V_{\bvec{\varepsilon}}^{\bvec{s}} )$)}
\put(-1, 16){\rotatebox{90}{\footnotesize symmetric capacity $I( V_{\bvec{\varepsilon}}^{\bvec{s}} )$}}
\put(0, 70){\footnotesize [nats]}
\end{overpic}
}
\hfill
\subfloat[input alphabet size $q = 512 \ (= 2^{9})$ \label{subfig:maec_512}]{
\begin{overpic}[width = 0.45\hsize, clip]{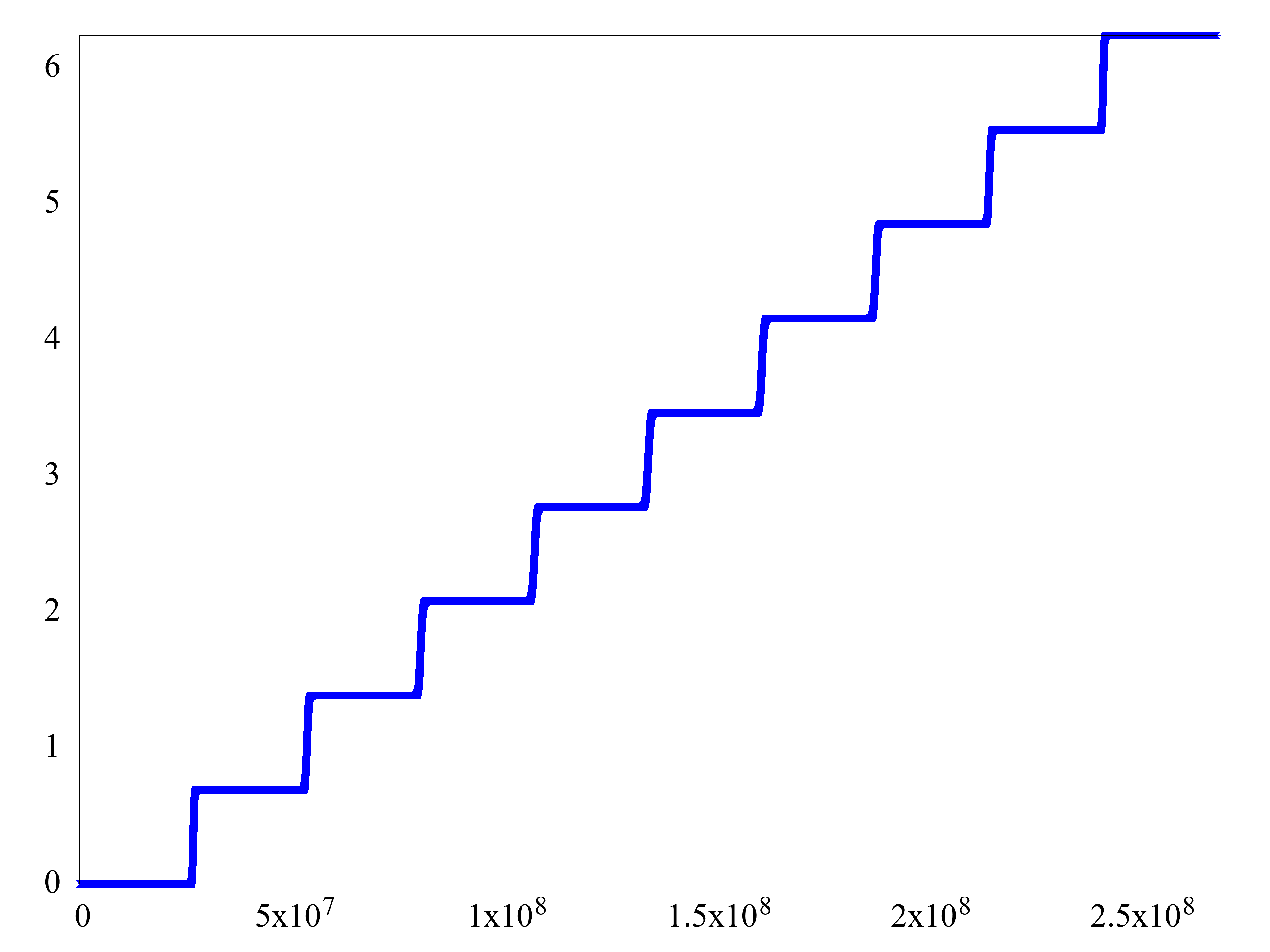}
\put(13, -1){\footnotesize indices of $\bvec{s}$ (sorted in increasing order of $I( V_{\bvec{\varepsilon}}^{\bvec{s}} )$)}
\put(-1, 16){\rotatebox{90}{\footnotesize symmetric capacity $I( V_{\bvec{\varepsilon}}^{\bvec{s}} )$}}
\put(-6, 70){\footnotesize [nats]}
\end{overpic}
}
\caption{Plots of the symmetric capacities $\{ I( V_{\bvec{\varepsilon}}^{\bvec{s}} ) \mid \bvec{s} \in \{ -, + \}^{n} \}$ under $n=28$-steps of the polar transform for several MAECs defined in \defref{def:V}.
Each initial probability vector $\bvec{\varepsilon} = ( \varepsilon_{d} )_{d|q}$ is given as follows:
(a) $\bvec{\varepsilon} = ( \varepsilon_{1}, \varepsilon_{2}, \varepsilon_{3}, \varepsilon_{6} ) = (0, 3/10, 3/5, 1/10)$,
(b) $\bvec{\varepsilon} = ( \varepsilon_{1}, \varepsilon_{2}, \varepsilon_{3}, \varepsilon_{6} ) = (1/4, 1/4, 1/4, 1/4)$,
(c) $\bvec{\varepsilon} = ( \varepsilon_{1}, \varepsilon_{3}, \varepsilon_{5}, \varepsilon_{9}, \varepsilon_{15}, \varepsilon_{45} ) = ( 0, 0, 0, 1/3, 2/3, 0 )$, and
(d) $\bvec{\varepsilon} = (\varepsilon_{1}, \varepsilon_{2}, \varepsilon_{4}, \varepsilon_{8}, \varepsilon_{16}, \varepsilon_{32}, \varepsilon_{64}, \varepsilon_{128}, \varepsilon_{256}, \varepsilon_{512} ) = (1/10, 1/10, \dots, 1/10 )$.
Note that \figref{subfig:maec_512} is analogous to that of \cite[Fig.~2]{park_barg_it2013} with different $n$ (see also \exref{ex:oec}).}
\label{fig:example_recursive_formulas}
\end{figure*}

Consider the synthetic channels $V_{\bvec{\varepsilon}}^{\bvec{s}}$ induced by the polar transform for an MAEC $V_{\bvec{\varepsilon}}$.
It follows from \lemref{lem:degraded_mutual}, \propref{prop:I(V)}, and \corref{cor:recursive_V} that it suffices to propagate the probability vector $\bvec{\varepsilon}^{\bvec{s}}$ by the recursive formulas given in \eqref{def:eps_s} for calculating the $\alpha$-symmetric capacity $I_{\alpha}( V_{\bvec{\varepsilon}}^{\bvec{s}} )$.
Some numerical examples are plotted in \figref{fig:example_recursive_formulas}, which illustrates the symmetric capacities $I( V_{\bvec{\varepsilon}}^{\bvec{s}} )$ of the synthetic channels $V_{\bvec{\varepsilon}}^{\bvec{s}}$.
From \figref{fig:example_recursive_formulas}, we may naturally conjecture that the polar transform for MAECs induce multilevel channel polarization.
On the other hand, whereas \figref{subfig:maec_6_case1} appears to depict the multilevel channel polarization phenomenon with $q=6$, Figure~\ref{subfig:maec_6_case2} seems to imply that two-level channel polarization occurs with $q=6$.
Moreover, in \figref{subfig:maec_45}, we may conjecture that the limiting proportion of almost useless synthetic channels $V_{\bvec{\varepsilon}}^{\bvec{s}}$ (i.e., $I( V_{\bvec{\varepsilon}}^{\bvec{s}} ) \approx 0$) approaches zero as $n \to \infty$, while the limiting proportion of almost noiseless synthetic channels $V_{\bvec{\varepsilon}}^{\bvec{s}}$ (i.e., $I( V_{\bvec{\varepsilon}}^{\bvec{s}} ) \approx 1$) does not approach one as $n \to \infty$.
These questions are completely solved in the next section.

\section{Asymptotic Distributions of Multilevel Channel Polarization}
\label{sect:asymptotic_distribution_MAEC}

Let $q$ be an integer, and $\bvec{\varepsilon} = ( \varepsilon_{d} )_{d|q}$ an arbitrary probability vector.
In this section, we characterize the asymptotic distribution of multilevel channel polarization for $\mathrm{MAEC}_{q}( \bvec{\varepsilon} )$ defined in \defref{def:V}.
Define
\begin{align}
\mu_{d}^{(n)}
\coloneqq
\frac{ 1 }{ 2^{n} } \sum_{\bvec{s} \in \{ -, + \}^{n}} \varepsilon_{d}^{\bvec{s}}
\label{def:mu_d}
\end{align}
for each $d|q$ and $n \in \mathbb{N}$, where $\varepsilon_{d}^{\bvec{s}}$ can be calculated by \eqref{def:eps_s}.
Note that $\bvec{\varepsilon}^{\bvec{s}} = ( \varepsilon_{d}^{\bvec{s}} )_{d|q}$ is a probability vector for every $\bvec{s} \in \{ -, + \}^{\ast}$, and so is the vector $( \mu_{d}^{(n)} )_{d|q}$ for every $n \in \mathbb{N}$.
In addition, we define
\begin{align}
\mu_{d}^{(\infty)}
\coloneqq
\lim_{n \to \infty} \mu_{d}^{(n)}
\label{def:mu_d_infty}
\end{align}
for each $d|q$, provided that the limit exists.
As will be shown later, the limit $\mu_{d}^{(\infty)}$ exists for every $d|q$, and the probability vector $( \mu_{d}^{(\infty)} )_{d|q}$ coincides with the desired asymptotic distribution.
We summarize this fact in the following corollary.

\begin{corollary}
\label{cor:multilevel}
Let $q \ge 2$ be an integer, and $\bvec{\varepsilon} = ( \varepsilon_{d} )_{d|q}$ a probability vector.
For any fixed $0 < \delta < \log(q/(q-1))$, it holds that
\begin{align}
\!\!
\frac{ 1 }{ 2^{n} } \left| \left\{ \bvec{s} \in \{ -, + \}^{n} \, \middle| \!
\begin{array}{l}
|I(V_{\bvec{\varepsilon}}^{\bvec{s}}) - \log d | < \delta ,
\\[5pt]
| I(V_{\bvec{\varepsilon}}^{\bvec{s}}[\ker \varphi_{d}]) - \log d | < \delta
\end{array}\!\!\!
\right\} \right|
\to
\mu_{d}^{(\infty)}
\label{eq:asymptotic_distribution}
\end{align}
as $n \to \infty$ for every $d|q$,
where $V_{\bvec{\varepsilon}}^{\bvec{s}}[\ker \varphi_{d}]$ denotes the homomorphism channel of $V_{\bvec{\varepsilon}}^{\bvec{s}}$ defined as in \eqref{def:homomorphism}, the function $\varphi_{d} : x \mapsto (x + d\mathbb{Z})$ denotes the natural projection, and $\ker \varphi_{d} \coloneqq \{ x \in \mathbb{Z}/q\mathbb{Z} \mid \varphi_{d}( x ) = d \mathbb{Z} \}$ denotes the kernel of $\varphi_{d}$.
\end{corollary}

\corref{cor:multilevel} is a direct consequence of \thref{th:polarization} that will be stated in \sectref{sect:asymptotic_distribution}.
A formal proof of \corref{cor:multilevel} is given in \appref{app:multilevel}.
It follows from \corref{cor:multilevel} that
\begin{align}
\sum_{d|q} \lim_{n \to \infty} \frac{ 1 }{ 2^{n} } \left| \left\{ \bvec{s} \in \{ -, + \}^{n} \, \middle| \!
\begin{array}{l}
|I(V_{\bvec{\varepsilon}}^{\bvec{s}}) - \log d | < \delta ,
\\[5pt]
| I(V_{\bvec{\varepsilon}}^{\bvec{s}}[\ker \varphi_{d}]) - \log d | < \delta
\end{array}\!\!\!
\right\} \right|
=
1 ,
\label{eq:multilevel_sum_V}
\end{align}
which is an analogue of \eqref{eq:multilevel}.
Therefore, \corref{cor:multilevel} characterizes each term in the sum of \eqref{eq:multilevel} for every MAEC.
Based on \corref{cor:multilevel}, we regard the probability vector $( \mu_{d}^{(\infty)} )_{d|q}$ as the \emph{asymptotic distribution} of multilevel channel polarization for $\mathrm{MAEC}_{q}( \bvec{\varepsilon} )$.
An algorithm of calculating the asymptotic distribution $( \mu_{d}^{(\infty)} )_{d|q}$ will be described in \thref{th:mu_d} of \sectref{sect:composite}.

\subsection{Special Case: The Input Alphabet Size $q = p^{r}$ is a Prime Power}
\label{sect:primepower}

Let $q = p^{r}$ be a prime power for some prime number $p$ and some positive integer $r$.
Note that in this case, an MAEC is equivalent to an OEC (see \exref{ex:oec}), and the probability vector $\bvec{\varepsilon} = ( \varepsilon_{d} )_{d|q}$ can be written as $\bvec{\varepsilon} = ( \varepsilon_{p^{i}} )_{i = 0}^{r}$.

\begin{proposition}
\label{prop:primepower_conservation}
Let $q$ be a prime power.
For any probability vectors $\bvec{\varepsilon} = ( \varepsilon_{d} )_{d|q}$ and $\bvec{\varepsilon}^{\prime} = ( \varepsilon_{d}^{\prime} )_{d|q}$, it holds that
\begin{align}
\varepsilon_{d}^{\boxast} + \varepsilon_{d}^{\varoast}
=
\varepsilon_{d} + \varepsilon_{d}^{\prime}
\label{eq:primepower_conservation}
\end{align}
for every $d|q$, where $\varepsilon_{d}^{\boxast}$ and $\varepsilon_{d}^{\varoast}$ are defined in \eqref{def:eps_minus} and \eqref{def:eps_plus}, respectively.
\end{proposition}

\begin{IEEEproof}[Proof of \propref{prop:primepower_conservation}]
See \appref{app:primepower_conservation}.
\end{IEEEproof}

If $\bvec{\varepsilon}$ and $\bvec{\varepsilon}^{\prime}$ are the same, then \propref{prop:primepower_conservation} can be readily specialized to the identity
\begin{align}
\frac{ 1 }{ 2 } \Big( \varepsilon_{d}^{\bvec{s}-} + \varepsilon_{d}^{\bvec{s}+} \Big)
& =
\varepsilon_{d}^{\bvec{s}}
\label{eq:martingale_primepower}
\end{align}
for every $d|q$ and every $\bvec{s} \in \{ -, + \}^{\ast}$, which can be thought of as a martingale-like property%
\footnote{Strictly speaking, when we consider the pass of polar transform $\bvec{s} \in \{ -, + \}^{n}$ as a sequence of independent and uniformly distributed Bernoulli r.v.'s $B_{1}, \dots, B_{n}$, we may think of \eqref{eq:martingale_primepower} as a martingale property.
See \cite[Section~IV]{arikan_it2009} for details of the polarization process.}
of the recursive formulas stated in \eqref{def:eps_s} with respect to the polarization process.
Indeed, we observe from \eqref{eq:martingale_primepower} that
\begin{align}
\mu_{d}^{(n)}
=
\varepsilon_{d}
\label{eq:primepower_conservation_mu_d}
\end{align}
for every $d|q$ and every $n \in \mathbb{N}$, where $\mu_{d}^{(n)}$ is defined in \eqref{def:mu_d}.
Equation~\eqref{eq:primepower_conservation_mu_d} implies the following theorem.

\begin{theorem}
\label{th:primepower}
If $q$ is a prime power, then
$
\mu_{d}^{(\infty)}
=
\varepsilon_{d}
$
for every $d|q$.
\end{theorem}

Therefore, the asymptotic distribution $( \mu_{d}^{(\infty)} )_{d|q}$ coincides with an initial probability vector $\bvec{\varepsilon} = ( \varepsilon_{d} )_{d|q}$, provided that $q = p^{r}$ is a prime power.
Hence, we can verify that the asymptotic distribution of \figref{subfig:maec_512} is given by $\mu_{d}^{(\infty)} = 1/10$ for every $d|q$.

In the following, we give another proof of \thref{th:primepower}.
This alternative proof can be considered as a digression of our discussion.
It gives us, however, some ideas to solve for the asymptotic distribution $( \mu_{d}^{(\infty)} )_{d|q}$ when $q$ is not a prime power.

For each integer $a \ge 1$ and each sequence $\bvec{s} \in \{ -, + \}^{\ast}$, we define
\begin{align}
T^{\bvec{s}}( a )
& \coloneqq
\sum_{i = a}^{r} \varepsilon_{p^{i}}^{\bvec{s}} ,
\label{def:T} \\
B^{\bvec{s}}( a )
& \coloneqq
\sum_{i = 0}^{a-1} \varepsilon_{p^{i}}^{\bvec{s}} ,
\label{def:B}
\end{align}
where $\bvec{\varepsilon}^{\bvec{s}} = ( \varepsilon_{d}^{\bvec{s}} )_{d|q} = ( \varepsilon_{p^{i}} )_{i = 0}^{r}$ is recursively defined in \eqref{def:eps_s}.
If the sequence $\bvec{s} = \varnothing$ is empty, then we omit the superscripts $\bvec{s}$ in $T^{\bvec{s}}( a )$ and $B^{\bvec{s}}( a )$ and denote these quantities respectively as $T( a )$ and $B( a )$.
Clearly, it holds that
\begin{align}
T^{\bvec{s}}( a ) + B^{\bvec{s}}( a )
=
\sum_{i = 0}^{r} \varepsilon_{p^{i}}^{\bvec{s}}
=
1
\label{eq:sum_TB}
\end{align}
for each $a \ge 1$ and each $\bvec{s} \in \{ -, + \}^{\ast}$.

\begin{lemma}
\label{lem:recursive_TB}
For each integer $a \ge 1$ and each sequence $\bvec{s} \in \{ -, + \}^{\ast}$, it holds that
\begin{align}
T^{\bvec{s}-}( a )
& =
T^{\bvec{s}}( a )^{2} ,
\label{eq:T_minus} \\
B^{\bvec{s}-}( a )
& =
2 \, B^{\bvec{s}}( a ) \, T^{\bvec{s}}( a ) + B^{\bvec{s}}( a )^{2} ,
\label{eq:B_minus} \\
T^{\bvec{s}+}( a )
& =
2 \, B^{\bvec{s}}( a ) \, T^{\bvec{s}}( a ) + T^{\bvec{s}}( a )^{2} ,
\\
B^{\bvec{s}+}( a )
& =
B^{\bvec{s}}( a )^{2} .
\end{align}
\end{lemma}

\begin{IEEEproof}[Proof of \lemref{lem:recursive_TB}]
See \appref{app:recursive_TB}.
\end{IEEEproof}

One can see from \lemref{lem:recursive_TB} that the pair of partial sums $T^{\bvec{s}}( a )$ and $B^{\bvec{s}}( a )$ behave similarly to the polar transform for BECs (see \sectref{sect:bec}).
The following lemma is a straightforward consequence of \lemref{lem:recursive_TB}.

\begin{lemma}
\label{lem:martingale_primepower}
For each integer $a \ge 1$ and each sequence $\bvec{s} \in \{ -, + \}^{\ast}$, it holds that
\begin{align}
\frac{ 1 }{ 2 } \Big( T^{\bvec{s}-}( a ) + T^{\bvec{s}+}( a ) \Big)
& =
T^{\bvec{s}}( a ) ,
\\
\frac{ 1 }{ 2 } \Big( B^{\bvec{s}-}( a ) + B^{\bvec{s}+}( a ) \Big)
& =
B^{\bvec{s}}( a ) .
\end{align}
Consequently, it holds that
\begin{align}
\frac{ 1 }{ 2^{n} } \sum_{\bvec{s} \in \{ -, + \}^{n}} T^{\bvec{s}}( a )
& =
T( a ) ,
\\
\frac{ 1 }{ 2^{n} } \sum_{\bvec{s} \in \{ -, + \}^{n}} B^{\bvec{s}}( a )
& =
B( a )
\end{align}
for every integers $n \ge 1$ and $a \ge 1$.
\end{lemma}

\lemref{lem:martingale_primepower} presents a martingale-like property for two partial sums $T^{\bvec{s}}( a )$ and $B^{\bvec{s}}( a )$ with respect to the polarization process.
Employing this martingale-like property, we can give an alternative proof of \thref{th:primepower} by induction.

\begin{IEEEproof}[Alternative Proof of \thref{th:primepower}]
As a counterpart of \eqref{eq:primepower_conservation_mu_d}, it suffices to verify that
\begin{align}
\mu_{p^{i}}^{(n)}
=
\varepsilon_{p^{i}}
\label{eq:induction_hypothesis}
\end{align}
for every $n \in \mathbb{N}$ and every $i = 0, 1, \dots, r$.
We prove \eqref{eq:induction_hypothesis} by induction.
We observe that
\begin{align}
\mu_{1}^{(n)}
& =
\frac{ 1 }{ 2^{n} } \sum_{\bvec{s} \in \{ -, + \}^{n}} \varepsilon_{1}^{\bvec{s}}
\notag \\
& =
\frac{ 1 }{ 2^{n} } \sum_{\bvec{s} \in \{ -, + \}^{n}} B^{\bvec{s}}( 1 )
\notag \\
& \overset{\mathclap{\text{(a)}}}{=}
B( 1 )
\notag \\
& =
\varepsilon_{1}
\end{align}
for every $n \in \mathbb{N}$, where (a) follows from \lemref{lem:martingale_primepower}.
This implies \eqref{eq:induction_hypothesis} with $i = 0$.
Let $0 \le k < r$ be an integer.
Suppose that \eqref{eq:induction_hypothesis} holds for every $n \in \mathbb{N}$ and every $i = 0, 1, \dots, k$.
Then, we have
\begin{align}
\mu_{p^{k+1}}^{(n)}
& =
\frac{ 1 }{ 2^{n} } \sum_{\bvec{s} \in \{ -, + \}^{n}} \varepsilon_{p^{k+1}}^{\bvec{s}}
\notag \\
& =
\frac{ 1 }{ 2^{n} } \sum_{\bvec{s} \in \{ -, + \}^{n}} \sum_{i = 0}^{k+1} \varepsilon_{p^{k+1}}^{\bvec{s}} - \sum_{j = 0}^{k} \varepsilon_{p^{j}}
\notag \\
& =
\frac{ 1 }{ 2^{n} } \sum_{\bvec{s} \in \{ -, + \}^{n}} B^{\bvec{s}}( k+1 ) - \sum_{j = 0}^{k} \varepsilon_{p^{j}}
\notag \\
& \overset{\mathclap{\text{(a)}}}{=}
B( k+1 ) - \sum_{j = 0}^{k} \varepsilon_{p^{j}}
\notag \\
& =
\sum_{i = 0}^{k+1} \varepsilon_{p^{i}} - \sum_{j = 0}^{k} \varepsilon_{p^{j}}
\notag \\
& =
\varepsilon_{p^{k+1}}
\end{align}
for every $n \in \mathbb{N}$, where (a) follows from \lemref{lem:martingale_primepower}.
This completes the proof of \thref{th:primepower}.
\end{IEEEproof}

Even if $q$ is not a prime factor, we can deduce some martingale-like properties in the probability vector $( \varepsilon_{d}^{\bvec{s}} )_{d|q}$ by considering \emph{four} partial sums of $( \varepsilon_{d}^{\bvec{s}} )_{d|q}$, instead of $T^{\bvec{s}}( \cdot )$ and $B^{\bvec{s}}( \cdot )$.
Such martingale-like properties as well as the above alternative proof are useful to solve the asymptotic distribution $( \mu_{d}^{(\infty)} )_{d|q}$.
In the next subsection, we define four such partial sums and explore these properties.

\subsection{General Case: The Input Alphabet Size $q = p_{1}^{r_{1}} p_{2}^{r_{2}} \cdots p_{m}^{r_{m}}$ is a Composite Number}
\label{sect:composite}

Henceforth, assume that the input alphabet size $q$ can be factorized as%
\footnote{Even if $q$ has only one prime factor $q = p_{1}^{r_{1}}$, in this subsection, we write $q = p_{1}^{r_{1}} p_{2}^{r_{2}} \cdots p_{m}^{r_{m}}$ for some $m \ge 2$ by setting $r_{2} = \cdots = r_{m} = 0$.
Doing so, the analyses in \sectref{sect:composite} can specialize to the case where $q$ is a prime power.}
$q = p_{1}^{r_{1}} p_{2}^{r_{2}} \cdots p_{m}^{r_{m}}$ with $m$ distinct prime factors $p_{1}^{r_{1}}$, $p_{2}^{r_{2}}$, $\ldots$, and $p_{m}^{r_{m}}$.
If a positive integer $d$ of $q$ can be factorized by $d = p_{1}^{t_{1}} p_{2}^{t_{2}} \cdots p_{m}^{t_{m}}$, then we write it as $d = \langle \bvec{t} \rangle$ for the sake of brevity, where $\bvec{t} = (t_{1}, t_{2}, \dots, t_{m})$.
Namely, defining a partial order $\bvec{t} \le \bvec{u}$ between two $m$-tuples $\bvec{t}$ and $\bvec{u}$ by $t_{i} \le u_{i}$ for every $i = 1, 2, \dots, m$, we observe that $d$ divides $q$ if and only if $\bvec{0} \le \bvec{t} \le \bvec{r}$ with $d = \langle \bvec{t} \rangle$ and $q = \langle \bvec{r} \rangle$, where $\bvec{0} = (0, \dots, 0)$ denotes the all zeros vector.
As in \eqref{def:T} and \eqref{def:B}, the key idea of our analyses is that for each pair of integers $i$ and $j$ satisfying $1 \le i < j \le m$, we combine the probability masses $( \varepsilon_{\langle \bvec{t} \rangle}^{\bvec{s}} )_{\bvec{0} \le \bvec{t} \le \bvec{r}}$ into the following four quantities:%
\footnote{Note that \eqref{def:theta}--\eqref{def:beta} are well-defined even if $0 \le j < i \le r$, and it holds that $\lambda_{i, j}(a, b) = \rho_{j, i}(b, a)$.}%
\begin{align}
\theta_{i, j}^{\bvec{s}}(a, b)
& \coloneqq
\sum_{\substack{ \bvec{t} : \bvec{0} \le \bvec{t} \le \bvec{r} , \\ t_{i} \ge a, t_{j} \ge b } } \varepsilon_{\langle \bvec{t} \rangle}^{\bvec{s}} ,
\label{def:theta} \\
\lambda_{i, j}^{\bvec{s}}(a, b)
& \coloneqq
\sum_{\substack{ \bvec{t} : \bvec{0} \le \bvec{t} \le \bvec{r} , \\ t_{i} \ge a, t_{j} < b } }  \varepsilon_{\langle \bvec{t} \rangle}^{\bvec{s}} ,
\label{def:lambda} \\
\rho_{i, j}^{\bvec{s}}(a, b)
& \coloneqq
\sum_{\substack{ \bvec{t} : \bvec{0} \le \bvec{t} \le \bvec{r} , \\ t_{i} < a, t_{j} \ge b } }  \varepsilon_{\langle \bvec{t} \rangle}^{\bvec{s}} ,
\label{def:rho} \\
\beta_{i, j}^{\bvec{s}}(a, b)
& \coloneqq
\sum_{\substack{ \bvec{t} : \bvec{0} \le \bvec{t} \le \bvec{r} , \\ t_{i} < a, t_{j} < b } }  \varepsilon_{\langle \bvec{t} \rangle}^{\bvec{s}} ,
\label{def:beta} 
\end{align}
where $a, b \ge 1$ are integers, $\bvec{s} \in \{ -, + \}^{\ast}$ a sequence, and $\bvec{\varepsilon}^{\bvec{s}} = ( \varepsilon_{d}^{\bvec{s}} )_{d|q} = ( \varepsilon_{\langle \bvec{t} \rangle}^{\bvec{s}} )_{\bvec{0} \le \bvec{t} \le \bvec{r}}$ is recursively defined in \eqref{def:eps_s} with an initial probability vector $\bvec{\varepsilon} = ( \varepsilon_{d} )_{d|q}$.
If the sequence $\bvec{s} = \varnothing$ is empty, then we omit the superscripts $\bvec{s}$ in the notations in \eqref{def:theta}--\eqref{def:beta} and write them as $\theta_{i, j}(a, b)$, $\lambda_{i, j}(a, b)$, $\rho_{i, j}(a, b)$, and $\beta_{i, j}(a, b)$.
Note that
\begin{align}
\theta_{i, j}^{\bvec{s}}(a, b) + \lambda_{i, j}^{\bvec{s}}(a, b) + \rho_{i, j}^{\bvec{s}}(a, b) + \beta_{i, j}^{\bvec{s}}(a, b)
& =
\sum_{d|q} \varepsilon_{d}^{\bvec{s}}
=
1
\label{eq:sum_theta_lambda_rho_beta}
\end{align}
for each $1 \le i < j \le m$, each $a, b \ge 1$, and each $\bvec{s} \in \{ -, + \}^{\ast}$.
Some examples of \eqref{def:theta}--\eqref{def:beta} are given as follows:

\begin{example}
\label{ex:theta_lambda_rho_beta_6}
Consider the case where $q = 6$ (see Examples~\ref{ex:sahebi_pradhan} and~\ref{ex:recursive_6}).
Set $m = 2$, $(p_{1}, p_{2}) = (2, 3)$, and $(r_{1}, r_{2}) = (1, 1)$.
Let $( \varepsilon_{d} )_{d|q} = (\varepsilon_{1}, \varepsilon_{2}, \varepsilon_{3}, \varepsilon_{6})$ be an initial four-dimensional probability vector.
Since $m = 2$, it suffices to consider the case where $(i, j) = (1, 2)$.
For every $a, b \ge 2$ and every $\bvec{s} \in \{ -, + \}^{\ast}$, we observe that
\begin{align}
& \left\{
\begin{array}{l}
\theta_{1, 2}^{\bvec{s}}(1, 1)
=
\varepsilon_{6}^{\bvec{s}} ,
\\[5pt]
\lambda_{1, 2}^{\bvec{s}}(1, 1)
=
\varepsilon_{2}^{\bvec{s}} ,
\\[5pt]
\rho_{1, 2}^{\bvec{s}}(1, 1)
=
\varepsilon_{3}^{\bvec{s}} ,
\\[5pt]
\beta_{1, 2}^{\bvec{s}}(1, 1)
=
\varepsilon_{1}^{\bvec{s}} .
\end{array}
\right.
\label{eq:theta_lambda_rho_beta_6} \\[5pt]
& \left\{
\begin{array}{l}
\theta_{1, 2}^{\bvec{s}}(a, 1)
=
0 ,
\\[5pt]
\lambda_{1, 2}^{\bvec{s}}(a, 1)
=
0 ,
\\[5pt]
\rho_{1, 2}^{\bvec{s}}(a, 1)
=
\varepsilon_{3}^{\bvec{s}} + \varepsilon_{6}^{\bvec{s}} ,
\\[5pt]
\beta_{1, 2}^{\bvec{s}}(a, 1)
=
\varepsilon_{1}^{\bvec{s}} + \varepsilon_{2}^{\bvec{s}} .
\end{array}
\right.
\\[5pt]
& \left\{
\begin{array}{l}
\theta_{1, 2}^{\bvec{s}}(1, b)
=
0 ,
\\[5pt]
\lambda_{1, 2}^{\bvec{s}}(1, b)
=
\varepsilon_{2}^{\bvec{s}} + \varepsilon_{6}^{\bvec{s}} ,
\\[5pt]
\rho_{1, 2}^{\bvec{s}}(1, b)
=
0 ,
\\[5pt]
\beta_{1, 2}^{\bvec{s}}(1, b)
=
\varepsilon_{1}^{\bvec{s}} + \varepsilon_{3}^{\bvec{s}} .
\end{array}
\right.
\\[5pt]
& \left\{
\begin{array}{l}
\theta_{1, 2}^{\bvec{s}}(a, b)
=
0 ,
\\[5pt]
\lambda_{1, 2}^{\bvec{s}}(a, b)
=
0 ,
\\[5pt]
\rho_{1, 2}^{\bvec{s}}(a, b)
=
0 ,
\\[5pt]
\beta_{1, 2}^{\bvec{s}}(a, b)
=
\varepsilon_{1}^{\bvec{s}} + \varepsilon_{2}^{\bvec{s}} + \varepsilon_{3}^{\bvec{s}} + \varepsilon_{6}^{\bvec{s}} = 1 .
\end{array}
\right.
\end{align}
\end{example}

\begin{figure}
\centering
\subfloat[when $(a, b) = (1, 2)$]{
\begin{tikzpicture}
\node (n00) at (0, 0) {\textbullet};
\node (n10) at (-1, 1) {\textbullet};
\node (n01) at (1, 1) {\textbullet};
\node (n20) at (-2, 2) {\textbullet};
\node (n11) at (0, 2) {\textbullet};
\node (n02) at (2, 2) {\textbullet};
\node (n30) at (-3, 3) {\textbullet};
\node (n21) at (-1, 3) {\textbullet};
\node (n12) at (1, 3) [red] {\textbullet};
\node (n31) at (-2, 4) {\textbullet};
\node (n22) at (0, 4) {\textbullet};
\node (n32) at (-1, 5) {\textbullet};
\draw [thick] (n00) -- (n01);
\draw [thick] (n00) -- (n10);
\draw [thick] (n01) -- (n02);
\draw [thick] (n01) -- (n11);
\draw [thick] (n10) -- (n11);
\draw [thick] (n10) -- (n20);
\draw [thick] (n02) -- (n12);
\draw [thick] (n11) -- (n12);
\draw [thick] (n11) -- (n21);
\draw [thick] (n20) -- (n21);
\draw [thick] (n20) -- (n30);
\draw [thick] (n12) -- (n22);
\draw [thick] (n21) -- (n22);
\draw [thick] (n21) -- (n31);
\draw [thick] (n30) -- (n31);
\draw [thick] (n22) -- (n32);
\draw [thick] (n31) -- (n32);
\draw (n00) node [below] {$1$};
\draw (n10) node [below] {$p_{1}$};
\draw (n01) node [below] {$p_{2}$};
\draw (n02) node [right] {$p_{2}^{2}$};
\draw (n11) node [right] {$p_{1} p_{2}$};
\draw (n12) node [right, red] {$p_{1} p_{2}^{2}$};
\draw (n20) node [below] {$p_{1}^{2}$};
\draw (n30) node [below] {$p_{1}^{3}$};
\draw (n21) node [right] {$p_{1}^{2} p_{2}$};
\draw (n22) node [right] {$p_{1}^{2} p_{2}^{2}$};
\draw (n31) node [right] {$p_{1}^{3} p_{2}$};
\draw (n32) node [right] {$p_{1}^{3} p_{2}^{2}$};
\node (theta) at (2, 5) [draw, ellipse] {$\theta_{1, 2}^{\bvec{s}}(a, b)$};
\draw [rounded corners] (1, 2.6) -- (-1.4, 5) -- (-1, 5.4) -- (1.4, 3) -- cycle;
\draw [-latex, very thick, rounded corners] (theta) -- (0.5, 4.75) -- (0, 4.4);
\node (lambda) at (-3, 5.5) [draw, ellipse] {$\lambda_{1, 2}^{\bvec{s}}(a, b)$};
\draw [rounded corners] (-1, 0.6) -- (-3.4, 3) -- (-2, 4.4) -- (0.4, 2) -- cycle;
\draw [-latex, very thick, rounded corners] (lambda) -- (-3, 4.5) -- (-2.7, 3.8);
\node (beta) at (-2.75, 0.5) [draw, ellipse] {$\beta_{1, 2}^{\bvec{s}}(a, b)$};
\draw [rounded corners] (0, -0.4) -- (-0.4, 0) -- (1, 1.4) -- (1.4, 1) -- cycle;
\draw [-latex, very thick, rounded corners] (beta) -- (-1, -0.5) -- (-0.3, -0.2);
\node (rho) at (2, -0.5) [draw, ellipse] {$\rho_{1, 2}^{\bvec{s}}(a, b)$};
\draw [rounded corners] (2, 1.6) -- (1.6, 2) -- (2, 2.4) -- (2.4, 2) -- cycle;
\draw [-latex, very thick, rounded corners] (rho) -- (2, 1.6);
\end{tikzpicture}
}\hfill%
\subfloat[when $(a, b) = (2, 1)$]{
\begin{tikzpicture}
\node (n00) at (0, 0) {\textbullet};
\node (n10) at (-1, 1) {\textbullet};
\node (n01) at (1, 1) {\textbullet};
\node (n20) at (-2, 2) {\textbullet};
\node (n11) at (0, 2) {\textbullet};
\node (n02) at (2, 2) {\textbullet};
\node (n30) at (-3, 3) {\textbullet};
\node (n21) at (-1, 3) [red] {\textbullet};
\node (n12) at (1, 3) {\textbullet};
\node (n31) at (-2, 4) {\textbullet};
\node (n22) at (0, 4) {\textbullet};
\node (n32) at (-1, 5) {\textbullet};
\draw [thick] (n00) -- (n01);
\draw [thick] (n00) -- (n10);
\draw [thick] (n01) -- (n02);
\draw [thick] (n01) -- (n11);
\draw [thick] (n10) -- (n11);
\draw [thick] (n10) -- (n20);
\draw [thick] (n02) -- (n12);
\draw [thick] (n11) -- (n12);
\draw [thick] (n11) -- (n21);
\draw [thick] (n20) -- (n21);
\draw [thick] (n20) -- (n30);
\draw [thick] (n12) -- (n22);
\draw [thick] (n21) -- (n22);
\draw [thick] (n21) -- (n31);
\draw [thick] (n30) -- (n31);
\draw [thick] (n22) -- (n32);
\draw [thick] (n31) -- (n32);
\draw (n00) node [below] {$1$};
\draw (n10) node [below] {$p_{1}$};
\draw (n01) node [below] {$p_{2}$};
\draw (n02) node [right] {$p_{2}^{2}$};
\draw (n11) node [right] {$p_{1} p_{2}$};
\draw (n12) node [right] {$p_{1} p_{2}^{2}$};
\draw (n20) node [below] {$p_{1}^{2}$};
\draw (n30) node [below] {$p_{1}^{3}$};
\draw (n21) node [right, red] {$p_{1}^{2} p_{2}$};
\draw (n22) node [right] {$p_{1}^{2} p_{2}^{2}$};
\draw (n31) node [right] {$p_{1}^{3} p_{2}$};
\draw (n32) node [right] {$p_{1}^{3} p_{2}^{2}$};
\node (theta) at (2, 5) [draw, ellipse] {$\theta_{1, 2}^{\bvec{s}}(a, b)$};
\draw [rounded corners] (-1, 2.6) -- (0.4, 4) -- (-1, 5.4) -- (-2.4, 4) -- cycle;
\draw [-latex, very thick, rounded corners] (theta) -- (0.25, 4.75) -- (0, 4.4);
\node (lambda) at (-3, 5.5) [draw, ellipse] {$\lambda_{1, 2}^{\bvec{s}}(a, b)$};
\draw [rounded corners] (-2, 1.6) -- (-1.6, 2) -- (-3, 3.4) -- (-3.4, 3) -- cycle;
\draw [-latex, very thick, rounded corners] (lambda) -- (-4, 3.75) -- (-3.2, 3.2);
\node (beta) at (-2.75, 0.5) [draw, ellipse] {$\beta_{1, 2}^{\bvec{s}}(a, b)$};
\draw [rounded corners] (0, -0.4) -- (0.4, 0) -- (-1, 1.4) -- (-1.4, 1) -- cycle;
\draw [-latex, very thick, rounded corners] (beta) -- (-1, 0) -- (-0.7, 0.3);
\node (rho) at (2, -0.5) [draw, ellipse] {$\rho_{1, 2}^{\bvec{s}}(a, b)$};
\draw [rounded corners] (1, 0.6) -- (2.4, 2) -- (1, 3.4) -- (-0.4, 2) -- cycle;
\draw [-latex, very thick, rounded corners] (rho) -- (2, 0.5) -- (1.7, 1.3);
\end{tikzpicture}
}
\caption{Hasse diagram of the positive divisors of $q = p_{1}^{3} \, p_{2}^{2}$ (see \exref{ex:Hasse}).}
\label{fig:Hasse}
\end{figure}

\begin{example}
\label{ex:Hasse}
Let $q = p_{1}^{3} p_{2}^{2}$, where $p_{1}$ and $p_{2}$ are distinct prime numbers (e.g., $q = 72 = 2^{3} \cdot 3^{2}$).
Then, we see that
\begin{align}
&
\left\{
\begin{array}{l}
\theta_{1, 2}^{\bvec{s}}(1, 2)
=
\varepsilon_{p_{1} p_{2}^{2}}^{\bvec{s}} + \varepsilon_{p_{1}^{2} p_{2}}^{\bvec{s}} + \varepsilon_{p_{1}^{3} p_{2}^{2}}^{\bvec{s}} ,
\\[7pt]
\lambda_{1, 2}^{\bvec{s}}(1, 2)
=
\varepsilon_{p_{1}}^{\bvec{s}} + \varepsilon_{p_{1} p_{2}}^{\bvec{s}} + \varepsilon_{p_{1}^{2}}^{\bvec{s}} + \varepsilon_{p_{1}^{2} p_{2}}^{\bvec{s}} + \varepsilon_{p_{1}^{3}}^{\bvec{s}} + \varepsilon_{p_{1}^{3} p_{2}}^{\bvec{s}} ,
\\[7pt]
\rho_{1, 2}^{\bvec{s}}(1, 2)
=
\varepsilon_{p_{2}^{2}}^{\bvec{s}} ,
\\[7pt]
\beta_{1, 2}^{\bvec{s}}(1, 2)
=
\varepsilon_{1}^{\bvec{s}} + \varepsilon_{p_{2}}^{\bvec{s}} .
\end{array}
\right.
\\[7pt]
&
\left\{
\begin{array}{l}
\theta_{1, 2}^{\bvec{s}}(2, 1)
=
\varepsilon_{p_{1}^{2} p_{2}}^{\bvec{s}} + \varepsilon_{p_{1}^{2} p_{2}^{2}}^{\bvec{s}} + \varepsilon_{p_{1}^{3} p_{2}}^{\bvec{s}} + \varepsilon_{p_{1}^{3} p_{2}^{2}}^{\bvec{s}} ,
\\[7pt]
\lambda_{1, 2}^{\bvec{s}}(2, 1)
=
\varepsilon_{p_{1}^{2}}^{\bvec{s}} + \varepsilon_{p_{1}^{3}}^{\bvec{s}} ,
\\[7pt]
\rho_{1, 2}^{\bvec{s}}(2, 1)
=
\varepsilon_{p_{2}}^{\bvec{s}} + \varepsilon_{p_{2}^{2}}^{\bvec{s}} + \varepsilon_{p_{1} p_{2}}^{\bvec{s}} + \varepsilon_{p_{1} p_{2}^{2}}^{\bvec{s}} ,
\\[7pt]
\beta_{1, 2}^{\bvec{s}}(2, 1)
=
\varepsilon_{1}^{\bvec{s}} + \varepsilon_{p_{1}}^{\bvec{s}} .
\end{array}
\right.
\end{align}
Graphical interpretations of these partial sums via Hasse diagrams are plotted in \figref{fig:Hasse}.
\end{example}

In the following, to characterize the asymptotic distribution $( \mu_{d}^{(\infty)} )_{d|q}$, we state some technical lemmas concerning these four partial sums $\theta_{i, j}^{\bvec{s}}$, $\lambda_{i, j}^{\bvec{s}}$, $\rho_{i, j}^{\bvec{s}}$, and $\beta_{i, j}^{\bvec{s}}$.
Firstly, we provide recursive formulas of these partial sums under the polar transform as follows:

\begin{lemma}
\label{lem:formulas}
For any $\bvec{s} \in \{ -, + \}^{\ast}$, $1 \le i < j \le m$, and $a, b \ge 1$, it holds that
\begin{align}
& \left\{
\begin{array}{l}
\theta_{i, j}^{\bvec{s}-}(a, b)
=
\theta_{i, j}^{\bvec{s}}(a, b)^{2} ,
\\[7pt]
\lambda_{i, j}^{\bvec{s}-}(a, b)
=
\lambda_{i, j}^{\bvec{s}}(a, b) \, \big[ \lambda_{i, j}^{\bvec{s}}(a, b) + 2 \, \theta_{i, j}^{\bvec{s}}(a, b) \big] ,
\\[7pt]
\rho_{i, j}^{\bvec{s}-}(a, b)
=
\rho_{i, j}^{\bvec{s}}(a, b) \, \big[ \rho_{i, j}^{\bvec{s}}(a, b) + 2 \, \theta_{i, j}^{\bvec{s}}(a, b) \big] ,
\\[7pt]
\beta_{i, j}^{\bvec{s}-}(a, b)
=
\beta_{i, j}^{\bvec{s}}(a, b) \, \big[ 2 - \beta_{i, j}^{\bvec{s}}(a, b) \big] + 2 \, \lambda_{i, j}^{\bvec{s}}(a, b) \, \rho_{i, j}^{\bvec{s}}(a, b) ,
\end{array}
\right.
\\[7pt]
& \left\{
\begin{array}{l}
\theta_{i, j}^{\bvec{s}+}(a, b)
=
\theta_{i, j}^{\bvec{s}}(a, b) \, \big[ 2 - \theta_{i, j}^{\bvec{s}}(a, b) \big] + 2 \, \lambda_{i, j}^{\bvec{s}}(a, b) \, \rho_{i, j}^{\bvec{s}}(a, b) ,
\\[7pt]
\lambda_{i, j}^{\bvec{s}+}(a, b)
=
\lambda_{i, j}^{\bvec{s}}(a, b) \, \big[ \lambda_{i, j}^{\bvec{s}}(a, b) + 2 \, \beta_{i, j}^{\bvec{s}}(a, b) \big] ,
\\[7pt]
\rho_{i, j}^{\bvec{s}+}(a, b)
=
\rho_{i, j}^{\bvec{s}}(a, b) \, \big[ \rho_{i, j}^{\bvec{s}}(a, b) + 2 \, \beta_{i, j}^{\bvec{s}}(a, b) \big] ,
\\[7pt]
\beta_{i, j}^{\bvec{s}+}(a, b)
=
\beta_{i, j}^{\bvec{s}}(a, b)^{2} .
\end{array}
\right.
\end{align}
\end{lemma}

\begin{IEEEproof}[Proof of \lemref{lem:formulas}]
See \appref{app:formulas}.
\end{IEEEproof}

Similar to \lemref{lem:martingale_primepower}, as shown in the following lemma, \lemref{lem:formulas} characterizes the average value of \eqref{def:theta}--\eqref{def:beta} when the polar transform is applied once.

\begin{lemma}
\label{lem:quasi-conservation}
For any $\bvec{s} \in \{ -, + \}^{\ast}$, $1 \le i < j \le m$, and $a, b \ge 1$, it holds that
\begin{align}
\frac{ 1 }{ 2 } \Big[ \theta_{i, j}^{\bvec{s}-}(a, b) + \theta_{i, j}^{\bvec{s}+}(a, b) \Big]
& =
\theta_{i, j}^{\bvec{s}}(a, b) + \lambda_{i, j}^{\bvec{s}}(a, b) \, \rho_{i, j}^{\bvec{s}}(a, b) ,
\\
\frac{ 1 }{ 2 } \Big[ \lambda_{i, j}^{\bvec{s}-}(a, b) + \lambda_{i, j}^{\bvec{s}+}(a, b) \Big]
& =
\lambda_{i, j}^{\bvec{s}}(a, b) \, \big[ 1 - \rho_{i, j}^{\bvec{s}}(a, b) \big] ,
\\
\frac{ 1 }{ 2 } \Big[ \rho_{i, j}^{\bvec{s}-}(a, b) + \rho_{i, j}^{\bvec{s}+}(a, b) \Big]
& =
\rho_{i, j}^{\bvec{s}}(a, b) \, \big[ 1 - \lambda_{i, j}^{\bvec{s}}(a, b) \big] ,
\\
\frac{ 1 }{ 2 } \Big[ \beta_{i, j}^{\bvec{s}-}(a, b) + \beta_{i, j}^{\bvec{s}+}(a, b) \Big]
& =
\beta_{i, j}^{\bvec{s}}(a, b) + \lambda_{i, j}^{\bvec{s}}(a, b) \, \rho_{i, j}^{\bvec{s}}(a, b) .
\end{align}
\end{lemma}

\begin{IEEEproof}[Proof of \lemref{lem:quasi-conservation}]
\lemref{lem:quasi-conservation} follows in a straightforward manner from \lemref{lem:formulas}.
\end{IEEEproof}

The idea of \lemref{lem:quasi-conservation} comes from the conservation property $[I(W^{-}) + I(W^{+})]/2 = I(W)$. 
Note that in general, these quantities are not conserved by the polar transform.
In fact, \lemref{lem:quasi-conservation} can be thought of as being equivalent to the following sub- and super-martingale-like properties with respect to the polarization process:
\begin{align}
\frac{ 1 }{ 2 } \Big[ \theta_{i, j}^{\bvec{s}-}(a, b) + \theta_{i, j}^{\bvec{s}+}(a, b) \Big]
& \ge
\theta_{i, j}^{\bvec{s}}(a, b) ,
\label{ineq:sub_theta} \\
\frac{ 1 }{ 2 } \Big[ \lambda_{i, j}^{\bvec{s}-}(a, b) + \lambda_{i, j}^{\bvec{s}+}(a, b) \Big]
& \le
\lambda_{i, j}^{\bvec{s}}(a, b) ,
\label{ineq:super_lambda} \\
\frac{ 1 }{ 2 } \Big[ \rho_{i, j}^{\bvec{s}-}(a, b) + \rho_{i, j}^{\bvec{s}+}(a, b) \Big]
& \le
\rho_{i, j}^{\bvec{s}}(a, b) ,
\label{ineq:super_rho} \\
\frac{ 1 }{ 2 } \Big[ \beta_{i, j}^{\bvec{s}-}(a, b) + \beta_{i, j}^{\bvec{s}+}(a, b) \Big]
& \ge
\beta_{i, j}^{\bvec{s}}(a, b) .
\label{ineq:sub_beta}
\end{align}

The following lemma states a property between $\lambda_{i, j}^{\bvec{s}}(a, b)$ and $\rho_{i, j}^{\bvec{s}}(a, b)$; it shows that the inequality between $\lambda_{i, j}^{\bvec{s}}(a, b)$ and $\rho_{i, j}^{\bvec{s}}(a, b)$ is invariant under one pass of the polar transform $\bvec{s} \in \{ -, + \}^{\ast}$.

\begin{lemma}
\label{lem:ineq}
For each $1 \le i < j \le m$ and $a, b \ge 1$, it holds that $\lambda_{i, j}^{\bvec{s}}(a, b) \le \rho_{i, j}^{\bvec{s}}(a, b)$ for every  $\bvec{s} \in \{ -, + \}^{\ast}$ if and only if $\lambda_{i, j}(a, b) \le \rho_{i, j}(a, b)$.
\end{lemma}

\begin{IEEEproof}[Proof of \lemref{lem:ineq}]
See \appref{app:ineq}.
\end{IEEEproof}

We now define the average value of \eqref{def:theta}--\eqref{def:beta} as follows:
\begin{align}
\mu_{i, j}^{(n)}[\theta](a, b)
& \coloneqq
\frac{ 1 }{ 2^{n} } \sum_{\bvec{s} \in \{ -, + \}^{n}} \theta_{i, j}^{\bvec{s}}(a, b) ,
\label{def:mu_theta}
\\
\mu_{i, j}^{(n)}[\lambda](a, b)
& \coloneqq
\frac{ 1 }{ 2^{n} } \sum_{\bvec{s} \in \{ -, + \}^{n}} \lambda_{i, j}^{\bvec{s}}(a, b) ,
\label{def:mu_lambda}
\\
\mu_{i, j}^{(n)}[\rho](a, b)
& \coloneqq
\frac{ 1 }{ 2^{n} } \sum_{\bvec{s} \in \{ -, + \}^{n}} \rho_{i, j}^{\bvec{s}}(a, b) ,
\label{def:mu_rho}
\\
\mu_{i, j}^{(n)}[\beta](a, b)
& \coloneqq
\frac{ 1 }{ 2^{n} } \sum_{\bvec{s} \in \{ -, + \}^{n}} \beta_{i, j}^{\bvec{s}}(a, b) .
\label{def:mu_beta}
\end{align}
For convenience, when $n = 0$, we write $\mu_{i, j}^{(0)}[\theta](a, b) \coloneqq \theta_{i, j}(a, b)$, $\mu_{i, j}^{(0)}[\lambda](a, b) \coloneqq \lambda_{i, j}(a, b)$, $\mu_{i, j}^{(0)}[\rho](a, b) \coloneqq \rho_{i, j}(a, b)$, and $\mu_{i, j}^{(0)}[\beta](a, b) \coloneqq \beta_{i, j}(a, b)$.
Unlike the case when $q$ is a prime power (see \lemref{lem:martingale_primepower}), these quantities are not preserved under the polar transform.
However, as shown in the following lemma, the difference between $\mu_{i, j}^{(n)}[\lambda](a, b)$ and $\mu_{i, j}^{(n)}[\rho](a, b)$, and several addition between two quantities are preserved under the polar transform.

\begin{lemma}
\label{lem:martingale}
For any $n \ge 0$, $1 \le i < j \le m$, and $a, b \ge 1$, it holds that
\begin{align}
\!
\mu_{i, j}^{(n)}[\lambda](a, b) - \mu_{i, j}^{(n)}[\rho](a, b)
& =
\lambda_{i, j}(a, b) - \rho_{i, j}(a, b) ,
\label{eq:lambda_minus_rho} \\
\!
\mu_{i, j}^{(n)}[\theta](a, b) + \mu_{i, j}^{(n)}[\lambda](a, b)
& =
\theta_{i, j}(a, b)+ \lambda_{i, j}(a, b) ,
\label{eq:theta_plus_lambda} \\
\!
\mu_{i, j}^{(n)}[\theta](a, b) + \mu_{i, j}^{(n)}[\rho](a, b)
& =
\theta_{i, j}(a, b) + \rho_{i, j}(a, b) ,
\label{eq:theta_plus_rho} \\
\!
\mu_{i, j}^{(n)}[\beta](a, b) + \mu_{i, j}^{(n)}[\lambda](a, b)
& =
\beta_{i, j}(a, b) + \lambda_{i, j}(a, b) ,
\label{eq:beta_plus_lambda} \\
\!
\mu_{i, j}^{(n)}[\beta](a, b) + \mu_{i, j}^{(n)}[\rho](a, b)
& =
\beta_{i, j}(a, b) + \rho_{i, j}(a, b) .
\label{eq:beta_plus_rho} 
\end{align}
\end{lemma}

\begin{IEEEproof}[Proof of \lemref{lem:martingale}]
See \appref{app:martingale}.
\end{IEEEproof}

\lemref{lem:martingale} implies that the left-hand sides of \eqref{eq:lambda_minus_rho}--\eqref{eq:beta_plus_rho} has martingale-like properties.
It is worth mentioning that \lemref{lem:martingale} is useful to prove the limits of \eqref{def:mu_theta}--\eqref{def:mu_beta} as $n \to \infty$.

\begin{lemma}
\label{lem:convergent}
For each $1 \le i < j \le m$ and $a, b \ge 1$, the four sequences $( \mu_{i, j}^{(n)}[\theta](a, b) )_{n=1}^{\infty}$, $( \mu_{i, j}^{(n)}[\lambda](a, b) )_{n=1}^{\infty}$, $( \mu_{i, j}^{(n)}[\rho](a, b) )_{n=1}^{\infty}$, and $( \mu_{i, j}^{(n)}[\beta](a, b) )_{n=1}^{\infty}$ are convergent.
\end{lemma}

\begin{IEEEproof}[Proof of \lemref{lem:convergent}]
See \appref{app:convergent}.
\end{IEEEproof}

By \lemref{lem:convergent}, we can define the following limits:
\begin{align}
\mu_{i, j}^{(\infty)}[\theta](a, b)
& \coloneqq
\lim_{n \to \infty} \mu_{i, j}^{(n)}[\theta](a, b) ,
\\
\mu_{i, j}^{(\infty)}[\lambda](a, b)
& \coloneqq
\lim_{n \to \infty} \mu_{i, j}^{(n)}[\lambda](a, b) ,
\\
\mu_{i, j}^{(\infty)}[\rho](a, b)
& \coloneqq
\lim_{n \to \infty} \mu_{i, j}^{(n)}[\rho](a, b) ,
\\
\mu_{i, j}^{(\infty)}[\beta](a, b)
& \coloneqq
\lim_{n \to \infty} \mu_{i, j}^{(n)}[\beta](a, b) .
\end{align}
The following theorem shows that these limits can be evaluated easily in terms of the initial probability vector $\bvec{\varepsilon} = ( \varepsilon_{d} )_{d|q}$.

\begin{lemma}
\label{lem:fujisaki}
For any $1 \le i < j \le m$ and $a, b \ge 1$, it holds that
\begin{align}
\mu_{i, j}^{(\infty)}[\theta](a, b)
& =
\theta_{i, j}(a, b) + \min\{ \lambda_{i, j}(a, b), \rho_{i, j}(a, b) \} ,
\label{eq:theta_inf} \\
\mu_{i, j}^{(\infty)}[\lambda](a, b)
& =
\big| \lambda_{i, j}(a, b) - \rho_{i, j}(a, b) \big|^{+} ,
\label{eq:lambda_inf} \\
\mu_{i, j}^{(\infty)}[\rho](a, b)
& =
\big| \rho_{i, j}(a, b) - \lambda_{i, j}(a, b) \big|^{+} ,
\label{eq:rho_inf} \\
\mu_{i, j}^{(\infty)}[\beta](a, b)
& =
\beta_{i, j}(a, b) + \min\{ \lambda_{i, j}(a, b), \rho_{i, j}(a, b) \} ,
\label{eq:beta_inf} 
\end{align}
where $| c |^{+} \coloneqq \max\{ 0, c \}$ for $c \in \mathbb{R}$.
\end{lemma}

\begin{IEEEproof}[Proof of \lemref{lem:fujisaki}]
See \appref{app:fujisaki}.
\end{IEEEproof}

If $q$ is a semiprime, i.e., if $q = p_{1} p_{2}$ for some distinct prime numbers $p_{1}$ and $p_{2}$, then \lemref{lem:fujisaki} immediately yields the asymptotic distribution $( \mu_{d}^{(\infty)} )_{d|q}$ defined in \eqref{def:mu_d_infty}, as shown in the following example.

\begin{example}
\label{ex:asymptotic_distribution_6}
Let $q = 6 = 2 \cdot 3$ (see Examples~\ref{ex:sahebi_pradhan}--\ref{ex:theta_lambda_rho_beta_6}).
It follows from \eqref{eq:theta_lambda_rho_beta_6} of \exref{ex:theta_lambda_rho_beta_6} and \lemref{lem:fujisaki} that
\begin{align}
\left\{
\begin{array}{l}
\mu_{6}^{(\infty)}
=
\mu_{1, 2}^{(\infty)}[\theta](1, 1) = \varepsilon_{6} + \min\{ \varepsilon_{2}, \varepsilon_{3} \} ,
\\[5pt]
\mu_{2}^{(\infty)}
=
\mu_{1, 2}^{(\infty)}[\lambda](1, 1) = | \varepsilon_{2} - \varepsilon_{3} |^{+} ,
\\[5pt]
\mu_{3}^{(\infty)}
=
\mu_{1, 2}^{(\infty)}[\rho](1, 1) = | \varepsilon_{3} - \varepsilon_{2} |^{+} ,
\\[5pt]
\mu_{1}^{(\infty)}
=
\mu_{1, 2}^{(\infty)}[\beta](1, 1) = \varepsilon_{1} + \min\{ \varepsilon_{2}, \varepsilon_{3} \}
\end{array}
\right.
\end{align}
for every initial probability vector $\bvec{\varepsilon} = ( \varepsilon_{d} )_{d|q} = (\varepsilon_{1}, \varepsilon_{2}, \varepsilon_{3}, \varepsilon_{6})$.
Therefore, the asymptotic distribution of \figref{subfig:maec_6_case1} is given by $( \mu_{1}^{(\infty)}, \mu_{2}^{(\infty)}, \mu_{3}^{(\infty)}, \mu_{6}^{(\infty)} ) = ( 3/10, 0, 3/10, 2/5 )$, and \figref{subfig:maec_6_case2} by $( \mu_{1}^{(\infty)}, \mu_{2}^{(\infty)}, \mu_{3}^{(\infty)}, \mu_{6}^{(\infty)} ) = ( 1/2, 0, 0, 1/2 )$.
\end{example}

The following theorem shows that the limit $\mu_{d}^{(\infty)}$ defined in \eqref{def:mu_d_infty} always exists for each $d|q$, and the asymptotic distribution $( \mu_{d}^{(\infty)} )_{d|q}$ can be calculated algorithmically and exactly for every composite number $q$ having two or more distinct prime factors.

\begin{algorithm}[t]
\DontPrintSemicolon
\KwData{An initial probability vector $\bvec{\varepsilon} = ( \varepsilon_{d} )_{d|q}$}
\KwResult{The asymptotic distribution $( \mu_{d}^{(\infty)} )_{d|q}$}
	Initialize $( \mu_{d}^{(\infty)} )_{d|q}$ by the zero vector $(0, \dots, 0)$\;
	$\xi \longleftarrow 0$\;
	$\bvec{t} = (t_{1}, \dots, t_{m}) \longleftarrow (0, \dots, 0)$\;
	\While{$0 \le \xi < 1$}{
		$(i, j) \longleftarrow (1, 2)$\;
		\While{$j \le m$}{
			\uIf{$\lambda_{i, j}(t_{i}+1, t_{j}+1) \le \rho_{i, j}(t_{i}+1, t_{j}+1)$}{
				$(k, l) \longleftarrow (j, i)$\;
				$(i, j) \longleftarrow (k, j+1)$\;
			}
			\Else{
				$(k, l) \longleftarrow (i, i)$\;
				$j \longleftarrow j + 1$\;
			}
		}
		$\mu_{\langle \bvec{t} \rangle}^{(\infty)} \longleftarrow \beta_{l, m}(t_{l}+1, t_{m}+1) + \min\{ \lambda_{l, m}(t_{l}+1, t_{m}+1), \lambda_{l, m}(t_{l}+1, t_{m}+1) \} - \xi$\;
		$\xi \longleftarrow \xi + \mu_{\langle \bvec{t} \rangle}^{(\infty)}$\;
		$t_{k} \longleftarrow t_{k} + 1$\;
	}
\caption{Solving asymptotic distribution}
\label{alg:main}
\end{algorithm}

\begin{theorem}
\label{th:mu_d}
The asymptotic distribution $( \mu_{d}^{(\infty)} )_{d|q}$ can be calculated by Algorithm~\ref{alg:main} in%
\footnote{While $\mathrm{O}( \cdot )$ stands for the Big-O notation used to denote the computational complexity of a certain procedure, note that $\omega_{\mathrm{NT}}( \cdot )$ and $\Omega_{\mathrm{NT}}( \cdot )$ are number theoretic notations, i.e., these notations do not stand for the little-omega and Big-Omega notations, respectively, of the computational complexity.}
time $\mathrm{O}( \omega_{\mathrm{NT}}(q) \, \Omega_{\mathrm{NT}}(q) \, \tau( q ) )$, where $\omega_{\mathrm{NT}}( q ) = m$ denotes the number of distinct prime factors of $q$, $\Omega_{\mathrm{NT}}(q) \coloneqq \sum_{i = 1}^{m} r_{i}$ denotes the number of prime factors of $q$ with multiplicity, and $\tau( q ) \coloneqq \prod_{i = 1}^{m} (r_{i}+1)$ denotes the number of positive divisors of $q$.
\end{theorem}

\begin{IEEEproof}[Proof of \thref{th:mu_d}]
See \appref{app:mu_d}.
\end{IEEEproof}

By \thref{th:mu_d}, we can immediately observe the following corollary.

\begin{corollary}
\label{cor:mu_d}
For any initial probability vector $\bvec{\varepsilon} = ( \varepsilon_{d} )_{d|q}$, there exists a sequence $( \bvec{t}^{(h)} = (t_{1}^{(h)}, \dots, t_{m}^{(h)}) )_{h = 0}^{m}$ satisfying (i) $\bvec{0} = \bvec{t}^{(0)} \le \bvec{t}^{(1)} \le \cdots \le \bvec{t}^{(m)} = \bvec{r}$ and (ii) $\mu_{\langle \bvec{t} \rangle}^{(\infty)} > 0$ only if $\bvec{t} = \bvec{t}^{(h)}$ for some $0 \le h \le m$.
Consequently, the asymptotic distribution $( \mu_{d}^{(\infty)} )_{d|q}$ has at most $\Omega_{\mathrm{NT}}( q ) + 1$ positive probability masses.
\end{corollary}

By Algorithm~\ref{alg:main}, we can solve for the asymptotic distribution of \figref{subfig:maec_45} as
\begin{align}
( \mu_{1}^{(\infty)}, \mu_{3}^{(\infty)}, \mu_{5}^{(\infty)}, \mu_{9}^{(\infty)}, \mu_{15}^{(\infty)}, \mu_{45}^{(\infty)} )
=
(0, 0, 1/3, 0, 1/3, 1/3) .
\end{align}
A more complicated example of Algorithm~\ref{alg:main} is given as follows:

\begin{example}
\label{ex:mu_d}
Consider an MAEC $V_{\bvec{\varepsilon}}$ defined in \defref{def:V} with an initial probability vector $\bvec{\varepsilon} = (\varepsilon_{d})_{d|q}$ as follows:
The input alphabet size is $q = 4500 = 2^{2} \cdot 3^{2} \cdot 5^{3}$, where note that the set of positive divisors $d$ of $q$ is $\{ 1, \allowbreak 2, \allowbreak 3, \allowbreak 4, \allowbreak 5, \allowbreak 6, \allowbreak 9, \allowbreak 10, \allowbreak 12, \allowbreak 15, \allowbreak 18, \allowbreak 20, \allowbreak 25, \allowbreak 30, \allowbreak 36, \allowbreak 45, \allowbreak 50, \allowbreak 60, \allowbreak 75, \allowbreak 90, \allowbreak 100, \allowbreak 125, \allowbreak 150, \allowbreak 180, \allowbreak 225, \allowbreak 250, \allowbreak 300, \allowbreak 375, \allowbreak 450, \allowbreak 500, \allowbreak 750, \allowbreak 900, \allowbreak 1125, \allowbreak 1500, \allowbreak 2250, \allowbreak 4500 \}$.
The initial probability vector $(\varepsilon_{d})_{d|q}$ is given by%
\footnote{The elements $\varepsilon_{d}$ of $(\varepsilon_{d})_{d|q}$ are sorted in increasing order of divisors $d$.}
$(\varepsilon_{d})_{d|q} = (1/150) \times (0, \allowbreak 1, \allowbreak 2, \allowbreak 3, \allowbreak 4, \allowbreak 5, \allowbreak 6, \allowbreak 7, \allowbreak 8, \allowbreak 9, \allowbreak 0, \allowbreak 1, \allowbreak 2, \allowbreak 3, \allowbreak 4, \allowbreak 5, \allowbreak 6, \allowbreak 7, \allowbreak 8, \allowbreak 9, \allowbreak 0, \allowbreak 1, \allowbreak 2, \allowbreak 3, \allowbreak 4, \allowbreak 5, \allowbreak 6, \allowbreak 7, \allowbreak 8, \allowbreak 9, \allowbreak 0, \allowbreak 1, \allowbreak 2, \allowbreak 3, \allowbreak 4, \allowbreak 5)$.
Then, Algorithm~\ref{alg:main} solves the asymptotic distribution $(\mu_{d}^{(\infty)})_{d|q} = (29/150, \allowbreak 0, \allowbreak 0, \allowbreak 0, \allowbreak 1/15, \allowbreak 0, \allowbreak 0, \allowbreak 0, \allowbreak 0, \allowbreak 11/150, \allowbreak 0, \allowbreak 0, \allowbreak 0, \allowbreak 9/50, \allowbreak 0, \allowbreak 0, \allowbreak 0, \allowbreak 0, \allowbreak 0, \allowbreak 0, \allowbreak 0, \allowbreak 0, \allowbreak 11/75, \allowbreak 0, \allowbreak 0, \allowbreak 0, \allowbreak 0, \allowbreak 0, \allowbreak 1/150, \allowbreak 0, \allowbreak 0, \allowbreak 7/75, \allowbreak 0, \allowbreak 0, \allowbreak 0, \allowbreak 6/25)$.
We summarize this result in \tabref{table:mu_d}.
\end{example}

\begin{table*}[!t]
\caption{Example of Algorithm~\ref{alg:main} with the setting of \exref{ex:mu_d} (see also \figref{fig:mu_d}).
The input alphabet size is $q = 4500 = 2^{2} \cdot 3^{2} \cdot 5^{3}$.
An initial probability vector $( \varepsilon_{d} )_{d|q}$ and its resultant asymptotic distribution $( \mu_{d}^{(\infty)} )_{d|q}$ are summarized in the table.}
\centering
\label{table:mu_d}
\begin{tabular}{rrrrrrrrrrrrr}
\toprule
divisor $d$ & $1$ & $2$ & $3$ & $4$ & $5$ & $6$ & $9$ & $10$ & $12$ & $15$ & $18$ & $20$ \\
\midrule
$( \varepsilon_{d} )_{d|q}$ & $0$ & $1/150$ & $2/150$ & $3/150$ & $4/150$ & $5/150$ & $6/150$ & $7/150$ & $8/150$ & $9/150$ & $0$ & $1/150$ \\[3pt]
$( \mu_{d}^{(\infty)} )_{d|q}$ & $29/150$ & $0$ & $0$ & $0$ & $1/15$ & $0$ & $0$ & $0$ & $0$ & $11/150$ & $0$ & $0$ \\
\bottomrule
\toprule
divisor $d$ & $25$ & $30$ & $36$ & $45$ & $50$ & $60$ & $75$ & $90$ & $100$ & $125$ & $150$ & $180$ \\
\midrule
$( \varepsilon_{d} )_{d|q}$ & $2/150$ & $3/150$ & $4/150$ & $5/150$ & $6/150$ & $7/150$ & $8/150$ & $9/150$ & $0$ & $1/150$ & $2/150$ & $3/150$ \\[3pt]
$( \mu_{d}^{(\infty)} )_{d|q}$ & $0$ & $9/50$ & $0$ & $0$ & $0$ & $0$ & $0$ & $0$ & $0$ & $0$ & $11/75$ & $0$ \\
\bottomrule
\toprule
divisor $d$ & $225$ & $250$ & $300$ & $375$ & $450$ & $500$ & $750$ & $900$ & $1125$ & $1500$ & $2250$ & $4500$ \\
\midrule
$( \varepsilon_{d} )_{d|q}$ & $4/150$ & $5/150$ & $6/150$ & $7/150$ & $8/150$ & $9/150$ & 0 & $1/150$ & $2/150$ & $3/150$ & $4/150$ & $5/150$ \\[3pt]
$( \mu_{d}^{(\infty)} )_{d|q}$ & $0$ & $0$ & $0$ & $0$ & $1/150$ & $0$ & $0$ & $7/75$ & $0$ & $0$ & $0$ & $6/25$ \\
\bottomrule
\end{tabular}
\end{table*}

Figure~\ref{fig:mu_d} in \sectref{sect:contribution} shows an example of multilevel channel polarization for the MAEC given in \exref{ex:mu_d} (see also \tabref{table:mu_d}), where note that \figref{fig:mu_d} is calculated and plotted by employing \propref{prop:I(V)} and the recursive formulas stated in \eqref{def:eps_s} of \corref{cor:recursive_V}.

In this subsection, we have given an algorithm for calculating the asymptotic distribution $( \mu_{d}^{(\infty)} )_{d|q}$.
In the next subsection, we will show that $( \mu_{d}^{(\infty)} )_{d|q}$ is, in a rigorous sense, the asymptotic distribution of multilevel channel polarization for $\mathrm{MAEC}_{q}( \bvec{\varepsilon} )$.

\subsection{Formal Statement of Asymptotic Distribution}
\label{sect:asymptotic_distribution}

The following theorem states that $( \varepsilon_{d}^{\bvec{s}} )_{d|q}$ uniformly tends to a unit vector $(0, \dots, 0, 1, 0, \dots, 0)$ as $n$ goes to infinity for each sequence of polarization process $( \bvec{s} = s_{1} s_{2} \cdots s_{n} )_{n=1}^{\infty}$, and the limiting proportions are exactly characterized by the asymptotic distribution $( \mu_{d}^{(\infty)} )_{d|q}$.

\begin{theorem}
\label{th:polarization}
For any fixed $\delta \in (0, 1)$, it holds that
\begin{align}
\lim_{n \to \infty} \frac{ 1 }{ 2^{n} } \Big| \Big\{ \bvec{s} \in \{ -, + \}^{n} \ \Big| \ \delta \le \varepsilon_{d}^{\bvec{s}} \le 1 - \delta \Big\} \Big|
& =
0 ,
\label{eq:proportion0} \\
\lim_{n \to \infty} \frac{ 1 }{ 2^{n} } \Big| \Big\{ \bvec{s} \in \{ -, + \}^{n} \ \Big| \ \varepsilon_{d}^{\bvec{s}} > 1 - \delta \Big\} \Big|
& =
\mu_{d}^{(\infty)}
\label{eq:proportion1}
\end{align}
for every $d|q$, where $(\mu_{d}^{(\infty)})_{d|q}$ can be calculated by Algorithm~\ref{alg:main} (cf. \thref{th:mu_d}).
\end{theorem}

\begin{IEEEproof}[Proof of \thref{th:polarization}]
See \appref{app:polarization}.
\end{IEEEproof}

\thref{th:polarization} immediately proves \corref{cor:multilevel} of \sectref{sect:asymptotic_distribution_MAEC};
this corollary formally characterizes the asymptotic distribution of multilevel channel polarization for MAECs.
See \appref{app:multilevel} for the proof of \corref{cor:multilevel}.

\section{Concluding Remarks}
\label{sect:conclusion}

We have proposed a general type of erasure-like channels called \emph{modular arithmetic erasure channels} (MAECs).
Similar to the well-known recursive formulas of the polar transform for a BEC, in \thref{th:recursive_V} and \corref{cor:recursive_V}, we derived the recursive formulas for an MAEC.
Hence, the MAEC is a simple toy model to study the phenomenon of multilevel channel polarization.
In \sectref{sect:asymptotic_distribution_MAEC}, we exactly characterized the \emph{asymptotic distribution} of multilevel channel polarization for an MAEC.
In particular, we also established an algorithm to calculate the asymptotic distribution in Algorithm~\ref{alg:main}.
This partially solves an open problem in the study of non-binary polar coding (cf.\ \cite[Section~9.2.1]{nasser_PhD}).

An interesting future work is to generalize the results in \sectref{sect:asymptotic_distribution_MAEC} from MAECs to general DMCs.
On the other hand, it is also interesting to generalize the requirement of working over a ring $\mathbb{Z}/q\mathbb{Z}$ to weaker algebraic structures.
Recently, the present authors \cite{isit2019} generalized the results of this study to the case in which the polar transform is defined on a group with infinite order and where the polar coding is used in the context of source coding.
In short, this follow-up work \cite{isit2019} generalizes the results in \sectref{sect:asymptotic_distribution_MAEC} from a cyclic group $(\mathbb{Z}/q\mathbb{Z}, +)$ to a locally cyclic group.
It is worth pointing out that generalizing the algebraic structure is important to deal with the \emph{multiple access channel polarization.}
Abbe and Telatar \cite{abbe_telatar_it2012} considered the polar transform for $m$-user multiple-access channels over an elementary abelian group $\mathbb{F}_{2}^{m} =  \mathbb{F}_{2} \times \mathbb{F}_{2} \times \dots \times \mathbb{F}_{2}$.
Nasser and Telatar \cite[Section~VII]{nasser_telatar_it2016} and Nasser \cite{nasser_it2017_fourier} also considered polar transforms defined on an arbitrary finite abelian group.
Specifically, Nasser and Telatar \cite[Section~VIII]{nasser_telatar_it2016} studied another erasure-like channel called a \emph{combination of $l$ linear channels} in which the input alphabet is given as an elementary abelian group.
On the other hand, it is difficult to exactly characterize the asymptotic distribution of such an erasure-like channel (cf. \cite[Section~VIII-B]{nasser_telatar_it2016}).
Generalizing our results to an abelian, or a non-abelian, group is of interests to gain deeper understanding of multilevel channel polarization or polarization in network information theory problems (such as the multiple-access channel).

\section*{Acknowledgement}

The authors would like to thank Prof.~Vincent~Y.~F.~Tan for greatly improving the presentation of this paper.
The authors are also grateful to Prof.\ Krishna Narayanan and anonymous reviewers for their helpful comments.
Particularly, one of the anonymous reviewers corrected issues in the proof of \corref{cor:multilevel} (\appref{app:multilevel}) and simplified the proof of \lemref{lem:formulas} (\appref{app:formulas}).

\appendices

\section{Proof of \lemref{lem:invariant_degradedness}}
\label{app:invariant_degradedness}

By \defref{def:output_equiv}, there exist two channels $Q_{1} : \mathcal{Z}_{1} \to \mathcal{Y}_{1}$ and $Q_{2} : \mathcal{Z}_{2} \to \mathcal{Y}_{2}$ satisfying
\begin{align}
W_{1}(y_{1} \mid x_{1})
& =
\sum_{z_{1} \in \mathcal{Z}_{1}} Q_{1}(y_{1} \mid z_{1}) \, \tilde{W}_{1}(z_{1} \mid x_{1}) ,
\\
W_{2}(y_{2} \mid x_{2})
& =
\sum_{z_{2} \in \mathcal{Z}_{2}} Q_{2}(y_{2} \mid z_{2}) \, \tilde{W}_{2}(z_{2} \mid x_{2}) .
\end{align}
For each $(u_{1}, y_{1}, y_{2}) \in \mathbb{Z}/q\mathbb{Z} \times \mathcal{Y}_{1} \times \mathcal{Y}_{2}$, we have
\begin{align}
(W_{1} \boxast W_{2})(y_{1}, y_{2} \mid u_{1})
& =
\sum_{u_{2}^{\prime} \in \mathbb{Z}/q\mathbb{Z}} \frac{1}{q} W_{1}(y_{1} \mid u_{1} + \gamma \cdot u_{2}^{\prime}) \, W_{2}(y_{2} \mid u_{2}^{\prime})
\notag \\
& =
\sum_{u_{2}^{\prime} \in \mathbb{Z}/q\mathbb{Z}} \frac{1}{q} \left( \sum_{z_{1} \in \mathcal{Z}_{1}} Q_{1}(y_{1} \mid z_{1}) \, \tilde{W}_{1}(z_{1} \mid u_{1} + \gamma \cdot u_{2}^{\prime}) \right) \left( \sum_{z_{2} \in \mathcal{Z}_{2}} Q_{2}(y_{2} \mid z_{2}) \, \tilde{W}_{2}(z_{2} \mid u_{2}^{\prime}) \right)
\notag \\
& =
\sum_{(z_{1}, z_{2}) \in \mathcal{Z}_{1} \times \mathcal{Z}_{2}} Q_{1}(y_{1} \mid z_{1}) \, Q_{2}(y_{2} \mid z_{2}) + \sum_{u_{2}^{\prime} \in \mathbb{Z}/q\mathbb{Z}} \frac{1}{q} \tilde{W}_{1}(z_{1} \mid u_{1} + \gamma \cdot u_{2}^{\prime}) \, \tilde{W}_{2}(z_{2} \mid u_{2}^{\prime})
\notag \\
& =
\sum_{(z_{1}, z_{2}) \in \mathcal{Z}_{1} \times \mathcal{Z}_{2}} Q_{1, 2}(y_{1}, y_{2} \mid z_{1}, z_{2}) \, (\tilde{W}_{1} \boxast \tilde{W}_{2})(z_{1}, z_{2} \mid u_{1}) ,
\end{align}
which implies that $W_{1} \boxast W_{2} \preceq \tilde{W}_{1} \boxast \tilde{W}_{2}$, where the product channel $Q_{1, 2} : \mathcal{Z}_{1} \times \mathcal{Z}_{2} \to \mathcal{Y}_{1} \times \mathcal{Y}_{2}$ is given by
\begin{align}
Q_{1, 2}(y_{1}, y_{2} \mid z_{1}, z_{2})
& =
Q_{1}(y_{1} \mid z_{1}) \, Q_{2}(y_{2} \mid z_{2}) .
\end{align}
Similarly, for each $(u_{1}, u_{2}, y_{1}, y_{2}) \in (\mathbb{Z}/q\mathbb{Z})^{2} \times \mathcal{Y}_{1} \times \mathcal{Y}_{2}$, we see that
\begin{align}
(W_{1} \varoast W_{2})(y_{1}, y_{2}, u_{1} \mid u_{2})
& =
\frac{ 1 }{ q } W_{1}(y_{1} \mid u_{1} + \gamma \cdot u_{2}) \, W_{2}(y_{2} \mid u_{2})
\notag \\
& =
\frac{ 1 }{ q } \left( \sum_{z_{1} \in \mathcal{Z}_{1}} Q_{1}(y_{1} \mid z_{1}) \, \tilde{W}_{1}(z_{1} \mid u_{1} + \gamma \cdot u_{2}) \right) \, \left( \sum_{z_{2} \in \mathcal{Z}_{2}} Q_{2}(y_{2} \mid z_{2}) \, \tilde{W}_{2}(z_{2} \mid u_{2}) \right)
\notag \\
& =
\sum_{(z_{1}, z_{2}) \in \mathcal{Z}_{1} \times \mathcal{Z}_{2}} Q_{1}(y_{1} \mid z_{1}) \, Q_{2}(y_{2} \mid z_{2}) \, \left( \frac{ 1 }{ q } \tilde{W}_{1}(z_{1} \mid u_{1} + \gamma \cdot u_{2}) \, \tilde{W}_{2}(z_{2} \mid u_{2}) \right)
\notag \\
& =
\sum_{(z_{1}, z_{2}) \in \mathcal{Z}_{1} \times \mathcal{Z}_{2}} Q_{1, 2}(y_{1}, y_{2} \mid z_{1}, z_{2}) \, (\tilde{W}_{1} \varoast \tilde{W}_{2})(z_{1}, z_{2}, u_{1} \mid u_{2})
\notag \\
& =
\sum_{(z_{1}, z_{2}, u_{1}^{\prime}) \in \mathcal{Z}_{1} \times \mathcal{Z}_{2} \times \mathcal{X}} \hat{Q}_{1, 2}(y_{1}, y_{2}, u_{1} \mid z_{1}, z_{2}, u_{1}^{\prime}) \, (\tilde{W}_{1} \varoast \tilde{W}_{2})(z_{1}, z_{2}, u_{1}^{\prime} \mid u_{2}) ,
\end{align}
which implies that $W_{1} \varoast W_{2} \preceq \tilde{W}_{1} \varoast \tilde{W}_{2}$, where the channel $\hat{Q}_{1, 2} : \mathcal{Z}_{1} \times \mathcal{Z}_{2} \times \mathbb{Z}/q\mathbb{Z} \to \mathcal{Y}_{1} \times \mathcal{Y}_{2} \times \mathbb{Z}/q\mathbb{Z}$ is given by
\begin{align}
\hat{Q}_{1, 2}(y_{1}, y_{2}, u_{1} \mid z_{1}, z_{2}, u_{1}^{\prime})
=
\begin{cases}
Q_{1, 2}(y_{1}, y_{2} \mid z_{1}, z_{2})
& \mathrm{if} \ u_{1} = u_{1}^{\prime} ,
\\
0
& \mathrm{if} \ u_{1} \neq u_{1}^{\prime} .
\end{cases}
\end{align}
This completes the proof of \lemref{lem:invariant_degradedness}.
\hfill\IEEEQEDhere

\section{Proof of \propref{prop:I(V)}}
\label{app:proof_prop:I(V)}

A direct calculation shows that
\begin{align}
I_{\alpha}( V_{\bvec{\varepsilon}} )
& =
\frac{ \alpha }{ \alpha - 1 } \log \left( \sum_{y \in \mathcal{Y}} \left( \sum_{x \in \mathcal{X}} \frac{ 1 }{ q } V_{\bvec{\varepsilon}}(y \mid x)^{\alpha} \right)^{1/\alpha} \right)
\notag \\
& =
\frac{ \alpha }{ \alpha - 1 } \log \left( \sum_{d|q} \sum_{y \in \mathbb{Z}/d\mathbb{Z}} \left( \sum_{x \in \mathbb{Z}/q\mathbb{Z}} \frac{ 1 }{ q } V_{\bvec{\varepsilon}}(y \mid x)^{\alpha} \right)^{1/\alpha} \right)
\notag \\
& =
\frac{ \alpha }{ \alpha - 1 } \log \left( \sum_{d|q} \sum_{y \in \mathbb{Z}/d\mathbb{Z}} \left( \sum_{\substack{ x \in \mathbb{Z}/q\mathbb{Z} : \\ x \equiv y \ (\mathrm{mod} \, d) }} \frac{ 1 }{ q } \, \varepsilon_{d}^{\alpha} \right)^{1/\alpha} \right)
\notag \\
& =
\frac{ \alpha }{ \alpha - 1 } \log \left( \sum_{d|q} \sum_{y \in \mathbb{Z}/d\mathbb{Z}} \left( \frac{ q }{ d } \frac{ 1 }{ q } \, \varepsilon_{d}^{\alpha} \right)^{1/\alpha} \right)
\notag \\
& =
\frac{ \alpha }{ \alpha - 1 } \log \left( \sum_{d|q} \sum_{y \in \mathbb{Z}/d\mathbb{Z}} \frac{ \varepsilon_{d} }{ d^{1/\alpha} } \right)
\notag \\
& =
\frac{ \alpha }{ \alpha - 1 } \log \left( \sum_{d|q} \frac{ \varepsilon_{d} }{ d^{(1/\alpha) - 1} } \right)
\notag \\
& =
\frac{ \alpha }{ \alpha - 1 } \log \left( \sum_{d|q} d^{(\alpha - 1)/\alpha} \, \varepsilon_{d} \right)
\end{align}
for each $\alpha \in (0, 1) \cup (1, \infty)$.
The rest of formulas can be verified as follows:
\begin{align}
I_{0}( V_{\bvec{\varepsilon}} )
& =
\min_{y \in \mathcal{Y}} \left( \log \frac{ q }{ | \{ x \in \mathcal{X} \mid V_{\bvec{\varepsilon}}(y \mid x) > 0 \} | } \right)
\notag \\
& =
\min_{d|q} \min_{y \in \mathbb{Z}/d\mathbb{Z}} \left( \log \frac{ q }{ | \{ x \in \mathcal{X} \mid V_{\bvec{\varepsilon}}(y \mid x) > 0 \} | } \right)
\notag \\
& =
\min_{d|q : \varepsilon_{d} > 0} \min_{y \in \mathbb{Z}/d\mathbb{Z}} \left( \log \frac{ q }{ (q/d) } \right)
\notag \\
& =
\min_{d|q : \varepsilon_{d} > 0} \Big( \log d \Big) ,
\\
I( V_{\bvec{\varepsilon}} )
& =
\sum_{y \in \mathcal{Y}} \sum_{x \in \mathcal{X}} \frac{ 1 }{ q } V_{\bvec{\varepsilon}}(y \mid x) \log \frac{ V_{\bvec{\varepsilon}}(y \mid x) }{ \sum_{x^{\prime} \in \mathcal{X}} (1/q) V_{\bvec{\varepsilon}}(y \mid x^{\prime}) }
\notag \\
& =
\sum_{d|q} \sum_{y \in \mathbb{Z}/d\mathbb{Z}} \sum_{x \in \mathbb{Z}/q\mathbb{Z}} \frac{ 1 }{ q } V_{\bvec{\varepsilon}}(y \mid x) \log \frac{ V_{\bvec{\varepsilon}}(y \mid x) }{ \sum_{x^{\prime} \in \mathbb{Z}/q\mathbb{Z}} (1/q) V_{\bvec{\varepsilon}}(y \mid x^{\prime}) }
\notag \\
& =
\sum_{d|q} \sum_{y \in \mathbb{Z}/d\mathbb{Z}} \sum_{\substack{ x \in \mathbb{Z}/q\mathbb{Z} : \\ x \equiv y \ (\mathrm{mod} \, d) }} \frac{ 1 }{ q } \, \varepsilon_{d} \left( \log \varepsilon_{d} - \log \sum_{\substack{ x^{\prime} \in \mathbb{Z}/q\mathbb{Z} : \\ x^{\prime} \equiv y \ (\mathrm{mod} \, d) }} \frac{ 1 }{ q } \, \varepsilon_{d} \right)
\notag \\
& =
\sum_{d|q} d \, \frac{ q }{ d } \, \frac{ 1 }{ q } \, \varepsilon_{d} \log d
\notag \\
& =
\sum_{d|q} (\log d) \, \varepsilon_{d} ,
\\
I_{\infty}( V_{\bvec{\varepsilon}} )
& =
\log \left( \sum_{y \in \mathcal{Y}} \max_{x \in \mathcal{X}} V_{\bvec{\varepsilon}}(y \mid x) \right)
\notag \\
& =
\log \left( \sum_{d|q} \sum_{y \in \mathbb{Z}/d\mathbb{Z}} \max_{x \in \mathbb{Z}/q\mathbb{Z}} V_{\bvec{\varepsilon}}(y \mid x) \right)
\notag \\
& =
\log \left( \sum_{d|q} \sum_{y \in \mathbb{Z}/d\mathbb{Z}} \varepsilon_{d} \right)
\notag \\
& =
\log \left( \sum_{d|q} d \, \varepsilon_{d} \right) .
\end{align}
This completes the proof of \propref{prop:I(V)}.
\hfill\IEEEQEDhere

\section{Proof of \thref{th:recursive_V}}
\label{app:recursive_V}

Given $a, b \in \mathbb{Z}/q\mathbb{Z}$ and $d|q$, define the congruence between $a$ and $b$ modulo $d$ as
\begin{align}
a \equiv b \pmod{d}
\overset{\text{def}}{\iff}
a + d \mathbb{Z}
=
b + d \mathbb{Z} .
\label{def:congruence}
\end{align}
To prove \thref{th:recursive_V}, we employ the following well-known result in elementary number theory.

\begin{lemma}[A variant of Chinese Remainder Theorem]
\label{lem:CRT}
Let $d_{1}, d_{2} \in \mathbb{N}$.
For every $a$ and $b$, the system of two congruences
\begin{align}
z
& \equiv
a
\pmod{d_{1}} ,
\label{eq:congruence_d1} \\
z
& \equiv
b
\pmod{d_{2}}
\label{eq:congruence_d2} 
\end{align}
has a solution $z$ if and only if
\begin{align}
a
\equiv
b
\pmod{\gcd(d_{1}, d_{2})} .
\label{eq:condition_gcd}
\end{align}
In particular, when the solution $z$ exists, it is unique modulo $\lcm(d_{1}, d_{2})$.
\end{lemma}

We now introduce two useful notations.
Let $P$ be a probability distribution on $\mathcal{X}$ and $W : \mathcal{X} \to \mathcal{Y}$ a channel.
Then, the output distribution $PW$ on $\mathcal{Y}$ is defined by
\begin{align}
PW( y )
\coloneqq
\sum_{x \in \mathcal{X}} P( x ) \, W(y \mid x)
\label{def:output}
\end{align}
for each $y \in \mathcal{Y}$.
In addition, the backward channel $\overline{W}_{P} : \mathcal{Y} \to \mathcal{X}$ is defined by
\begin{align}
\overline{W}_{P}(x \mid y)
\coloneqq
\frac{ P( x ) \, W(y \mid x) }{ PW( y ) }
\label{def:backward}
\end{align}
for each $(x, y) \in \mathcal{X} \times \mathcal{Y}$.
Throughout this proof, assume that every input distribution $P$ is uniform, and we drop the subscript, writing $\overline{W}_{P}$ as $\overline{W}$ for brevity.

Recall that MAECs are defined in \defref{def:V}.
Let $q \ge 2$ be an integer.
Consider two probability vectors $\bvec{\varepsilon} = ( \varepsilon_{d} )_{d|q}$ and $\bvec{\varepsilon}^{\prime} = ( \varepsilon_{d}^{\prime} )_{d|q}$.
By the construction of the output alphabet $\mathcal{Y}$ (see \eqref{def:alphabetY}), note that each output symbol $y \in \mathcal{Y}$ can be written by $y = z + d \mathbb{Z}$ for some $z \in \mathbb{Z}/q\mathbb{Z}$ and some divisor $d|q$.
It follows from \eqref{eq:V} and \eqref{def:output} that
\begin{align}
PV_{\bvec{\varepsilon}}( z + d \mathbb{Z} )
& =
\sum_{x \in \mathbb{Z}/q\mathbb{Z}} \frac{ 1 }{ q } \, V_{\bvec{\varepsilon}}(z + d \mathbb{Z} \mid x)
\notag \\
& =
\sum_{\substack{ x \in \mathbb{Z}/q\mathbb{Z} : \\ x \equiv z \ (\mathrm{mod} \, d) }} \frac{ \varepsilon_{d} }{ q }
\notag \\
& =
\frac{ \varepsilon_{d} }{ d }
\label{eq:output_V}
\end{align}
for each $z \in \mathbb{Z}/q\mathbb{Z}$ and $d|q$.
In addition, it follows from \eqref{eq:V}, \eqref{eq:output_V}, and \eqref{def:backward} that
\begin{align}
\overline{V_{\bvec{\varepsilon}}}(x \mid z + d\mathbb{Z})
& =
\frac{ 1 }{ q } \frac{ V_{\bvec{\varepsilon}}(z + d\mathbb{Z} \mid x) }{ PV_{\bvec{\varepsilon}}( z + d\mathbb{Z} ) }
\notag \\
& =
\frac{ d }{ q } \frac{ V_{\bvec{\varepsilon}}(z + d\mathbb{Z} \mid x) }{ \varepsilon_{d} }
\notag \\
& =
\begin{dcases}
\frac{ d }{ q }
& \mathrm{if} \ x \equiv z \pmod{d} ,
\\
0
& \mathrm{otherwise} ,
\end{dcases}
\label{eq:backward_V}
\end{align}
provided that $\varepsilon_{d} > 0$, for each $x, z \in \mathbb{Z}/q\mathbb{Z}$ and $d|q$.
Similarly, one has
\begin{align}
PV_{\bvec{\varepsilon}^{\prime}}( z + d \mathbb{Z} )
=
\frac{ \varepsilon_{d}^{\prime} }{ d }
\label{eq:output_V_prime}
\end{align}
and
\begin{align}
\overline{V_{\bvec{\varepsilon}^{\prime}}}(x \mid z + d \mathbb{Z})
=
\begin{dcases}
\frac{ d }{ q }
& \mathrm{if} \ x \equiv z \pmod{d} ,
\\
0
& \mathrm{otherwise} ,
\end{dcases}
\label{eq:backward_V_prime}
\end{align}
provided that $\varepsilon_{d}^{\prime} > 0$, for each $z \in \mathbb{Z}/q\mathbb{Z}$ and $d|q$.

Given a unit $\gamma \in \mathbb{Z}/q\mathbb{Z}$, consider the worse channel $V_{\bvec{\varepsilon}} \boxast V_{\bvec{\varepsilon}^{\prime}}$ and the better channel $V_{\bvec{\varepsilon}} \varoast V_{\bvec{\varepsilon}^{\prime}}$ defined in \eqref{def:minus} and \eqref{def:plus}, respectively.
We first prove the assertion of \thref{th:recursive_V} for the worse channel.

\subsection{Proof for the Worse Channel $V_{\bvec{\varepsilon}} \boxast V_{\bvec{\varepsilon}^{\prime}}$}

It follows from \eqref{def:minus} and \eqref{def:output} that
\begin{align}
P(V_{\bvec{\varepsilon}} \boxast V_{\bvec{\varepsilon}^{\prime}})(y_{1}, y_{2})
& =
\sum_{u_{1} \in \mathcal{X}} \frac{ 1 }{ q } (V_{\bvec{\varepsilon}} \boxast V_{\bvec{\varepsilon}^{\prime}})(y_{1}, y_{2} \mid u_{1})
\notag \\
& =
\sum_{u_{1} \in \mathcal{X}} \frac{ 1 }{ q } \sum_{u_{2}^{\prime} \in \mathcal{X}} \frac{ 1 }{ q } V_{\bvec{\varepsilon}}(y_{1} \mid u_{1} + \gamma \cdot u_{2}^{\prime}) \, V_{\bvec{\varepsilon}^{\prime}}(y_{2} \mid u_{2}^{\prime})
\notag \\
& =
\left( \sum_{u_{1} \in \mathcal{X}} \frac{ 1 }{ q } V_{\bvec{\varepsilon}}(y_{1} \mid u_{1}) \right) \left( \sum_{u_{2}^{\prime} \in \mathcal{X}} \frac{ 1 }{ q } V_{\bvec{\varepsilon}^{\prime}}(y_{2} \mid u_{2}^{\prime}) \right)
\notag \\
& =
PV_{\bvec{\varepsilon}}( y_{1} ) \, PV_{\bvec{\varepsilon}^{\prime}}( y_{2} )
\label{eq:output_Vminus}
\end{align}
for each $y_{1}, y_{2} \in \mathcal{Y}$, where the third equality follows from the fact that the map $a \mapsto a + \gamma \cdot b$ is a bijection on $\mathbb{Z}/q\mathbb{Z}$ for each $b \in \mathbb{Z}/q\mathbb{Z}$.
Moreover, it follows from \eqref{def:minus}, \eqref{def:backward}, and \eqref{eq:output_Vminus} that
\begin{align}
\overline{(V_{\bvec{\varepsilon}} \boxast V_{\bvec{\varepsilon}^{\prime}})}(u_{1} \mid y_{1}, y_{2})
& =
\frac{ 1 }{ q } \frac{ (V_{\bvec{\varepsilon}} \boxast V_{\bvec{\varepsilon}^{\prime}})(y_{1}, y_{2} \mid u_{1}) }{ P(V_{\bvec{\varepsilon}} \boxast V_{\bvec{\varepsilon}^{\prime}})(y_{1}, y_{2}) }
\notag \\
& =
\frac{ 1 }{ q } \frac{ (V_{\bvec{\varepsilon}} \boxast V_{\bvec{\varepsilon}^{\prime}})(y_{1}, y_{2} \mid u_{1}) }{ PV_{\bvec{\varepsilon}}( y_{1} ) \, PV_{\bvec{\varepsilon}^{\prime}}( y_{2} ) }
\notag \\
& =
\sum_{u_{2}^{\prime} \in \mathcal{X}} \left( \frac{ 1 }{ q } \frac{ V_{\bvec{\varepsilon}}(y_{1} \mid u_{1} + \gamma \cdot u_{2}^{\prime}) }{ PV_{\bvec{\varepsilon}}( y_{1} ) } \right) \left( \frac{ 1 }{ q } \frac{ V_{\bvec{\varepsilon}^{\prime}}(y_{2} \mid u_{2}^{\prime}) }{ PV_{\bvec{\varepsilon}^{\prime}}( y_{2} ) } \right)
\notag \\
& =
\sum_{u_{2}^{\prime} \in \mathcal{X}} \overline{V_{\bvec{\varepsilon}}}(u_{1} + \gamma \cdot u_{2}^{\prime} \mid y_{1}) \, \overline{V_{\bvec{\varepsilon}^{\prime}}}(u_{2}^{\prime} \mid y_{2}) ,
\label{eq:backward_Vminus}
\end{align}
provided that $PV_{\bvec{\varepsilon}}( y_{1} ) \, PV_{\bvec{\varepsilon}^{\prime}}( y_{2} ) > 0$, for each $(u_{1}, y_{1}, y_{2}) \in \mathcal{X} \times \mathcal{Y}^{2}$.
Furthermore, it follows from \eqref{eq:backward_V} and \eqref{eq:backward_V_prime} that
\begin{align}
\overline{V_{\bvec{\varepsilon}}}(u_{1} + \gamma \cdot u_{2}^{\prime} \mid z_{1} + d_{1} \mathbb{Z}) \, \overline{V_{\bvec{\varepsilon}^{\prime}}}(u_{2}^{\prime} \mid z_{2} + d_{2} \mathbb{Z})
=
\begin{dcases}
\frac{ d_{1} d_{2} }{ q^{2} }
& \mathrm{if} \ u_{1} + \gamma \cdot u_{2}^{\prime} \equiv z_{1} \pmod{ d_{1} } ,
\\
& \qquad\qquad u_{2}^{\prime} \equiv z_{2} \pmod{d_{2}} ,
\\
0
& \mathrm{otherwise} ,
\end{dcases}
\label{eq:bV1_times_bV2}
\end{align}
provided that $\varepsilon_{d_{1}} \, \varepsilon_{d_{2}}^{\prime} > 0$, for each $u_{1}, u_{2}^{\prime}, z_{1}, z_{2} \in \mathbb{Z}/q\mathbb{Z}$ and $d_{1}, d_{2}|q$.
Note that in \eqref{eq:bV1_times_bV2}, the system of two congruences
\begin{align}
u_{1} + \gamma \cdot u_{2}^{\prime}
& \equiv
z_{1} \pmod{d_{1}} ,
\\
u_{2}^{\prime}
& \equiv
z_{2} \pmod{d_{2}}
\end{align}
can be rewritten as
\begin{align}
u_{2}^{\prime}
& \equiv
\gamma^{-1} \cdot ( z_{1} - u_{1} ) \pmod{d_{1}} ,
\label{equiv:system1} \\
u_{2}^{\prime}
& \equiv
z_{2} \pmod{d_{2}} ;
\label{equiv:system2} 
\end{align}
and thus, it follows from \lemref{lem:CRT} that this system has a unique solution $u_{2}^{\prime} \in \mathbb{Z}/\lcm(d_{1}, d_{2})\mathbb{Z}$ if and only if
\begin{align}
\gamma^{-1} \cdot ( z_{1} - u_{1} ) 
\equiv
z_{2} \pmod{\gcd( d_{1}, d_{2})} ,
\label{eq:solution_condition}
\end{align}
which is equivalent to
\begin{align}
u_{1}
& \equiv
z_{1} - \gamma \cdot z_{2}
\pmod{ \gcd(d_{1}, d_{2}) } .
\end{align}
Therefore, for every $u_{1}, u_{2}^{\prime}, z_{1}, z_{2} \in \mathbb{Z}/q\mathbb{Z}$ and $d_{1}, d_{2}|q$ satisfying $\varepsilon_{d_{1}} \varepsilon_{d_{2}}^{\prime} > 0$, there exists a representative
\begin{align}
r
\in
(\gamma^{-1} \cdot (z_{1} - u_{1}) + d_{1} \mathbb{Z}) \cap (z_{2} + d_{2} \mathbb{Z})
\end{align}
such that
\begin{align}
\overline{V_{\bvec{\varepsilon}}}(u_{1} + \gamma \cdot u_{2}^{\prime} \mid z_{1} + d_{1} \mathbb{Z}) \, \overline{V_{\bvec{\varepsilon}^{\prime}}}(u_{2}^{\prime} \mid z_{2} + d_{2} \mathbb{Z})
=
\begin{dcases}
\frac{ d_{1} d_{2} }{ q^{2} }
& \mathrm{if} \ u_{1} \equiv z_{1} - \gamma \cdot z_{2} \pmod{ \gcd(d_{1}, d_{2}) } ,
\\
& \quad u_{2}^{\prime} \equiv r \pmod{\lcm(d_{1}, d_{2})} ,
\\
0
& \mathrm{otherwise} .
\end{dcases}
\label{eq:product_bV1_bV2}
\end{align}
Hence, we have from \eqref{eq:backward_Vminus} and \eqref{eq:product_bV1_bV2} that
\begin{align}
\overline{(V_{\bvec{\varepsilon}} \boxast V_{\bvec{\varepsilon}^{\prime}})}(u_{1} \mid z_{1} + d_{1}\mathbb{Z}, z_{2} + d_{2}\mathbb{Z})
& =
\sum_{u_{2}^{\prime} \in \mathbb{Z}/q\mathbb{Z}} \overline{V_{\bvec{\varepsilon}}}(u_{1} + \gamma \cdot u_{2}^{\prime} \mid z_{1} + d_{1}\mathbb{Z}) \, \overline{V_{\bvec{\varepsilon}^{\prime}}}(u_{2}^{\prime} \mid z_{2} + d_{2}\mathbb{Z})
\notag \\
& =
\begin{dcases}
\sum_{\substack{ u_{2}^{\prime} \in \mathbb{Z}/q\mathbb{Z} : \\ u_{2}^{\prime} \equiv r \ (\mathrm{mod} \, \lcm(d_{1}, d_{2})) }} \frac{ d_{1} d_{2} }{ q^{2} }
& \mathrm{if} \ u_{1} \equiv z_{1} - \gamma \cdot z_{2} \pmod{ \gcd(d_{1}, d_{2}) } ,
\\
0
& \mathrm{otherwise}
\end{dcases}
\notag \\
& =
\begin{dcases}
\frac{ q }{ \lcm(d_{1}, d_{2}) } \frac{ d_{1} d_{2} }{ q^{2} }
& \mathrm{if} \ u_{1} \equiv z_{1} - \gamma \cdot z_{2} \pmod{ \gcd(d_{1}, d_{2}) } ,
\\
0
& \mathrm{otherwise}
\end{dcases}
\notag \\
& =
\begin{dcases}
\frac{ \gcd(d_{1}, d_{2}) }{ q }
& \mathrm{if} \ u_{1} \equiv z_{1} - \gamma \cdot z_{2} \pmod{ \gcd(d_{1}, d_{2}) } ,
\\
0
& \mathrm{otherwise} ,
\end{dcases}
\label{eq:bV_minus}
\end{align}
provided that $\varepsilon_{d_{1}} \, \varepsilon_{d_{2}}^{\prime} > 0$, for each $u_{1}, z_{1}, z_{2} \in \mathbb{Z}/q\mathbb{Z}$ and $d_{1}, d_{2}|q$.
Therefore, it follows from \eqref{def:backward}, \eqref{eq:backward_V}, \eqref{eq:backward_V_prime}, \eqref{eq:output_Vminus}, and \eqref{eq:bV_minus} that
\begin{align}
(V_{\bvec{\varepsilon}} \boxast V_{\bvec{\varepsilon}^{\prime}})(z_{1} + d_{1} \mathbb{Z}, z_{2} + d_{2} \mathbb{Z} \mid u_{1})
& =
q \, \overline{(V_{\bvec{\varepsilon}} \boxast V_{\bvec{\varepsilon}^{\prime}})}(u_{1} \mid z_{1} + d_{1} \mathbb{Z}, z_{2} + d_{2} \mathbb{Z}) \, P(V_{\bvec{\varepsilon}} \boxast V_{\bvec{\varepsilon}^{\prime}})(z_{1} + d_{1} \mathbb{Z}, z_{2} + d_{2} \mathbb{Z})
\notag \\
& =
q \, \overline{(V_{\bvec{\varepsilon}} \boxast V_{\bvec{\varepsilon}^{\prime}})}(u_{1} \mid z_{1} + d_{1} \mathbb{Z}, z_{2} + d_{2} \mathbb{Z}) \, PV_{\bvec{\varepsilon}}( z_{1} + d_{1} \mathbb{Z} ) \, PV_{\bvec{\varepsilon}^{\prime}}(z_{2} + d_{2} \mathbb{Z})
\notag \\
& =
q \, \overline{(V_{\bvec{\varepsilon}} \boxast V_{\bvec{\varepsilon}^{\prime}})}(u_{1} \mid z_{1} + d_{1} \mathbb{Z}, z_{2} + d_{2} \mathbb{Z}) \, \frac{ \varepsilon_{d_{1}} }{ d_{1} } \frac{ \varepsilon_{d_{2}}^{\prime} }{ d_{2} }
\notag \\
& =
\begin{dcases}
\frac{ \varepsilon_{d_{1}} \, \varepsilon_{d_{2}}^{\prime} }{ \lcm(d_{1}, d_{2}) } 
& \mathrm{if} \ u_{1} \equiv z_{1} - \gamma \cdot z_{2} \pmod{ \gcd(d_{1}, d_{2}) } ,
\\
0
& \mathrm{otherwise}
\end{dcases}
\label{eq:Vminus}
\end{align}
for each $u_{1}, z_{1}, z_{2} \in \mathbb{Z}/q\mathbb{Z}$ and $d_{1}, d_{2}|q$.

Finally, to prove the equivalence between $V_{\bvec{\varepsilon}} \boxast V_{\bvec{\varepsilon}^{\prime}}$ and $V_{\bvec{\varepsilon} \boxast \bvec{\varepsilon}^{\prime}}$ with underlying probability vector $\bvec{\varepsilon} \boxast \bvec{\varepsilon}^{\prime} = ( \varepsilon_{d}^{\boxast} )_{d|q}$ given in \eqref{def:eps_minus}, it suffices to show the existence of two intermediate channels $Q_{1} : \mathcal{Y}^{2} \to \mathcal{Y}$ and $Q_{2} : \mathcal{Y} \to \mathcal{Y}^{2}$ ensuring that $V_{\bvec{\varepsilon} \boxast \bvec{\varepsilon}^{\prime}} \preceq V_{\bvec{\varepsilon}} \boxast V_{\bvec{\varepsilon}^{\prime}}$ and $V_{\bvec{\varepsilon}} \boxast V_{\bvec{\varepsilon}^{\prime}} \preceq V_{\bvec{\varepsilon} \boxast \bvec{\varepsilon}^{\prime}}$, respectively.
Define the channel $Q_{1} : \mathcal{Y}^{2} \to \mathcal{Y}$ by
\begin{align}
Q_{1}(z + d \mathbb{Z} \mid z_{1} + d_{1} \mathbb{Z}, z_{2} + d_{2} \mathbb{Z})
=
\begin{cases}
1
& \mathrm{if} \ \gcd(d_{1}, d_{2}) = d
\\
& \mathrm{and} \ z_{1} - \gamma \cdot z_{2} \equiv z \pmod{d} ,
\\
0
& \mathrm{otherwise}
\end{cases}
\label{def:Q1}
\end{align}
for each $z, z_{1}, z_{2} \in \mathbb{Z}/q\mathbb{Z}$ and $d, d_{1}, d_{2} | q$.
Then, a direct calculation shows that
\begin{align}
\sum_{y_{1}, y_{2} \in \mathcal{Y}} (V_{\bvec{\varepsilon}} \boxast V_{\bvec{\varepsilon}^{\prime}})(y_{1}, y_{2} \mid u_{1}) \, Q_{1}(z + d \mathbb{Z} \mid y_{1}, y_{2})
& =
\sum_{d_{1}|q} \sum_{y_{1} \in \mathbb{Z}/d_{1}\mathbb{Z}} \sum_{d_{2}|q} \sum_{y_{2} \in \mathbb{Z}/d_{2}\mathbb{Z}} (V_{\bvec{\varepsilon}} \boxast V_{\bvec{\varepsilon}^{\prime}})(y_{1}, y_{2} \mid u_{1}) \, Q_{1}(z + d \mathbb{Z} \mid y_{1}, y_{2})
\notag \\
& \overset{\mathclap{\text{(a)}}}{=}
\sum_{\substack{ d_{1}|q, d_{2}|q : \\ \gcd( d_{1}, d_{2} ) = d }} \sum_{\substack{ y_{1} \in \mathbb{Z}/d_{1}\mathbb{Z} , \\ y_{2} \in \mathbb{Z}/d_{2}\mathbb{Z} : \\ y_{1} - \gamma \cdot y_{2} \equiv z \pmod{d} }} (V_{\bvec{\varepsilon}} \boxast V_{\bvec{\varepsilon}^{\prime}})(y_{1}, y_{2} \mid u_{1})
\notag \\
& \overset{\mathclap{\text{(b)}}}{=}
\begin{dcases}
\sum_{\substack{ d_{1}|q, d_{2}|q : \\ \gcd( d_{1}, d_{2} ) = d }} \sum_{\substack{ y_{1} \in \mathbb{Z}/d_{1}\mathbb{Z} , \\ y_{2} \in \mathbb{Z}/d_{2}\mathbb{Z} : \\ y_{1} - \gamma \cdot y_{2} \equiv z \pmod{d} }} \frac{ \varepsilon_{d_{1}} \, \varepsilon_{d_{2}}^{\prime} }{ \lcm(d_{1}, d_{2}) }
& \mathrm{if} \ u_{1} \equiv z \pmod{ d } ,
\notag \\
0
& \mathrm{otherwise}
\end{dcases}
\notag \\
& =
\begin{dcases}
\sum_{\substack{ d_{1}|q, d_{2}|q : \\ \gcd( d_{1}, d_{2} ) = d }} \frac{ d_{1} d_{2} }{ \gcd(d_{1}, d_{2}) } \frac{ \varepsilon_{d_{1}} \, \varepsilon_{d_{2}}^{\prime} }{ \lcm(d_{1}, d_{2}) }
& \mathrm{if} \ u_{1} \equiv z \pmod{ d } ,
\notag \\
0
& \mathrm{otherwise}
\end{dcases}
\notag \\
& =
\begin{dcases}
\sum_{\substack{ d_{1}|q, d_{2}|q : \\ \gcd( d_{1}, d_{2} ) = d }} \varepsilon_{d_{1}} \, \varepsilon_{d_{2}}^{\prime}
& \mathrm{if} \ u_{1} \equiv z \pmod{ d } ,
\notag \\
0
& \mathrm{otherwise}
\end{dcases}
\notag \\
& \overset{\mathclap{\text{(c)}}}{=}
\begin{dcases}
\varepsilon_{d}^{\boxast}( \bvec{\varepsilon}, \bvec{\varepsilon}^{\prime} )
& \mathrm{if} \ u_{1} \equiv z \pmod{ d } ,
\notag \\
0
& \mathrm{otherwise}
\end{dcases}
\notag \\
& \overset{\mathclap{\text{(d)}}}{=}
V_{\bvec{\varepsilon} \boxast \bvec{\varepsilon}^{\prime}}(z + d \mathbb{Z} \mid u_{1})
\label{eq:equivalence_condition_minus1}
\end{align}
for every $u_{1}, z \in \mathbb{Z}/q\mathbb{Z}$ and $d|q$, where
\begin{itemize}
\item
(a) follows from \eqref{def:Q1},
\item
(b) follows from \eqref{eq:Vminus},
\item
(c) follows from \eqref{def:eps_minus}, and
\item
(d) follows from \eqref{eq:V}.
\end{itemize}
Similarly, define the DMC $Q_{2} : \mathcal{Y} \to \mathcal{Y}^{2}$ by
\begin{align}
Q_{2}(z_{1} + d_{1} \mathbb{Z}, z_{2} + d_{2} \mathbb{Z} \mid z + d \mathbb{Z})
& =
\begin{dcases}
\frac{ \varepsilon_{d_{1}} \, \varepsilon_{d_{2}}^{\prime} }{ \varepsilon_{d}^{\boxast} \, \lcm(d_{1}, d_{2}) }
& \mathrm{if} \ \gcd( d_{1}, d_{2} ) = d
\\
& \mathrm{and} \ z_{1} - \gamma \cdot z_{2} \equiv z \pmod{d} ,
\\
0
& \mathrm{otherwise}
\end{dcases}
\label{def:Q2}
\end{align}
for each $z, z_{1}, z_{2} \in \mathbb{Z}/q\mathbb{Z}$ and $d, d_{1}, d_{2} | q$.
Then a simple calculation yields that
\begin{align}
\sum_{y \in \mathcal{Y}} V_{\bvec{\varepsilon} \boxast \bvec{\varepsilon}^{\prime}}(y \mid u_{1}) \, Q_{2}(z_{1} + d_{1} \mathbb{Z}, z_{2} + d_{2} \mathbb{Z} \mid y)
& =
\sum_{d|q} \sum_{y \in \mathbb{Z}/d\mathbb{Z}} V_{\bvec{\varepsilon} \boxast \bvec{\varepsilon}^{\prime}}(y \mid u_{1}) \, Q_{2}(z_{1} + d_{1} \mathbb{Z}, z_{2} + d_{2} \mathbb{Z} \mid y)
\notag \\
& \overset{\mathclap{\text{(a)}}}{=}
\sum_{\substack{ d|q : \\ d = \gcd(d_{1}, d_{2}) }} \sum_{\substack{ y \in \mathbb{Z}/d\mathbb{Z} : \\ y \equiv z_{1} - \gamma \cdot z_{2} \ (\mathrm{mod} \, d) }} V_{\bvec{\varepsilon} \boxast \bvec{\varepsilon}^{\prime}}(y \mid u_{1}) \, \frac{ \varepsilon_{d_{1}} \, \varepsilon_{d_{2}}^{\prime} }{ \varepsilon_{d}^{\boxast} \, \lcm(d_{1}, d_{2}) }
\notag \\
& =
V_{\bvec{\varepsilon} \boxast \bvec{\varepsilon}^{\prime}}((z_{1} - \gamma \cdot z_{2}) + \gcd(d_{1}, d_{2}) \mathbb{Z} \mid u_{1}) \, \frac{ \varepsilon_{d_{1}} \, \varepsilon_{d_{2}}^{\prime} }{ \varepsilon_{\gcd(d_{1}, d_{2})}^{\boxast} \, \lcm(d_{1}, d_{2}) }
\notag \\
& \overset{\mathclap{\text{(b)}}}{=}
\begin{dcases}
\frac{ \varepsilon_{d_{1}} \, \varepsilon_{d_{2}}^{\prime} }{ \lcm(d_{1}, d_{2}) }
& \mathrm{if} \ u_{1} \equiv z_{1} - \gamma \cdot z_{2} \pmod{ \gcd(d_{1}, d_{2}) } ,
\\
0
& \mathrm{otherwise}
\end{dcases}
\notag \\
& \overset{\mathclap{\text{(c)}}}{=}
(V_{\bvec{\varepsilon}} \boxast V_{\bvec{\varepsilon}^{\prime}})(z_{1} + d_{1} \mathbb{Z}, z_{2} + d_{2} \mathbb{Z} \mid u_{1})
\label{eq:equivalence_condition_minus2}
\end{align}
for every $u_{1}, z_{1}, z_{2} \in \mathbb{Z}/q\mathbb{Z}$ and $d_{1}, d_{2} | q$, where
\begin{itemize}
\item
(a) follows from \eqref{def:Q2},
\item
(b) follows from \eqref{eq:V}, and
\item
(c) follows from \eqref{eq:Vminus}.
\end{itemize}
Therefore, we observe from \eqref{eq:equivalence_condition_minus1} and \eqref{eq:equivalence_condition_minus2} that $V_{\bvec{\varepsilon}} \boxast V_{\bvec{\varepsilon}^{\prime}}$ is equivalent to $V_{\bvec{\varepsilon} \boxast \bvec{\varepsilon}^{\prime}}$.
This completes the proof of \eqref{eq:maec_minus_recursive} in \thref{th:recursive_V}.

\subsection{Proof for the Better Channel $V_{\bvec{\varepsilon}} \varoast V_{\bvec{\varepsilon}^{\prime}}$}

After some algebra, we get
\begin{align}
P(V_{\bvec{\varepsilon}} \varoast V_{\bvec{\varepsilon}^{\prime}})( y_{1}, y_{2}, u_{1} )
& \overset{\mathclap{\text{(a)}}}{=}
\sum_{u_{2}^{\prime} \in \mathcal{X}} \frac{ 1 }{ q } (V_{\bvec{\varepsilon}} \varoast V_{\bvec{\varepsilon}^{\prime}})( y_{1}, y_{2}, u_{1} \mid u_{2}^{\prime} )
\notag \\
& \overset{\mathclap{\text{(b)}}}{=}
\sum_{u_{2}^{\prime} \in \mathcal{X}} \frac{ 1 }{ 	q^{2} } V_{\bvec{\varepsilon}}( y_{1} \mid u_{1} + \gamma \cdot u_{2}^{\prime} ) \, V_{\bvec{\varepsilon}^{\prime}}( y_{2} \mid u_{2}^{\prime} )
\notag \\
& = \,
PV_{\bvec{\varepsilon}}( y_{1} ) \, PV_{\bvec{\varepsilon}^{\prime}}( y_{2} ) \sum_{u_{2}^{\prime} \in \mathcal{X}} \left( \frac{ 1 }{ q } \frac{ V_{\bvec{\varepsilon}}( y_{1} \mid u_{1} + \gamma \cdot u_{2}^{\prime} ) }{ PV_{\bvec{\varepsilon}}( y_{1} ) } \right) \left( \frac{ 1 }{ q } \frac{ V_{\bvec{\varepsilon}^{\prime}}( y_{2} \mid u_{2}^{\prime} ) }{ PV_{\bvec{\varepsilon}^{\prime}}( y_{2} ) } \right)
\notag \\
& \overset{\mathclap{\text{(c)}}}{=}
PV_{\bvec{\varepsilon}}( y_{1} ) \, PV_{\bvec{\varepsilon}^{\prime}}( y_{2} ) \sum_{u_{2}^{\prime} \in \mathcal{X}} \overline{V_{\bvec{\varepsilon}}}( y_{1} \mid u_{1} + \gamma \cdot u_{2}^{\prime} ) \, \overline{V_{\bvec{\varepsilon}^{\prime}}}( y_{2} \mid u_{2}^{\prime} )
\notag \\
& \overset{\mathclap{\text{(d)}}}{=}
PV_{\bvec{\varepsilon}}( y_{1} ) \, PV_{\bvec{\varepsilon}^{\prime}}( y_{2} ) \, \overline{(V_{\bvec{\varepsilon}} \boxast V_{\bvec{\varepsilon}^{\prime}})}(u_{1} \mid y_{1}, y_{2}) ,
\label{eq:output_Vplus}
\end{align}
provided that $PV_{\bvec{\varepsilon}}( y_{1} ) \, PV_{\bvec{\varepsilon}^{\prime}}( y_{2} ) > 0$, for each $u_{1} \in \mathbb{Z}/q\mathbb{Z}$ and $y_{1}, y_{2} \in \mathcal{Y}$, where
\begin{itemize}
\item
(a) follows from \eqref{def:output},
\item
(b) follows from \eqref{def:plus},
\item
(c) follows from \eqref{def:backward}, and
\item
(d) follows from \eqref{eq:backward_Vminus}.
\end{itemize}
Moreover, noting that
\begin{align}
PV_{\bvec{\varepsilon}}( z_{1} + d_{1}\mathbb{Z} ) > 0
& \overset{\eqref{eq:output_V}}{\iff}
\varepsilon_{d_{1}} > 0,
\label{eq:condition1} \\
PV_{\bvec{\varepsilon}^{\prime}}( z_{2} + d_{2}\mathbb{Z} ) > 0
& \overset{\eqref{eq:output_V_prime}}{\iff}
\varepsilon_{d_{2}}^{\prime} > 0,
\label{eq:condition2} \\
\overline{(V_{\bvec{\varepsilon}} \boxast V_{\bvec{\varepsilon}^{\prime}})}(u_{1} \mid z_{1} + d_{1}\mathbb{Z}, z_{2} + d_{2}\mathbb{Z}) > 0
& \overset{\eqref{eq:bV_minus}}{\iff}
z_{1} - \gamma \cdot z_{2} \equiv u_{1} \pmod{\gcd(d_{1}, d_{2})} ,
\label{eq:condition3} 
\end{align}
we have
\begin{align}
\overline{(V_{\bvec{\varepsilon}} \varoast V_{\bvec{\varepsilon}^{\prime}})}(u_{2} \mid y_{1}, y_{2}, u_{1})
& \overset{\mathclap{\text{(a)}}}{=}
\frac{ 1 }{ q } \frac{ (V_{\bvec{\varepsilon}} \varoast V_{\bvec{\varepsilon}^{\prime}})(y_{1}, y_{2}, u_{1} \mid u_{2}) }{ P(V_{\bvec{\varepsilon}} \varoast V_{\bvec{\varepsilon}^{\prime}})( y_{1}, y_{2}, u_{1} ) }
\notag \\
& \overset{\mathclap{\text{(b)}}}{=}
\frac{ 1 }{ q } \frac{ (V_{\bvec{\varepsilon}} \varoast V_{\bvec{\varepsilon}^{\prime}})(y_{1}, y_{2}, u_{1} \mid u_{2}) }{ PV_{\bvec{\varepsilon}}( y_{1} ) \, PV_{\bvec{\varepsilon}^{\prime}}( y_{2} ) \, \overline{(V_{\bvec{\varepsilon}} \boxast V_{\bvec{\varepsilon}^{\prime}})}(u_{1} \mid y_{1}, y_{2}) }
\notag \\
& \overset{\mathclap{\text{(c)}}}{=}
\frac{ 1 }{ q^{2} } \frac{ V_{\bvec{\varepsilon}}( y_{1} \mid u_{1} + \gamma \cdot u_{2} ) \, V_{\bvec{\varepsilon}^{\prime}}( y_{2} \mid u_{2} ) }{ PV_{\bvec{\varepsilon}}( y_{1} ) \, PV_{\bvec{\varepsilon}^{\prime}}( y_{2} ) \, \overline{(V_{\bvec{\varepsilon}} \boxast V_{\bvec{\varepsilon}^{\prime}})}(u_{1} \mid y_{1}, y_{2}) }
\notag \\
& \overset{\mathclap{\text{(d)}}}{=}
\frac{ \overline{V_{\bvec{\varepsilon}}}( u_{1} + \gamma \cdot u_{2} \mid y_{1} ) \, \overline{V_{\bvec{\varepsilon}^{\prime}}}( u_{2} \mid y_{2} ) }{ \overline{(V_{\bvec{\varepsilon}} \boxast V_{\bvec{\varepsilon}^{\prime}})}(u_{1} \mid y_{1}, y_{2}) } ,
\label{eq:backward_Vplus}
\end{align}
provided that $PV_{\bvec{\varepsilon}}( y_{1} ) \, PV_{\bvec{\varepsilon}^{\prime}}( y_{2} ) \, \overline{(V_{\bvec{\varepsilon}} \boxast V_{\bvec{\varepsilon}^{\prime}})}(u_{1} \mid y_{1}, y_{2}) > 0$, for each $(u_{1}, u_{2}, y_{1}, y_{2}) \in \mathcal{X}^{2} \times \mathcal{Y}^{2}$, where
\begin{itemize}
\item
(a) follows from \eqref{def:backward},
\item
(b) follows from \eqref{eq:output_Vplus},
\item
(c) follows from \eqref{def:plus}, and
\item
(d) follows from \eqref{def:backward}.
\end{itemize}
Referring to the conditions in \eqref{eq:condition1}--\eqref{eq:condition2}, for every $u_{1}, u_{2}, z_{1}, z_{2} \in \mathbb{Z}/q\mathbb{Z}$ and $d_{1}, d_{2}|q$ satisfying $\varepsilon_{d_{1}} \varepsilon_{d_{2}}^{\prime} > 0$ and $z_{1} - \gamma \cdot z_{2} \equiv u_{1} \pmod{\gcd(d_{1}, d_{2})}$, we observe that
\begin{align}
\overline{(V_{\bvec{\varepsilon}} \varoast V_{\bvec{\varepsilon}^{\prime}})}(u_{2} \mid z_{1} + d_{1}\mathbb{Z}, z_{2} + d_{2}\mathbb{Z}, u_{1} )
& \overset{\mathclap{\text{(a)}}}{=}
\frac{ \overline{V_{\bvec{\varepsilon}}}( u_{1} + \gamma \cdot u_{2} \mid z_{1} + d_{1}\mathbb{Z} ) \, \overline{V_{\bvec{\varepsilon}^{\prime}}}( u_{2} \mid z_{2} + d_{2}\mathbb{Z} ) }{ \overline{(V_{\bvec{\varepsilon}} \boxast V_{\bvec{\varepsilon}^{\prime}})}(u_{1} \mid z_{1} + d_{1}\mathbb{Z}, z_{2} + d_{2}\mathbb{Z}) }
\notag \\
& \overset{\mathclap{\text{(b)}}}{=} \,
\frac{ q }{ \gcd(d_{1}, d_{2}) } \, \overline{V_{\bvec{\varepsilon}}}( u_{1} + \gamma \cdot u_{2} \mid z_{1} + d_{1}\mathbb{Z} ) \, \overline{V_{\bvec{\varepsilon}^{\prime}}}( u_{2} \mid z_{2} + d_{2}\mathbb{Z} )
\notag \\
& \overset{\mathclap{\text{(c)}}}{=}
\begin{dcases}
\frac{ \lcm(d_{1}, d_{2}) }{ q }
& \mathrm{if} \ u_{1} + \gamma \cdot u_{2} \equiv z_{1} \pmod{d_{1}} ,
\\
& \qquad\quad\ \; u_{2} \equiv z_{2} \pmod{d_{2}} ,
\\
0
& \mathrm{otherwise}
\end{dcases}
\notag \\
& \overset{\mathclap{\text{(d)}}}{=}
\begin{dcases}
\frac{ \lcm(d_{1}, d_{2}) }{ q }
& \mathrm{if} \ u_{2} \equiv r \pmod{\lcm(d_{1}, d_{2})} ,
\\
0
& \mathrm{otherwise} ,
\end{dcases}
\label{eq:backward_Vplus_complete}
\end{align}
where
\begin{itemize}
\item
(a) follows from \eqref{eq:backward_Vplus},
\item
(b) follows from \eqref{eq:bV_minus},
\item
(c) follows from \eqref{eq:product_bV1_bV2}, and
\item
(d) follows from \lemref{lem:CRT} with some solution $r \in (\gamma^{-1} (z_{1} - u_{1}) + d_{1} \mathbb{Z}) \cap (z_{2} + d_{2} \mathbb{Z})$ of the system of two congruences
\begin{align}
u_{1} + \gamma \cdot u_{2}
& \equiv
z_{1} \pmod{d_{1}} ,
\\
u_{2}
& \equiv
z_{2} \pmod{d_{2}}
\end{align}
with respect to $u_{2} \in \mathbb{Z}/q\mathbb{Z}$ for given $u_{1}, z_{1}, z_{2} \in \mathbb{Z}/q\mathbb{Z}$ and $d_{1}, d_{2} | q$.
\end{itemize}
Therefore, we have that
\begin{align}
&
(V_{\bvec{\varepsilon}} \varoast V_{\bvec{\varepsilon}^{\prime}})(z_{1} + d_{1}\mathbb{Z}, z_{2} + d_{2}\mathbb{Z}, u_{1} \mid u_{2})
\notag \\
& \qquad \qquad \overset{\mathclap{\text{(a)}}}{=}
q \, P(V_{\bvec{\varepsilon}} \varoast V_{\bvec{\varepsilon}^{\prime}})( z_{1} + d_{1}\mathbb{Z}, z_{2} + d_{2}\mathbb{Z}, u_{1} ) \, \overline{(V_{\bvec{\varepsilon}} \varoast V_{\bvec{\varepsilon}^{\prime}})}(u_{2} \mid z_{1} + d_{1}\mathbb{Z}, z_{2} + d_{2}\mathbb{Z}, u_{1} )
\notag \\
& \qquad \qquad \overset{\mathclap{\text{(b)}}}{=}
q \, PV_{\bvec{\varepsilon}}( z_{1} + d_{1}\mathbb{Z} ) \, PV_{\bvec{\varepsilon}^{\prime}}( z_{2} + d_{2}\mathbb{Z} ) \, \overline{(V_{\bvec{\varepsilon}} \boxast V_{\bvec{\varepsilon}^{\prime}})}(u_{1} \mid z_{1} + d_{1}\mathbb{Z}, z_{2} + d_{2}\mathbb{Z}) \, \overline{(V_{\bvec{\varepsilon}} \varoast V_{\bvec{\varepsilon}^{\prime}})}(u_{2} \mid z_{1} + d_{1}\mathbb{Z}, z_{2} + d_{2}\mathbb{Z}, u_{1} )
\notag \\
& \qquad \qquad \overset{\mathclap{\text{(c)}}}{=}
q \, \frac{ \varepsilon_{d_{1}} }{ d_{1} } \frac{ \varepsilon_{d_{2}}^{\prime} }{ d_{2} } \, \overline{(V_{\bvec{\varepsilon}} \boxast V_{\bvec{\varepsilon}^{\prime}})}(u_{1} \mid z_{1} + d_{1}\mathbb{Z}, z_{2} + d_{2}\mathbb{Z}) \, \overline{(V_{\bvec{\varepsilon}} \varoast V_{\bvec{\varepsilon}^{\prime}})}(u_{2} \mid z_{1} + d_{1}\mathbb{Z}, z_{2} + d_{2}\mathbb{Z}, u_{1} )
\notag \\
& \qquad \qquad \overset{\mathclap{\text{(d)}}}{=}
\begin{dcases}
\varepsilon_{d_{1}} \, \varepsilon_{d_{2}}^{\prime} \, \frac{ q }{ d_{1} d_{2} } \frac{ \gcd( d_{1}, d_{2} ) }{ q } \, \overline{(V_{\bvec{\varepsilon}} \varoast V_{\bvec{\varepsilon}^{\prime}})}(u_{2} \mid z_{1} + d_{1}\mathbb{Z}, z_{2} + d_{2}\mathbb{Z}, u_{1} )
& \mathrm{if} \ z_{1} - \gamma \cdot z_{2} \equiv u_{1} \pmod{ \gcd(d_{1}, d_{2}) } ,
\\
0
& \mathrm{otherwise}
\end{dcases}
\notag \\
& \qquad \qquad = \,
\begin{dcases}
\frac{ \varepsilon_{d_{1}} \, \varepsilon_{d_{2}}^{\prime} }{ \lcm(d_{1}, d_{2}) } \, \overline{(V_{\bvec{\varepsilon}} \varoast V_{\bvec{\varepsilon}^{\prime}})}(u_{2} \mid z_{1} + d_{1}\mathbb{Z}, z_{2} + d_{2}\mathbb{Z}, u_{1} )
& \mathrm{if} \ z_{1} - \gamma \cdot z_{2} \equiv u_{1} \pmod{ \gcd(d_{1}, d_{2}) } ,
\\
0
& \mathrm{otherwise}
\end{dcases}
\notag \\
& \qquad \qquad \overset{\mathclap{\text{(e)}}}{=}
\begin{dcases}
\frac{ \varepsilon_{d_{1}} \, \varepsilon_{d_{2}}^{\prime} }{ \lcm(d_{1}, d_{2}) } \frac{ \lcm(d_{1}, d_{2}) }{ q }
& \mathrm{if} \ z_{1} - \gamma \cdot z_{2} \equiv u_{1} \pmod{ \gcd(d_{1}, d_{2}) } ,
\\
& \qquad \qquad \ u_{2} \equiv r \ \, \pmod{ \lcm(d_{1}, d_{2}) }
\\
0
& \mathrm{otherwise}
\end{dcases}
\notag \\
& \qquad \qquad = \,
\begin{dcases}
\frac{ \varepsilon_{d_{1}} \, \varepsilon_{d_{2}}^{\prime} }{ q }
& \mathrm{if} \ z_{1} - \gamma \cdot z_{2} \equiv u_{1} \pmod{ \gcd(d_{1}, d_{2}) } ,
\\
& \qquad \qquad \ u_{2} \equiv r \ \, \pmod{ \lcm(d_{1}, d_{2}) } ,
\\
0
& \mathrm{otherwise}
\end{dcases}
\label{eq:Vplus}
\end{align}
for every $u_{1}, u_{2}, z_{1}, z_{2} \in \mathbb{Z}/q\mathbb{Z}$ and $d_{1}, d_{2}|q$ satisfying the right sides of the conditions \eqref{eq:condition1}--\eqref{eq:condition3}, where
\begin{itemize}
\item
(a) follows from \eqref{def:backward},
\item
(b) follows from \eqref{eq:output_Vplus},
\item
(c) follows from \eqref{eq:output_V} and \eqref{eq:output_V_prime},
\item
(d) follows from \eqref{eq:bV_minus}, and 
\item
(e) follows from \eqref{eq:backward_Vplus_complete}.
\end{itemize}
Note that by the definition stated in \eqref{eq:V}, we readily see that
\begin{align}
(V_{\bvec{\varepsilon}} \varoast V_{\bvec{\varepsilon}^{\prime}})(z_{1} + d_{1}\mathbb{Z}, z_{2} + d_{2}\mathbb{Z}, u_{1} \mid u_{2})
=
0
\label{eq:plus-zero}
\end{align}
for every $u_{1}, u_{2}, z_{1}, z_{2} \in \mathbb{Z}/q\mathbb{Z}$ and $d_{1}, d_{2} | q$ satisfying $\varepsilon_{d_{1}} \varepsilon_{d_{2}}^{\prime} = 0$.
Moreover, it follows from \lemref{lem:CRT} that \eqref{eq:plus-zero} also holds for every $u_{1}, u_{2}, z_{1}, z_{2} \in \mathbb{Z}/q\mathbb{Z}$ and $d_{1}, d_{2} | q$ in which
\begin{align}
z_{1} - \gamma \cdot z_{2}
\equiv
u_{1} \pmod{ \gcd(d_{1}, d_{2}) }
\end{align}
does not hold.
Hence, we conclude that \eqref{eq:Vplus} holds for every $u_{1}, u_{2}, z_{1}, z_{2} \in \mathbb{Z}/q\mathbb{Z}$ and $d_{1}, d_{2} | q$.

Finally, to prove the equivalence between $V_{\bvec{\varepsilon}} \varoast V_{\bvec{\varepsilon}^{\prime}}$ and $V_{\bvec{\varepsilon} \varoast \bvec{\varepsilon}^{\prime}}$ with underlying probability vector $\varepsilon \varoast \varepsilon^{\prime} = ( \varepsilon_{d}^{\varoast} )_{d|q}$ given in \eqref{def:eps_plus}, it suffices to show the existence of two intermediate channels $Q_{3} : \mathcal{Y}^{2} \times \mathbb{Z}/q\mathbb{Z} \to \mathcal{Y}$ and $Q_{4} : \mathcal{Y} \to \mathcal{Y}^{2} \times \mathbb{Z}/q\mathbb{Z}$ ensuring that $V_{\bvec{\varepsilon} \varoast \bvec{\varepsilon}^{\prime}} \preceq V_{\bvec{\varepsilon}} \varoast V_{\bvec{\varepsilon}^{\prime}}$ and $V_{\bvec{\varepsilon}} \varoast V_{\bvec{\varepsilon}^{\prime}} \preceq V_{\bvec{\varepsilon} \varoast \bvec{\varepsilon}^{\prime}}$, respectively.
Define the channel $Q_{3} : \mathcal{Y}^{2} \times \mathbb{Z}/q\mathbb{Z} \to \mathcal{Y}$ by
\begin{align}
Q_{3}(z + d\mathbb{Z} \mid z_{1} + d_{1}\mathbb{Z}, z_{2} + d_{2}\mathbb{Z}, u_{1})
=
\begin{cases}
1
& \mathrm{if} \ \lcm(d_{1}, d_{2}) = d
\\
& \mathrm{and} \ z \equiv r \pmod{ d }
\\
& \mathrm{and} \ z_{1} - \gamma \cdot z_{2} \equiv u_{1} \pmod{ \gcd(d_{1}, d_{2}) } ,
\\
0
& \mathrm{otherwise}
\end{cases}
\label{def:Q3}
\end{align}
for each $u_{1}, z, z_{1}, z_{2} \in \mathbb{Z}/q\mathbb{Z}$ and $d, d_{1}, d_{2} | q$.
Then, a direct calculation shows that
\begin{align}
&
\sum_{y_{1} \in \mathcal{Y}} \sum_{y_{2} \in \mathcal{Y}} \sum_{u_{1} \in \mathcal{X}} (V_{\bvec{\varepsilon}} \varoast V_{\bvec{\varepsilon}^{\prime}})(y_{1}, y_{2}, u_{1} \mid u_{2}) \, Q_{3}(z + d\mathbb{Z} \mid y_{1}, y_{2}, u_{1})
\notag \\
& \qquad \qquad \qquad \qquad \qquad =
\sum_{d_{1}|q} \sum_{y_{1} \in \mathbb{Z}/d_{1}\mathbb{Z}} \sum_{d_{2}|q} \sum_{y_{2} \in \mathbb{Z}/d_{2}\mathbb{Z}} \sum_{u_{1} \in \mathbb{Z}/q\mathbb{Z}} (V_{\bvec{\varepsilon}} \varoast V_{\bvec{\varepsilon}^{\prime}})(y_{1}, y_{2}, u_{1} \mid u_{2}) \, Q_{3}(z + d\mathbb{Z} \mid y_{1}, y_{2}, u_{1})
\notag \\
& \qquad \qquad \qquad \qquad \qquad \overset{\mathclap{\text{(a)}}}{=}
\sum_{\substack{ d_{1}|q, d_{2}|q : \\ \lcm(d_{1}, d_{2}) = d }} \sum_{\substack{ u_{1} \in \mathbb{Z}/q\mathbb{Z}, \\ y_{1} \in \mathbb{Z}/d_{1}\mathbb{Z} , \\ y_{2} \in \mathbb{Z}/d_{2}\mathbb{Z} : \\ y_{1} - \gamma \cdot y_{2} \equiv u_{1} \ (\mathrm{mod} \, \gcd(d_{1}, d_{2})) }} (V_{\bvec{\varepsilon}} \varoast V_{\bvec{\varepsilon}^{\prime}})(y_{1}, y_{2}, u_{1} \mid u_{2}) \, \mathbbm{1}[ z \equiv r \pmod{ d } ]
\notag \\
& \qquad \qquad \qquad \qquad \qquad \overset{\mathclap{\text{(b)}}}{=}
\begin{dcases}
\sum_{\substack{ d_{1}|q, d_{2}|q : \\ \lcm(d_{1}, d_{2}) = d }} \frac{ q }{ \lcm(d_{1}, d_{2}) } \frac{ d_{1} d_{2} }{ \gcd(d_{1}, d_{2}) } \frac{ \varepsilon_{d_{1}} \, \varepsilon_{d_{2}}^{\prime} }{ q }
& \mathrm{if} \ u_{2} \equiv z \pmod{d} ,
\\
0
& \mathrm{otherwise}
\end{dcases}
\notag \\
& \qquad \qquad \qquad \qquad \qquad =
\begin{dcases}
\sum_{\substack{ d_{1}|q, d_{2}|q : \\ \lcm(d_{1}, d_{2}) = d }} \varepsilon_{d_{1}} \, \varepsilon_{d_{2}}^{\prime}
& \mathrm{if} \ u_{2} \equiv z \pmod{d} ,
\\
0
& \mathrm{otherwise}
\end{dcases}
\notag \\
& \qquad \qquad \qquad \qquad \qquad \overset{\mathclap{\text{(c)}}}{=}
\begin{cases}
\varepsilon_{d}^{\varoast}(\bvec{\varepsilon}, \bvec{\varepsilon}^{\prime})
& \mathrm{if} \ u_{2} \equiv z \pmod{d} ,
\\
0
& \mathrm{otherwise}
\end{cases}
\notag \\
& \qquad \qquad \qquad \qquad \qquad \overset{\mathclap{\text{(d)}}}{=}
V_{\bvec{\varepsilon} \varoast \bvec{\varepsilon}^{\prime}}(z + d\mathbb{Z} \mid u_{2})
\label{eq:equivalence_condition_plus1}
\end{align}
for every $u_{2}, z \in \mathbb{Z}/q\mathbb{Z}$ and $d|q$, where
\begin{itemize}
\item
(a) follows from \eqref{def:Q3} and defining the indicator function of a condition $A$ as
\begin{align}
\mathbbm{1}[ A ]
\coloneqq
\begin{cases}
1
& \text{if $A$ is true} ,
\\
0
& \text{if $A$ is false} ,
\end{cases}
\end{align}
\item
(b) follows from \eqref{eq:Vplus},
\item
(c) follows from \eqref{def:eps_plus}, and
\item
(d) follows from \eqref{eq:V}.
\end{itemize}
Similarly, define the channel $Q_{4} : \mathcal{Y} \to \mathcal{Y}^{2} \times \mathbb{Z}/q\mathbb{Z}$ by
\begin{align}
Q_{4}(z_{1} + d_{1}\mathbb{Z}, z_{2} + d_{2}\mathbb{Z}, u_{1} \mid z + d\mathbb{Z})
& =
\begin{dcases}
\frac{ \varepsilon_{d_{1}} \, \varepsilon_{d_{2}}^{\prime} }{ q \, \varepsilon_{d}^{\varoast}( \bvec{\varepsilon} , \bvec{\varepsilon}^{\prime} ) }
& \mathrm{if} \ \lcm(d_{1}, d_{2}) = d
\\
& \mathrm{and} \ z \equiv r \pmod{ d }
\\
& \mathrm{and} \ z_{1} - \gamma z_{2} \equiv u_{1} \pmod{ \gcd(d_{1}, d_{2}) } ,
\\
0
& \mathrm{otherwise}
\end{dcases}
\label{def:Q4}
\end{align}
for each $u_{1}, z, z_{1}, z_{2} \in \mathbb{Z}/q\mathbb{Z}$ and $d, d_{1}, d_{2} | q$.
Then, a simple calculation yields that
\begin{align}
&
\sum_{y \in \mathcal{Y}} V_{\bvec{\varepsilon} \varoast \bvec{\varepsilon}^{\prime}}(y \mid u_{2}) \, Q_{4}(z_{1} + d_{1}\mathbb{Z}, z_{2} + d_{2}\mathbb{Z}, u_{1} \mid y)
\notag \\
& \qquad \qquad \qquad \qquad \qquad =
\sum_{d|q} \sum_{y \in \mathbb{Z}/d\mathbb{Z}} V_{\bvec{\varepsilon} \varoast \bvec{\varepsilon}^{\prime}}(y \mid u_{2}) \, Q_{4}(z_{1} + d_{1}\mathbb{Z}, z_{2} + d_{2}\mathbb{Z}, u_{1} \mid y)
\notag \\
& \qquad \qquad \qquad \qquad \qquad \overset{\mathclap{\text{(a)}}}{=}
\sum_{\substack{ d|q : \\ d = \lcm(d_{1}, d_{2}) }} \sum_{\substack{ y \in \mathbb{Z}/d\mathbb{Z} : \\ y \equiv r \ (\mathrm{mod} \, d) }} V_{\bvec{\varepsilon} \varoast \bvec{\varepsilon}^{\prime}}(y \mid u_{2}) \, \frac{ \varepsilon_{d_{1}} \, \varepsilon_{d_{2}}^{\prime} }{ q \, \varepsilon_{d}^{\varoast}( \bvec{\varepsilon} , \bvec{\varepsilon}^{\prime} ) } \, \mathbbm{1} [ z_{1} - \gamma z_{2} \equiv u_{1} \pmod{ \gcd(d_{1}, d_{2}) } ]
\notag \\
& \qquad \qquad \qquad \qquad \qquad =
V_{\bvec{\varepsilon} \varoast \bvec{\varepsilon}^{\prime}}(r + \lcm(d_{1}, d_{2}) \mathbb{Z} \mid u_{2}) \, \frac{ \varepsilon_{d_{1}} \, \varepsilon_{d_{2}}^{\prime} }{ q \, \varepsilon_{\lcm(d_{1}, d_{2})}^{\varoast}( \bvec{\varepsilon} , \bvec{\varepsilon}^{\prime} ) } \, \mathbbm{1} [ z_{1} - \gamma z_{2} \equiv u_{1} \pmod{ \gcd(d_{1}, d_{2}) } ]
\notag \\
& \qquad \qquad \qquad \qquad \qquad \overset{\mathclap{\text{(b)}}}{=}
\begin{dcases}
\frac{ \varepsilon_{d_{1}} \, \varepsilon_{d_{2}}^{\prime} }{ q }
& \mathrm{if} \ z_{1} - \gamma z_{2} \equiv u_{1} \pmod{ \gcd(d_{1}, d_{2}) } ,
\\
& \qquad \quad \ \, u_{2} \equiv r \pmod{ \lcm(d_{1}, d_{2}) }
\\
0
& \mathrm{otherwise}
\end{dcases}
\notag \\
& \qquad \qquad \qquad \qquad \qquad \overset{\mathclap{\text{(c)}}}{=}
(V_{\bvec{\varepsilon}} \varoast V_{\bvec{\varepsilon}^{\prime}})(z_{1} + d_{1}\mathbb{Z}, z_{2} + d_{2}\mathbb{Z}, u_{1} \mid u_{2})
\label{eq:equivalence_condition_plus2}
\end{align}
for every $u_{1}, u_{2}, z_{1}, z_{2} \in \mathbb{Z}/q\mathbb{Z}$ and $d_{1}, d_{2} | q$, where
\begin{itemize}
\item
(a) follows from \eqref{def:Q4},
\item
(b) follows from \eqref{eq:V}, and
\item
(c) follows from \eqref{eq:Vplus}.
\end{itemize}
Therefore, we observe from \eqref{eq:equivalence_condition_plus1} and \eqref{eq:equivalence_condition_plus2} that $V_{\bvec{\varepsilon}} \varoast V_{\bvec{\varepsilon}^{\prime}}$ is equivalent to $V_{\bvec{\varepsilon} \varoast \bvec{\varepsilon}^{\prime}}$.
This completes the proof of \eqref{eq:maec_minus_recursive} written in \thref{th:recursive_V};
and all assertions of \thref{th:recursive_V} are proved.
\hfill\IEEEQEDhere

\section{Proof of \corref{cor:multilevel}}
\label{app:multilevel}

Consider an MAEC $V_{\bvec{\varepsilon}} : \mathbb{Z}/q\mathbb{Z} \to \mathcal{Y}_{q}$ defined in \defref{def:V}, where recall from \eqref{def:alphabetY} that
\begin{align}
\mathcal{Y}_{q}
=
\bigcup_{d^{\prime} | q} \frac{ \mathbb{Z} }{ d^{\prime} \mathbb{Z} } .
\label{def:alphabetYq}
\end{align}
It follows from \eqref{def:homomorphism} that for each $d|q$, the homomorphism channel $V_{\bvec{\varepsilon}}[\ker \varphi_{d}] : (\mathbb{Z}/q\mathbb{Z})/(\ker \varphi_{d}) \to \mathcal{Y}_{q}$ is given by
\begin{align}
V_{\bvec{\varepsilon}}[\ker \varphi_{d}](y \mid x + \ker \varphi_{d} )
& =
\frac{ 1 }{ |\ker \varphi_{d}| } \sum_{u \in x + \ker \varphi_{d}} V_{\bvec{\varepsilon}}(y \mid u)
\notag \\
& =
\frac{ d }{ q } \sum_{\substack{ u \in x + \ker \varphi_{d} : \\ y = u + d^{\prime} \mathbb{Z} }} \varepsilon_{d^{\prime}}
\notag \\
& =
\begin{dcases}
\frac{ d \, \varepsilon_{d^{\prime}} }{ \lcm( d, d^{\prime} ) }
& \mathrm{if} \ y \in \left\{ x + w + d^{\prime}\mathbb{Z} \ \middle| \!
\begin{array}{l}
w \in \mathbb{Z}/q\mathbb{Z} , \\
w + d\mathbb{Z} = d\mathbb{Z}
\end{array}\!\!\!
\right\}
\\
0
& \mathrm{otherwise}
\end{dcases}
\label{eq:V-ker}
\end{align}
for every $x \in \mathbb{Z}/q\mathbb{Z}$, $d^{\prime}|q$, and $y \in \mathbb{Z}/d^{\prime}\mathbb{Z}$, where
\begin{align}
x + \ker \varphi_{d}
=
\left\{ x + w \ \middle| \ w \in \frac{ \mathbb{Z} }{ q\mathbb{Z} } \ \mathrm{and} \ w + d \mathbb{Z} = d \mathbb{Z} \right\}
\end{align}
and the last equality follows from \lemref{lem:CRT}.

Now, we shall verify that, after relabelling the input symbols in $(\mathbb{Z}/q\mathbb{Z})/(\ker \varphi_{d})$ appropriately, the homomorphism channel $V_{\bvec{\varepsilon}}[\ker \varphi_{d}]$ is equivalent to another MAEC $V_{\bar{\bvec{\varepsilon}}} : \mathbb{Z}/d\mathbb{Z} \to \mathcal{Y}_{d}$ with probability vector $\bar{\bvec{\varepsilon}} = ( \bar{\varepsilon}_{d^{\prime}} )_{d^{\prime}|d}$ given as
\begin{align}
\bar{\varepsilon}_{d_{1}}
\coloneqq
\sum_{\substack{ d_{2} | q : \\ \gcd(d_{2}, d) = d_{1} }} \varepsilon_{d_{2}}
\label{def:bar-epsilon}
\end{align}
for each $d_{1}|d$.
Since the first isomorphism theorem states that the quotient group $(\mathbb{Z} / q\mathbb{Z}) / (\ker \varphi_{d})$ is isomorphic to $\mathbb{Z}/d\mathbb{Z}$, instead of $V_{\bvec{\varepsilon}}[ \ker \varphi_{d} ] : (\mathbb{Z} / q\mathbb{Z}) / (\ker \varphi_{d}) \to \mathcal{Y}_{q}$, it suffices to consider the equivalent channel $U_{\varepsilon, d} : \mathbb{Z}/d\mathbb{Z} \to \mathcal{Y}_{q}$ given as
\begin{align}
U_{\bvec{\varepsilon}, d}(y \mid x)
& \coloneqq
V_{\bvec{\varepsilon}}[ \ker \varphi_{d} ](y \mid \varphi_{d}^{-1}( x ))
\notag \\
& \: =
\begin{dcases}
\frac{ d \, \varepsilon_{d^{\prime}} }{ \lcm( d, d^{\prime} ) }
& \mathrm{if} \ y + d \mathbb{Z} = x + d^{\prime} \mathbb{Z} ,
\\
0
& \mathrm{otherwise}
\end{dcases}
\label{eq:UV_NP}
\end{align}
for each $x \in \mathbb{Z}/d\mathbb{Z}$, $d^{\prime}|q$, and $y \in \mathbb{Z}/d^{\prime}\mathbb{Z}$, where $\varphi_{d}^{-1}$ stands for the preimage of $\varphi_{d}$.
Defining the channel $Q_{5} : \mathcal{Y}_{q} \to \mathcal{Y}_{d}$ as
\begin{align}
Q_{5}(y_{1} \mid y_{2})
\coloneqq
\begin{dcases}
1
& \mathrm{if} \ y_{1} = y_{2} + d \mathbb{Z} ,
\\
0
& \mathrm{otherwise} ,
\end{dcases}
\label{def:Q5}
\end{align}
we observe that for all $x \in \mathbb{Z}/d\mathbb{Z}$, $d_{1}|d$, and $y_{1} \in \mathbb{Z}/d_{1}\mathbb{Z}$,
\begin{align}
\sum_{y_{2} \in \mathcal{Y}_{q}} U_{\bvec{\varepsilon}, d}(y_{2} \mid x) \, Q_{5}(y_{1} \mid y_{2})
& =
\sum_{d_{2}|q} \sum_{y_{2} \in \mathbb{Z}/d_{2}\mathbb{Z}} U_{\bvec{\varepsilon}, d}(y_{2} \mid x) \, Q_{5}(y_{1} \mid y_{2})
\notag \\
& \overset{\mathclap{\text{(a)}}}{=}
\sum_{d_{2}|q} \sum_{\substack{ y_{2} \in \mathbb{Z}/d_{2}\mathbb{Z} : \\ y_{1} = y_{2} + d\mathbb{Z}}} U_{\bvec{\varepsilon}, d}(y_{2} \mid x)
\notag \\
& \overset{\mathclap{\text{(b)}}}{=}
\sum_{\substack{ d_{2}|q : \\ \gcd(d, d_{2}) = d_{1} }} \sum_{\substack{ y_{2} \in \mathbb{Z}/d_{2}\mathbb{Z} : \\ y_{1} = y_{2} + d_{1}\mathbb{Z} }} U_{\bvec{\varepsilon}, d}(y_{2} \mid x)
\notag \\
& \overset{\mathclap{\text{(c)}}}{=}
\sum_{\substack{ d_{2}|q : \\ \gcd(d, d_{2}) = d_{1} }} \sum_{\substack{ y_{2} \in \mathbb{Z}/d_{2}\mathbb{Z} : \\ y_{1} = y_{2} + d_{1}\mathbb{Z} , \\ y_{2} = x + d_{2}\mathbb{Z} }} \frac{ d \, \varepsilon_{d_{2}} }{ \lcm( d, d_{2} ) }
\notag \\
& \overset{\mathclap{\text{(d)}}}{=}
\begin{dcases}
\sum_{\substack{ d_{2}|q : \\ \gcd(d, d_{2}) = d_{1} }} \varepsilon_{d_{2}}
& \mathrm{if} \ y_{1} = x + d_{1} \mathbb{Z} ,
\\
0
& \mathrm{otherwise} ,
\end{dcases}
\label{eq:homomorphism_equivalence_1}
\end{align}
where
\begin{itemize}
\item
(a) follows by the definition of $Q_{5} : \mathcal{Y}_{q} \to \mathcal{Y}_{d}$ in \eqref{def:Q5},
\item
(b) follows from the fact that $d \mathbb{Z} + d_{2} \mathbb{Z} = \gcd(d, d_{2}) \mathbb{Z}$,
\item
(c) follows by the definition of $U_{\bvec{\varepsilon}, d} : \mathbb{Z}/d\mathbb{Z} \to \mathcal{Y}_{q}$ in  \eqref{eq:V-ker} and \eqref{eq:UV_NP}, and
\item
(d) follows from \lemref{lem:CRT}.
\end{itemize}
On the other hand, defining the channel $Q_{6} : \mathcal{Y}_{d} \to \mathcal{Y}_{q}$ as
\begin{align}
Q_{6}(y_{2} \mid y_{1})
\coloneqq
\begin{dcases}
\frac{ d \, \varepsilon_{d_{2}} }{ \bar{\varepsilon}_{d_{1}} \lcm( d, d_{2} ) }
& \mathrm{if} \ y_{1} = y_{2} + d\mathbb{Z}
\\
& \mathrm{and} \ \bar{\varepsilon}_{d_{1}} > 0 ,
\\
0
& \mathrm{otherwise}
\end{dcases}
\label{def:Q6}
\end{align}
for each $d_{1}|d$, $d_{2}|q$, $y_{1} \in \mathbb{Z}/d_{1}\mathbb{Z}$, and $y_{2} \in \mathbb{Z}/d_{2}\mathbb{Z}$, we obtain
\begin{align}
\sum_{y_{1} \in \mathcal{Y}_{d}} V_{\bar{\bvec{\varepsilon}}}(y_{1} \mid x ) \, Q_{6}(y_{2} \mid y_{1})
& =
\sum_{d_{1}|d} \sum_{y_{1} \in \mathbb{Z}/d_{1}\mathbb{Z}} V_{\bar{\bvec{\varepsilon}}}(y_{1} \mid x ) \, Q_{6}(y_{2} \mid y_{1})
\notag \\
& \overset{\mathclap{\text{(a)}}}{=}
\sum_{d_{1}|d} \bar{\varepsilon}_{d_{1}} \, Q_{6}(y_{2} \mid x + d_{1}\mathbb{Z})
\notag \\
& \overset{\mathclap{\text{(b)}}}{=}
\begin{dcases}
\frac{ d \, \varepsilon_{d_{1}} }{ \lcm( d, d_{1} ) }
& \mathrm{if} \ y_{2} + d \mathbb{Z} = x + d_{1} \mathbb{Z} ,
\\
0
& \mathrm{otherwise} .
\end{dcases}
\label{eq:homomorphism_equivalence_2}
\end{align}
for every $x \in \mathbb{Z}/d\mathbb{Z}$, $d_{2}|q$, and $y_{2} \in \mathbb{Z}/d_{2}\mathbb{Z}$, where
\begin{itemize}
\item
(a) follows by the definition of $V_{\bar{\bvec{\varepsilon}}} : \mathbb{Z}/d\mathbb{Z} \to \mathcal{Y}_{d}$ in \defref{def:V}, and
\item
(b) follows by the definition of $Q : \mathcal{Y}_{d} \to \mathcal{Y}_{q}$ in \eqref{def:Q6}.
\end{itemize}
Combining \eqref{eq:homomorphism_equivalence_1} and \eqref{eq:homomorphism_equivalence_2}, we conclude that $V_{\bvec{\varepsilon}}[\ker \varphi_{d}] : (\mathbb{Z}/q\mathbb{Z}) / (\ker \varphi_{d}) \to \mathcal{Y}_{q}$ is equivalent to $V_{\bar{\bvec{\varepsilon}}} : \mathbb{Z}/d\mathbb{Z} \to \mathcal{Y}_{d}$.

Therefore, it follows from \propref{prop:I(V)} that
\begin{align}
I( V_{\bvec{\varepsilon}}[\ker \varphi_{d}] )
=
\sum_{d_{1}|d} \left( \sum_{\substack{ d_{2} | q : \\ \gcd(d_{2}, d) = d_{1} }} \varepsilon_{d_{2}} \right) \log d_{1} .
\label{eq:homomorphism_I}
\end{align}
Finally, we observe that
\begin{align}
\mu_{d}^{(\infty)}
& \overset{\mathclap{\text{(a)}}}{=}
\lim_{n \to \infty} \frac{ 1 }{ 2^{n} } \Big| \Big\{ \bvec{s} \in \{ -, + \}^{n} \ \Big| \ \varepsilon_{d}^{\bvec{s}} > 1 - \delta \Big\} \Big|
\notag \\
& =
\lim_{n \to \infty} \frac{ 1 }{ 2^{n} } \left| \left\{ \bvec{s} \in \{ -, + \}^{n} \ \middle| \!
\begin{array}{l}
|I( V_{\bvec{\varepsilon}}^{\bvec{s}} ) - \log d| < \delta ,
\\[5pt]
|I( V_{\bvec{\varepsilon}}^{\bvec{s}}[ \ker \varphi_{d} ] ) - \log d| < \delta
\end{array}\!\!\!
\right\} \right|
\end{align}
for every $d|q$, where
\begin{itemize}
\item
(a) follows from \thref{th:polarization}, and
\item
(b) follows from \propref{prop:I(V)}, \eqref{eq:homomorphism_I}, and the fact that $\varepsilon_{d} > 1 - \delta$ implies that $\bar{\varepsilon}_{d} > 1 - \delta$ (see \eqref{def:bar-epsilon}).
\end{itemize}
This completes the proof of \corref{cor:multilevel}.
\hfill\IEEEQEDhere

\section{Proof of \propref{prop:primepower_conservation}}
\label{app:primepower_conservation}

For each $i = 0, 1, \dots, r$, we have
\begin{align}
\varepsilon_{p^{i}}^{\boxast}
& =
\sum_{\substack{ d_{1}|q, d_{2}|q : \\ \gcd(d_{1}, d_{2}) = p^{i} }} \varepsilon_{d_{1}} \, \varepsilon_{d_{2}}^{\prime}
\notag \\
& =
\sum_{j = 0}^{r} \sum_{k = 0}^{r} \varepsilon_{p^{j}} \, \varepsilon_{p^{k}}^{\prime} \, \mathbbm{1}[ \min\{ j, k \} = i ]
\notag \\
& =
\sum_{j = 0}^{r} \sum_{k = 0}^{r} \varepsilon_{p^{j}} \, \varepsilon_{p^{k}}^{\prime} \, \Big( \mathbbm{1}[ i = j \le k ] + \mathbbm{1}[ i = k < j ] \Big) ,
\\
\varepsilon_{p^{i}}^{\varoast}
& =
\sum_{\substack{ d_{1}|q, d_{2}|q : \\ \lcm(d_{1}, d_{2}) = p^{i} }} \varepsilon_{d_{1}} \, \varepsilon_{d_{2}}^{\prime}
\notag \\
& =
\sum_{j = 0}^{r} \sum_{k = 0}^{r} \varepsilon_{p^{j}} \, \varepsilon_{p^{k}}^{\prime} \, \mathbbm{1}[ \max\{ j, k \} = i ]
\notag \\
& =
\sum_{j = 0}^{r} \sum_{k = 0}^{r} \varepsilon_{p^{j}} \, \varepsilon_{p^{k}}^{\prime} \, \Big( \mathbbm{1}[ k < j = i ] + \mathbbm{1}[ j \le k = i ] \Big) .
\end{align}
Hence, for each $i = 0, 1, \dots, r$, it holds that
\begin{align}
\varepsilon_{p^{i}}^{\boxast} + \varepsilon_{p^{i}}^{\varoast}
& =
\sum_{j = 0}^{r} \sum_{k = 0}^{r} \varepsilon_{p^{j}} \, \varepsilon_{p^{k}}^{\prime} \, \Big( \mathbbm{1}[ i = j \le k ] + \mathbbm{1}[ i = k < j ] + \mathbbm{1}[ k < j = i ] + \mathbbm{1}[ j \le k = i ] \Big)
\notag \\
& =
\sum_{j = 0}^{r} \sum_{k = 0}^{r} \varepsilon_{p^{j}} \, \varepsilon_{p^{k}}^{\prime} \, \Big( \mathbbm{1}[ i = j ] + \mathbbm{1}[ i = k ] \Big)
\notag \\
& =
\varepsilon_{p^{i}} \sum_{k = 0}^{r} \varepsilon_{p^{k}}^{\prime} + \varepsilon_{p^{i}}^{\prime} \sum_{j = 0}^{r} \varepsilon_{p^{j}}
\notag \\
& =
\varepsilon_{p^{i}} + \varepsilon_{p^{i}}^{\prime} .
\end{align}
This completes the proof of \propref{prop:primepower_conservation}.
\hfill\IEEEQEDhere

\section{Proof of \lemref{lem:recursive_TB}}
\label{app:recursive_TB}

We now prove the assertion for the minus transform.
A straightforward calculation yields
\begin{align}
\varepsilon_{p^{i}}^{\bvec{s}-} \,
& =
\sum_{\substack{ d_{1}|p^{r}, d_{2}|p^{r} : \\ \gcd(d_{1}, d_{2}) = p^{i} }} \varepsilon_{d_{1}}^{\bvec{s}} \varepsilon_{d_{2}}^{\bvec{s}}
\notag \\
& =
\sum_{j = i}^{r} \sum_{k = i}^{r} \varepsilon_{p^{j}}^{\bvec{s}} \, \varepsilon_{p^{k}}^{\bvec{s}} \, \mathbbm{1}[ \min\{ j, k \} = i ]
\notag \\
& =
\varepsilon_{p^{i}}^{\bvec{s}} \, \left( \sum_{j = i}^{r} \varepsilon_{p^{j}}^{\bvec{s}} + \sum_{k = i + 1}^{r} \varepsilon_{p^{k}}^{\bvec{s}} \right)
\end{align}
for each $i = 0, 1, \dots, r$, where the first equality follows from \eqref{def:eps_s}.
Then, we have
\begin{align}
T^{\bvec{s}-}( a )
& =
\sum_{i = a}^{r} \varepsilon_{p^{i}}^{\bvec{s}-}
\notag \\
& =
\sum_{i = a}^{r} \varepsilon_{p^{i}}^{\bvec{s}} \, \left( \sum_{j = i}^{r} \varepsilon_{p^{j}}^{\bvec{s}} + \sum_{k = i + 1}^{r} \varepsilon_{p^{k}}^{\bvec{s}} \right)
\notag \\
& =
\sum_{i = a}^{r} \varepsilon_{p^{i}}^{\bvec{s}} \, \left( \sum_{j = i}^{r} \varepsilon_{p^{j}}^{\bvec{s}} + \sum_{k = a}^{i-1} \varepsilon_{p^{k}}^{\bvec{s}} \right)
\notag \\
& =
\sum_{i = a}^{r} \varepsilon_{p^{i}}^{\bvec{s}} \sum_{j = a}^{r} \varepsilon_{p^{j}}^{\bvec{s}}
\notag \\
& =
T^{\bvec{s}}( a )^{2} ,
\end{align}
where the third equality follows from the fact that $\mathbbm{1}[ a \le i \le r ] \mathbbm{1}[ i < k \le r ] = \mathbbm{1}[ a \le i < k \le r ] = \mathbbm{1}[ a \le k \le r ] \mathbbm{1}[ a \le i < k ]$.
This is indeed \eqref{eq:T_minus}.
Moreover, it follows from \eqref{eq:sum_TB} and \eqref{eq:T_minus} that
\begin{align}
B^{\bvec{s}-}( a )
& =
1 - T^{\bvec{s}-}( a )
\notag \\
& =
1 - T^{\bvec{s}}( a )^{2}
\notag \\
& =
\Big( 1 - T^{\bvec{s}}( a ) \Big) \Big( 1 + T^{\bvec{s}}( a ) \Big)
\notag \\
& =
\Big( T^{\bvec{s}}( a ) + B^{\bvec{s}}( a ) - T^{\bvec{s}}( a ) \Big) \Big( T^{\bvec{s}}( a ) + B^{\bvec{s}}( a ) + T^{\bvec{s}}( a ) \Big)
\notag \\
& =
B^{\bvec{s}}( a ) \Big( 2 \, T^{\bvec{s}}( a ) + B^{\bvec{s}}( a ) \Big)
\notag \\
& =
2 \, T^{\bvec{s}}( a ) \, B^{\bvec{s}}( a ) + B^{\bvec{s}}( a )^{2} ,
\end{align}
which is indeed \eqref{eq:B_minus}.
The assertion for the plus transform can be dually proved; and this completes the proof of \lemref{lem:recursive_TB}.
\hfill\IEEEQEDhere

\section{Proof of \lemref{lem:formulas}}
\label{app:formulas}

By symmetry, it suffices to prove the required statement for the minus transform.
Fix a sequence $\bvec{s} \in \{ -, + \}^{\ast}$, indices $1 \le i < j \le m$, and integers $a, b \ge 1$ arbitrarily.
It follows from \eqref{def:eps_s} that
\begin{align}
\varepsilon_{\langle \bvec{t} \rangle}^{\bvec{s}-}
& =
\sum_{\substack{ d_{1}|q, d_{2}|q : \\ \gcd(d_{1}, d_{2}) = \langle \bvec{t} \rangle }} \varepsilon_{d_{1}}^{\bvec{s}} \, \varepsilon_{d_{2}}^{\bvec{s}}
\notag \\
& =
\sum_{\bvec{u} : \bvec{0} \le \bvec{u} \le \bvec{r}} \sum_{\bvec{v} : \bvec{0} \le \bvec{v} \le \bvec{r}} \varepsilon_{\langle \bvec{u} \rangle}^{\bvec{s}} \, \varepsilon_{\langle \bvec{v} \rangle}^{\bvec{s}} \, \mathbbm{1}_{\{ \bvec{t} = \bvec{u} \wedge \bvec{v} \}}
\label{eq:eps_minus_proof}
\end{align}
for every $\bvec{0} \le \bvec{t} \le \bvec{r}$, where
\begin{align}
\bvec{u} \wedge \bvec{v}
\coloneqq
(\min\{ u_{1}, v_{1} \}, \min\{ u_{2}, v_{2} \}, \dots, \min\{ u_{m}, v_{m} \}) .
\end{align}
Defining an $m$-tuple $\bvec{c} = (c_{1}, \dots, c_{m})$ by
\begin{align}
c_{k}
=
\begin{cases}
a
& \mathrm{if} \ k = i ,
\\
b
& \mathrm{if} \ k = j ,
\\
0
& \mathrm{otherwise}
\end{cases}
\end{align}
for each $k = 1, 2, \dots, m$, we observe that
\begin{align}
\theta_{i, j}^{\bvec{s}-}(a, b)
& \overset{\mathclap{\text{(a)}}}{=}
\sum_{\bvec{t} : \bvec{c} \le \bvec{t} \le \bvec{r}} \varepsilon_{\langle \bvec{t} \rangle}^{\bvec{s}-}
\notag \\
& \overset{\mathclap{\text{(b)}}}{=}
\sum_{\bvec{t} : \bvec{c} \le \bvec{t} \le \bvec{r}} \sum_{\bvec{u} : \bvec{0} \le \bvec{u} \le \bvec{r}} \sum_{\bvec{v} : \bvec{0} \le \bvec{v} \le \bvec{r}} \varepsilon_{\langle \bvec{u} \rangle}^{\bvec{s}} \, \varepsilon_{\langle \bvec{v} \rangle}^{\bvec{s}} \, \mathbbm{1}_{\{ \bvec{t} = \bvec{u} \wedge \bvec{v} \}}
\notag \\
& \overset{\mathclap{\text{(c)}}}{=}
\sum_{\bvec{t} : \bvec{c} \le \bvec{t} \le \bvec{r}} \sum_{\bvec{u} : \bvec{c} \le \bvec{u} \le \bvec{r}} \sum_{\bvec{v} : \bvec{c} \le \bvec{v} \le \bvec{r}} \varepsilon_{\langle \bvec{u} \rangle}^{\bvec{s}} \, \varepsilon_{\langle \bvec{v} \rangle}^{\bvec{s}} \, \mathbbm{1}_{\{ \bvec{t} = \bvec{u} \wedge \bvec{v} \}}
\notag \\
& =
\sum_{\bvec{u} : \bvec{c} \le \bvec{u} \le \bvec{r}} \sum_{\bvec{v} : \bvec{c} \le \bvec{v} \le \bvec{r}} \varepsilon_{\langle \bvec{u} \rangle}^{\bvec{s}} \, \varepsilon_{\langle \bvec{v} \rangle}^{\bvec{s}} \sum_{\bvec{t} : \bvec{c} \le \bvec{t} \le \bvec{r}} \mathbbm{1}_{\{ \bvec{t} = \bvec{u} \wedge \bvec{v} \}}
\notag \\
& =
\sum_{\bvec{u} : \bvec{c} \le \bvec{u} \le \bvec{r}} \sum_{\bvec{v} : \bvec{c} \le \bvec{v} \le \bvec{r}} \varepsilon_{\langle \bvec{u} \rangle}^{\bvec{s}} \, \varepsilon_{\langle \bvec{v} \rangle}^{\bvec{s}}
\notag \\
& =
\theta_{i, j}^{\bvec{s}}(a, b)^{2}
\label{eq:formulas1_proof}
\end{align}
where
\begin{itemize}
\item
(a) follows by the definition of $\theta_{i, j}^{\bvec{s}}(a, b)$ in \eqref{def:theta},
\item
(b) follows from \eqref{eq:eps_minus_proof}, and
\item
(c) follows from the fact that $\bvec{c} \le \bvec{t} = \bvec{u} \wedge \bvec{v}$ imply that $\bvec{c} \le \bvec{u}$ and $\bvec{c} \le \bvec{v}$.
\end{itemize}
On the other hand, we have
\begin{align}
\lambda_{i, j}^{\bvec{s}-}(a, b)
& \overset{\mathclap{\text{(a)}}}{=}
\theta_{i, j}^{\bvec{s}-}(a, 0) - \theta_{i, j}^{\bvec{s}-}(a, b)
\notag \\
& \overset{\mathclap{\text{(b)}}}{=}
\theta_{i, j}^{\bvec{s}}(a, 0)^{2} - \theta_{i, j}^{\bvec{s}}(a, b)^{2}
\notag \\
& =
\Big( \theta_{i, j}^{\bvec{s}}(a, 0) - \theta_{i, j}^{\bvec{s}}(a, b) \Big) \, \Big( \theta_{i, j}^{\bvec{s}}(a, 0) + \theta_{i, j}^{\bvec{s}}(a, b) \Big)
\notag \\
& \overset{\mathclap{\text{(c)}}}{=}
\lambda_{i, j}^{\bvec{s}}(a, b) \, \Big( \lambda_{i, j}^{\bvec{s}}(a, b) + 2 \, \theta_{i, j}^{\bvec{s}}(a, b) \Big) ,
\label{eq:formulas2_proof}
\end{align}
where
\begin{itemize}
\item
(a) and (c) follow by the definition of $\lambda_{i, j}^{\bvec{s}}(a, b)$ in \eqref{def:lambda}, and
\item
(b) follows from \eqref{eq:formulas1_proof}.
\end{itemize}
Since $\lambda_{i, j}^{\bvec{s}}(a, b) = \rho_{j, i}^{\bvec{s}}(b, a)$, we readily see from \eqref{eq:formulas2_proof} that
\begin{align}
\rho_{i, j}^{\bvec{s}-}(a, b)
=
\rho_{i, j}^{\bvec{s}}(a, b)^{2} + 2 \, \rho_{i, j}^{\bvec{s}}(a, b) \, \theta_{i, j}^{\bvec{s}}(a, b) .
\label{eq:formulas3_proof}
\end{align}
Finally, as $\theta_{i, j}^{\bvec{s}}(a, b) + \lambda_{i, j}^{\bvec{s}}(a, b) + \rho_{i, j}^{\bvec{s}}(a, b) + \beta_{i, j}^{\bvec{s}}(a, b) = 1$ (see \eqref{eq:sum_theta_lambda_rho_beta}), it follows from \eqref{eq:formulas1_proof}--\eqref{eq:formulas3_proof} that
\begin{align}
\beta_{i, j}^{\bvec{s}-}(a, b)
& =
1 - \theta_{i, j}^{\bvec{s}-}(a, b) - \lambda_{i, j}^{\bvec{s}-}(a, b) - \rho_{i, j}^{\bvec{s}-}(a, b)
\notag \\
& =
1 - \theta_{i, j}^{\bvec{s}}(a, b)^{2} - \big[ \lambda_{i, j}^{\bvec{s}}(a, b)^{2} + 2 \, \lambda_{i, j}^{\bvec{s}}(a, b) \, \theta_{i, j}^{\bvec{s}}(a, b) \big] - \big[ \rho_{i, j}^{\bvec{s}}(a, b)^{2} + 2 \, \rho_{i, j}^{\bvec{s}}(a, b) \, \theta_{i, j}^{\bvec{s}}(a, b) \big]
\notag \\
& =
1 - \theta_{i, j}^{\bvec{s}}(a, b)^{2} - \big[ \lambda_{i, j}^{\bvec{s}}(a, b) + \rho_{i, j}^{\bvec{s}}(a, b) \big]^{2} - 2 \, \theta_{i, j}^{\bvec{s}}(a, b) \, \big[ \lambda_{i, j}^{\bvec{s}}(a, b) + \rho_{i, j}^{\bvec{s}}(a, b) \big] + 2 \, \lambda_{i, j}^{\bvec{s}}(a, b) \, \rho_{i, j}^{\bvec{s}}(a, b)
\notag \\
& =
1 - \big[ \theta_{i, j}^{\bvec{s}}(a, b) + \lambda_{i, j}^{\bvec{s}}(a, b) + \rho_{i, j}^{\bvec{s}}(a, b) \big]^{2} + 2 \, \lambda_{i, j}^{\bvec{s}}(a, b) \, \rho_{i, j}^{\bvec{s}}(a, b)
\notag \\
& =
\big[ 1 - \theta_{i, j}^{\bvec{s}}(a, b) + \lambda_{i, j}^{\bvec{s}}(a, b) + \rho_{i, j}^{\bvec{s}}(a, b) \big] \, \big[ 1 + \theta_{i, j}^{\bvec{s}}(a, b) + \lambda_{i, j}^{\bvec{s}}(a, b) + \rho_{i, j}^{\bvec{s}}(a, b) \big] + 2 \, \lambda_{i, j}^{\bvec{s}}(a, b) \, \rho_{i, j}^{\bvec{s}}(a, b)
\notag \\
& =
\beta_{i, j}^{\bvec{s}}(a, b) \big[ 1 + \theta_{i, j}^{\bvec{s}}(a, b) + \lambda_{i, j}^{\bvec{s}}(a, b) + \rho_{i, j}^{\bvec{s}}(a, b) \big] + 2 \, \lambda_{i, j}^{\bvec{s}}(a, b) \, \rho_{i, j}^{\bvec{s}}(a, b)
\notag \\
& =
\beta_{i, j}^{\bvec{s}}(a, b) \big[ 2 - \beta_{i, j}^{\bvec{s}}(a, b) \big] + 2 \, \lambda_{i, j}^{\bvec{s}}(a, b) \, \rho_{i, j}^{\bvec{s}}(a, b) .
\end{align}
This completes the proof of \lemref{lem:formulas}.
\hfill\IEEEQEDhere

\section{Proof of \lemref{lem:ineq}}
\label{app:ineq}

Let $1 \le i < j \le m$ and $a, b \ge 1$ be given.
By the symmetry $\lambda_{i, j}^{\bvec{s}}(a, b) = \rho_{j, i}^{\bvec{s}}(b, a)$, it suffices to prove the ``if'' part.
We prove the lemma by induction.
If the sequence $\bvec{s}$ is empty, then the lemma is obvious.
Hence, it suffices to show that if $\lambda_{i, j}^{\bvec{s}}(a, b) \le \rho_{i, j}^{\bvec{s}}(a, b)$, then both $\lambda_{i, j}^{\bvec{s}-}(a, b) \le \rho_{i, j}^{\bvec{s}-}(a, b)$ and $\lambda_{i, j}^{\bvec{s}+}(a, b) \le \rho_{i, j}^{\bvec{s}+}(a, b)$ hold.
It follows from \lemref{lem:formulas} that
\begin{align}
\lambda_{i, j}^{\bvec{s}-}(a, b)
& =
\lambda_{i, j}^{\bvec{s}}(a, b) \big[ \lambda_{i, j}^{\bvec{s}}(a, b) + 2 \, \theta_{i, j}^{\bvec{s}}(a, b) \big]
\notag \\
& \overset{\mathclap{\text{(a)}}}{\le}
\rho_{i, j}^{\bvec{s}}(a, b) \big[ \rho_{i, j}^{\bvec{s}}(a, b) + 2 \, \theta_{i, j}^{\bvec{s}}(a, b) \big]
\notag \\
& =
\rho_{i, j}^{\bvec{s}-}(a, b) ,
\label{eq:minus_ineq}
\end{align}
where (a) follows by the hypothesis $\lambda_{i, j}^{\bvec{s}}(a, b) \le \rho_{i, j}^{\bvec{s}}(a, b)$.
Similar to \eqref{eq:minus_ineq}, we also have
\begin{align}
\lambda_{i, j}^{\bvec{s}+}(a, b)
& =
\lambda_{i, j}^{\bvec{s}}(a, b) \big[ \lambda_{i, j}^{\bvec{s}}(a, b) + 2 \, \beta_{i, j}^{\bvec{s}}(a, b) \big]
\notag \\
& \le
\rho_{i, j}^{\bvec{s}}(a, b) \big[ \rho_{i, j}^{\bvec{s}}(a, b) + 2 \, \beta_{i, j}^{\bvec{s}}(a, b) \big]
\notag \\
& =
\rho_{i, j}^{\bvec{s}+}(a, b) .
\label{eq:plus_ineq}
\end{align}
This completes the proof of \lemref{lem:ineq}.
\hfill\IEEEQEDhere

\section{Proof of \lemref{lem:martingale}}
\label{app:martingale}

Let $1 \le i < j \le m$ and $a, b \ge 1$ be given.
For each $n \in \mathbb{N}_{0}$, we have
\begin{align}
\mu_{i, j}^{(n+1)}[\lambda](a, b) - \mu_{i, j}^{(n+1)}[\rho](a, b)
& =
\frac{ 1 }{ 2^{n+1} } \sum_{\bvec{s} \in \{ -, + \}^{n}} \Big( \lambda_{i, j}^{\bvec{s}-}(a, b) + \lambda_{i, j}^{\bvec{s}+}(a, b) \Big) - \frac{ 1 }{ 2^{n+1} } \sum_{\bvec{s} \in \{ -, + \}^{n}} \Big( \rho_{i, j}^{\bvec{s}-}(a, b) + \rho_{i, j}^{\bvec{s}+}(a, b) \Big)
\notag \\
& \overset{\mathclap{\text{(a)}}}{=}
\frac{ 1 }{ 2^{n} } \sum_{\bvec{s} \in \{ -, + \}^{n}} \lambda_{i, j}^{\bvec{s}}(a, b) \big[ 1 - \rho_{i, j}^{\bvec{s}}(a, b) \big] - \frac{ 1 }{ 2^{n} } \sum_{\bvec{s} \in \{ -, + \}^{n}} \rho_{i, j}^{\bvec{s}}(a, b) \big[ 1 - \lambda_{i, j}^{\bvec{s}}(a, b) \big]
\notag \\
& =
\frac{ 1 }{ 2^{n} } \sum_{\bvec{s} \in \{ -, + \}^{n}} \lambda_{i, j}^{\bvec{s}}(a, b) - \frac{ 1 }{ 2^{n} } \sum_{\bvec{s} \in \{ -, + \}^{n}} \rho_{i, j}^{\bvec{s}}(a, b)
\notag \\
& =
\mu_{i, j}^{(n)}[\lambda](a, b) - \mu_{i, j}^{(n)}[\rho](a, b) ,
\label{eq:proof_martingale}
\end{align}
where (a) follows by \lemref{lem:quasi-conservation}.
This proves \eqref{eq:lambda_minus_rho} by induction.
The rest of equalities \eqref{eq:theta_plus_lambda}--\eqref{eq:beta_plus_rho} can be similarly proved by \lemref{lem:quasi-conservation}, as in \eqref{eq:proof_martingale}.
This completes the proof of \lemref{lem:martingale}.
\hfill\IEEEQEDhere

\section{Proof of \lemref{lem:convergent}}
\label{app:convergent}

Let $1 \le i < j \le m$ and $a, b \ge 1$ be given.
It follows from \eqref{ineq:sub_theta}--\eqref{ineq:sub_beta} that
\begin{itemize}
\item
the number $\mu_{i, j}^{(n)}[\theta](a, b)$ is nondecreasing as $n$ increases,
\item
the number $\mu_{i, j}^{(n)}[\lambda](a, b)$ is nonincreasing as $n$ increases,
\item
the number $\mu_{i, j}^{(n)}[\rho](a, b)$ is nonincreasing as $n$ increases, and
\item
the number $\mu_{i, j}^{(n)}[\beta](a, b)$ is nondecreasing as $n$ increases.
\end{itemize}
Therefore, since these numbers are bounded as
\begin{align}
0
& \le
\mu_{i, j}^{(n)}[\theta](a, b)
\le
1 ,
\\
0
& \le
\mu_{i, j}^{(n)}[\lambda](a, b)
\le
1 ,
\\
0
& \le
\mu_{i, j}^{(n)}[\rho](a, b)
\le
1 ,
\\
0
& \le
\mu_{i, j}^{(n)}[\beta](a, b)
\le
1
\end{align}
for every $n \in \mathbb{N}_{0}$, we obtain the claim of \lemref{lem:convergent}.
\hfill\IEEEQEDhere

\section{Proof of \lemref{lem:fujisaki}}
\label{app:fujisaki}

Let $1 \le i < j \le m$ and $a, b \ge 1$ be given.
Since $\lambda_{i, j}^{\bvec{s}}(a, b) = \rho_{j, i}^{\bvec{s}}(b, a)$, we may assume without loss of generality that $\lambda_{i, j}(a, b) \le \rho_{i, j}(a, b)$.
A simple calculation yields
\begin{align}
\mu_{i, j}^{(n+1)}[\lambda](a, b)
& =
\frac{ 1 }{ 2^{n} } \sum_{\bvec{s} \in \{ -, + \}^{n}} \frac{ 1 }{ 2 } \Big( \lambda_{i, j}^{\bvec{s}-}(a, b) + \lambda_{i, j}^{\bvec{s}+}(a, b) \Big)
\notag \\
& \overset{\mathclap{\text{(a)}}}{=}
\frac{ 1 }{ 2^{n} } \sum_{\bvec{s} \in \{ -, + \}^{n}} \lambda_{i, j}^{\bvec{s}}(a, b) \big[ 1 - \rho_{i, j}^{\bvec{s}}(a, b) \big]
\notag \\
& \overset{\mathclap{\text{(b)}}}{\le}
\frac{ 1 }{ 2^{n} } \sum_{\bvec{s} \in \{ -, + \}^{n}} \lambda_{i, j}^{\bvec{s}}(a, b) \big[1 - \lambda_{i, j}^{\bvec{s}}(a, b) \big]
\notag \\
& \overset{\mathclap{\text{(c)}}}{=}
\mu_{i, j}^{(n)}[\lambda](a, b) - \nu_{i, j}^{(n)}[\lambda](a, b) ,
\label{eq:lambda_mu_nu}
\end{align}
where
\begin{itemize}
\item
(a) follows by \lemref{lem:quasi-conservation},
\item
(b) follows by \lemref{lem:ineq}, and
\item
(c) follows by the definition of the second moment:
\begin{align}
\nu_{i, j}^{(n)}[\lambda](a, b)
\coloneqq
\frac{ 1 }{ 2^{n} } \sum_{\bvec{s} \in \{ -, + \}^{n}} \lambda_{i, j}^{\bvec{s}}(a, b)^{2} .
\end{align}
\end{itemize}
It follows from \eqref{eq:lambda_mu_nu} that
\begin{align}
0
\le
\nu_{i, j}^{(n)}[\lambda](a, b)
\le
\mu_{i, j}^{(n)}[\lambda](a, b) - \mu_{i, j}^{(n+1)}[\lambda](a, b) ,
\label{eq:convergence_nu}
\end{align}
and the squeeze theorem shows that $\nu_{i, j}^{(n)}[\lambda](a, b) \to 0$ as $n \to \infty$, because $\mu_{i, j}^{(n)}[\lambda](a, b) - \mu_{i, j}^{(n+1)}[\lambda](a, b) \to 0$ as $n \to \infty$ (cf. \lemref{lem:convergent}).
On the other hand, we observe that
\begin{align}
\mu_{i, j}^{(n)}[\lambda](a, b)^{2}
& =
\left[ \frac{ 1 }{ 2^{n} } \sum_{\bvec{s} \in \{ -, + \}^{n}} \lambda_{i, j}^{\bvec{s}}(a, b) \right]^{2}
\notag \\
& =
\frac{ 1 }{ 2^{2n} } \sum_{\bvec{s}_{1} \in \{ -, + \}^{n}} \left[ \lambda_{i, j}^{\bvec{s}_{1}}(a, b)^{2}
\vphantom{\sum_{\substack{ \bvec{s}_{2} \in \{ -, + \}^{n} : \\ \bvec{s}_{2} \neq \bvec{s}_{1} }}} + \sum_{\substack{ \bvec{s}_{2} \in \{ -, + \}^{n} : \\ \bvec{s}_{2} \neq \bvec{s}_{1} }} \lambda_{i, j}^{\bvec{s}_{1}}(a, b) \, \lambda_{i, j}^{\bvec{s}_{2}}(a, b) \right]
\notag \\
& \le
\frac{ 1 }{ 2^{2n} } \sum_{\bvec{s}_{1} \in \{ -, + \}^{n}} \left[ \lambda_{i, j}^{\bvec{s}_{1}}(a, b)^{2}
\vphantom{\sum_{\substack{ \bvec{s}_{2} \in \{ -, + \}^{n} : \\ \lambda_{i, j}^{\bvec{s}_{2}}(a, b) \ge \lambda_{i, j}^{\bvec{s}_{1}}(a, b) }}} + \sum_{\substack{ \bvec{s}_{2} \in \{ -, + \}^{n} : \\ \lambda_{i, j}^{\bvec{s}_{2}}(a, b) \ge \lambda_{i, j}^{\bvec{s}_{1}}(a, b) }} \lambda_{i, j}^{\bvec{s}_{2}}(a, b)^{2} + \sum_{\substack{ \bvec{s}_{3} \in \{ -, + \}^{n} : \\ \lambda_{i, j}^{\bvec{s}_{3}}(a, b) < \lambda_{i, j}^{\bvec{s}_{1}}(a, b) }} \lambda_{i, j}^{\bvec{s}_{1}}(a, b)^{2} \right]
\notag \\
& \le
\frac{ 1 }{ 2^{2n} } \sum_{\bvec{s}_{1} \in \{ -, + \}^{n}} \left[ \lambda_{i, j}^{\bvec{s}_{1}}(a, b)^{2}
\vphantom{\sum_{\bvec{s}_{2} \in \{ -, + \}^{n}}} + \sum_{\bvec{s}_{2} \in \{ -, + \}^{n}} \lambda_{i, j}^{\bvec{s}_{2}}(a, b)^{2} + (2^{n}-1) \, \lambda_{i, j}^{\bvec{s}_{1}}(a, b)^{2} \right]
\notag \\
& =
2 \, \nu_{i, j}^{(n)}[\lambda](a, b) ,
\end{align}
which implies that
\begin{align}
0
\le
\mu_{i, j}^{(n)}[\lambda](a, b)
\le
\sqrt{ 2 \, \nu_{i, j}^{(n)}[\lambda](a, b) } .
\label{ineq:Holder}
\end{align}
Note that the second inequality of \eqref{ineq:Holder} can be seen as a version of H\"{o}lder's inequality.
Then, it also follows by the squeeze theorem that $\mu_{i, j}^{(\infty)}[\lambda](a, b) = 0$, because $\nu_{i, j}^{(n)}[\lambda](a, b) \to 0$ as $n \to \infty$ (cf. \eqref{eq:convergence_nu}).
Hence, we have
\begin{align}
\mu_{i, j}^{(\infty)}[\rho](a, b)
& =
\mu_{i, j}^{(\infty)}[\rho](a, b) - \mu_{i, j}^{(\infty)}[\lambda](a, b)
\notag \\
& =
\lim_{n \to \infty} \Big( \mu_{i, j}^{(n)}[\rho](a, b) - \mu_{i, j}^{(n)}[\lambda](a, b) \Big)
\notag \\
& \overset{\mathclap{\text{(a)}}}{=}
\rho_{i, j}(a, b) - \lambda_{i, j}(a, b) ,
\label{eq:deriving_mu-rho-infty}
\\
\mu_{i, j}^{(\infty)}[\theta](a, b)
& =
\mu_{i, j}^{(\infty)}[\theta](a, b) + \mu_{i, j}^{(\infty)}[\lambda](a, b)
\notag \\
& =
\lim_{n \to \infty} \Big( \mu_{i, j}^{(n)}[\theta](a, b) + \mu_{i, j}^{(n)}[\lambda](a, b) \Big)
\notag \\
& \overset{\mathclap{\text{(b)}}}{=}
\theta_{i, j}(a, b) + \lambda_{i, j}(a, b) ,
\\
\mu_{i, j}^{(\infty)}[\beta](a, b)
& =
\mu_{i, j}^{(\infty)}[\beta](a, b) + \mu_{i, j}^{(\infty)}[\lambda](a, b)
\notag \\
& =
\lim_{n \to \infty} \Big( \mu_{i, j}^{(n)}[\beta](a, b) + \mu_{i, j}^{(n)}[\lambda](a, b) \Big)
\notag \\
& \overset{\mathclap{\text{(c)}}}{=}
\beta_{i, j}(a, b) + \lambda_{i, j}(a, b) ,
\end{align}
where (a)--(c) follow by \lemref{lem:martingale}.
Considering the counterpart hypothesis $\lambda_{i, j}(a, b) \ge \rho_{i, j}(a, b)$, we have \eqref{eq:theta_inf}--\eqref{eq:beta_inf}.
This completes the proof of \lemref{lem:fujisaki}.
\hfill\IEEEQEDhere

\section{Proof of \thref{th:mu_d}}
\label{app:mu_d}

We will show in this proof that the while loop in Lines~4--15 of Algorithm~\ref{alg:main} is accomplished by the $\Omega_{\mathrm{NT}}( q )$-th round.
For each $1 \le h \le \Omega_{\mathrm{NT}}(q)$, denote by $\bvec{t}^{(h)} = (t_{1}^{(h)}, \dots, t_{m}^{(h)})$ the vector $\bvec{t}$ at the beginning of the $h$-th round of this while loop, where note from the initialization in Line~3 of Algorithm~\ref{alg:main} that $\bvec{t}^{(1)} = (0, \dots, 0)$.
Given a number $1 \le h \le \Omega_{\mathrm{NT}}(q)$, suppose that we have completed the while loop in Lines~4--15 of Algorithm~\ref{alg:main} until the $(h-1)$-th round.
That is, we now consider the beginning of the $h$-th round of this while loop.
In this proof, suppose that the variable $k$ appearing from Line~8 of Algorithm~\ref{alg:main} is initialized as $1$ at Line~5 of Algorithm~\ref{alg:main}.

Firstly, we shall verify the following claim.

\begin{claim}
\label{cl:spade}
The while loop in Lines~6--12 of Algorithm~\ref{alg:main} finds the number $k$ such that for each $1 \le c \le m$ satisfying $c \neq k$, it holds that $ \mu_{\langle \bvec{t} \rangle}^{(\infty)} = 0$ for every $\bvec{0} \le \bvec{t} \le \bvec{r}$ satisfying $0 \le t_{k} \le t_{k}^{(h)}$ and $t_{c}^{(h)} < t_{c} \le r_{c}$.
\end{claim}

Since the ``$1$'' is added to the variable $j$ at the end of each round of the while loop in Lines~6 and~12 of Algorithm~\ref{alg:main} (see Lines~9 and~12 of Algorithm~\ref{alg:main}), the total number of rounds of this while loop is just $m - 1$ (see Line~6 of Algorithm~\ref{alg:main}).
Consider the beginning of the $\jmath$-th round of this while loop for some $1 \le \jmath \le m - 1$.
To prove \clref{cl:spade}, we shall shall verify the following claim.

\begin{claim}
\label{cl:jmath}
For each $1 \le c \le \jmath$ satisfying $c \neq k$, it holds that $\mu_{\langle \bvec{t} \rangle}^{(\infty)} = 0$ for every $\bvec{0} \le \bvec{t} \le \bvec{r}$ satisfying $0 \le t_{k} \le t_{k}^{(h)}$ and $t_{c}^{(h)} < t_{c} \le r_{c}$.
\end{claim}

Note that \clref{cl:jmath} coincides with \clref{cl:spade} if $\jmath = m$.
We prove \clref{cl:jmath} by induction.
It is clear from the initialization $k = 1$ that \clref{cl:jmath} holds with $\jmath = 1$.
Now, suppose that $2 \le \jmath \le m - 1$ and \clref{cl:jmath} holds at the previous round before the $\jmath$-th round.
Consider the conditional branch in Lines~7 and~10 of Algorithm~\ref{alg:main} at the $\jmath$-th round.
Note that $(i, j) = (k, \jmath + 1)$ at the beginning of this conditional branch (see Lines~8, 9, 11, and~12 of Algorithm~\ref{alg:main}).
If $\lambda_{i, j}( t_{i}^{(h)} + 1, t_{j}^{(h)} + 1 ) \le \rho_{i, j}( t_{i}^{(h)} + 1, t_{j}^{(h)} + 1 )$, then it follows from \eqref{eq:lambda_inf} of \lemref{lem:fujisaki} that $\mu_{\langle \bvec{t} \rangle}^{(\infty)} = 0$ for every $\bvec{0} \le \bvec{t} \le \bvec{r}$ satisfying $t_{i}^{(h)} < t_{i} \le r_{i}$ and $0 \le t_{j} \le t_{j}^{(h)}$.
Similarly, if $\rho_{i, j}( t_{i}^{(h)} + 1, t_{j}^{(h)} + 1 ) < \lambda_{i, j}( t_{i}^{(h)} + 1, t_{j}^{(h)} + 1 )$, then it follows from \eqref{eq:rho_inf} of \lemref{lem:fujisaki} that $\mu_{\langle \bvec{t} \rangle}^{(\infty)} = 0$ for every $\bvec{0} \le \bvec{t} \le \bvec{r}$ satisfying $0 \le t_{i} \le t_{i}^{(h)}$ and $t_{j}^{(h)} < t_{j} \le r_{j}$.
Thus, \clref{cl:jmath} holds by replacing the variable $k$ as in Line~8 or~11 of Algorithm~\ref{alg:main} according to this conditional branch.
Therefore, \clref{cl:spade} is also proven by induction.

For each $1 \le h \le \Omega_{\mathrm{NT}}(q)$, denote by $\xi^{(h)}$ the variable $\xi$ at the beginning of the $h$-th round of the while loop in Lines~4--15 of Algorithm~\ref{alg:main}.
Secondly, we shall verify the following claim.

\begin{claim}
\label{cl:club}
After executing the operation in Line~15 of Algorithm~\ref{alg:main}, the desired value $\mu_{\langle \bvec{t} \rangle}^{(\infty)}$ is given as
\begin{align}
\mu_{\langle \bvec{t} \rangle}^{(\infty)}
& =
0
\qquad \mathrm{if} \ \bvec{t} \neq \bvec{t}^{(g)} \ \mathrm{for} \ \mathrm{all} \ 1 \le g \le h ,
\label{eq:clubsuit_1} \\
\mu_{\langle \bvec{t} \rangle}^{(\infty)}
& \ge
0
\qquad \mathrm{if} \ \bvec{t} = \bvec{t}^{(g)} \ \mathrm{for} \ \mathrm{some} \ 1 \le g \le h
\label{eq:clubsuit_2}
\end{align}
for every $\bvec{0} \le \bvec{t} \le \bvec{r}$ satisfying $0 \le t_{c} < t_{c}^{(h+1)}$ for some $1 \le c \le m$, and the next variable $\xi^{(h+1)}$ is given by
\begin{align}
\xi^{(h+1)}
& =
\sum_{g = 1}^{h} \mu_{\langle \bvec{t}^{(g)} \rangle}^{(\infty)}
\notag \\
& =
\sum_{\substack{ \bvec{t} : \bvec{0} \le \bvec{t} \le \bvec{r} , \\ 0 \le t_{c} < t_{c}^{(h+1)} \, \text{\emph{for some}} \, 1 \le c \le m }} \varepsilon_{\langle \bvec{t} \rangle} .
\label{eq:clubsuit_3}
\end{align}
\end{claim}

We prove \clref{cl:club} by induction.
Suppose that $h = 1$.
It follows from the initialization in Line~2 of Algorithm~\ref{alg:main} that $\xi^{(1)} = 0$.
Moreover, it follows from \eqref{def:mu_d}, \eqref{def:beta}, and \eqref{def:mu_beta} that
\begin{align}
\mu_{i, j}^{(n)}[\beta](a, b)
=
\sum_{\bvec{t} : \bvec{0} \le \bvec{t} \le \bvec{r}, t_{i} < a, t_{j} < b} \mu_{\langle \bvec{t} \rangle}^{(n)} .
\label{eq:beta_n}
\end{align}
Since the pair $(k, l)$ satisfies that $l = k$ if $k < m$ (see Line~11 of Algorithm~\ref{alg:main}), and $l < k$ if $k = m$ (see Line~9 of Algorithm~\ref{alg:main}), we observe from \eqref{eq:beta_n} and \clref{cl:spade} that
\begin{align}
\mu_{\langle \bvec{t}^{(1)} \rangle}^{(\infty)}
& =
\mu_{l, m}^{(\infty)}[\beta](t_{l}^{(1)}+1, t_{m}^{(1)}+1) - \xi^{(1)} .
\label{eq:sol_alg_1-st_1}
\end{align}
In addition, it follows from \eqref{eq:beta_inf} of \lemref{lem:fujisaki} that
\begin{align}
\mu_{l, m}^{(\infty)}[\beta](t_{l}^{(1)}+1, t_{m}^{(1)}+1)
& =
\beta_{l, m}(t_{l}^{(1)}+1, t_{m}^{(1)}+1) + \min\{ \lambda_{l, m}(t_{l}^{(1)}+1, t_{m}^{(1)}+1), \rho_{l, m}(t_{l}^{(1)}+1, t_{m}^{(1)}+1) \} .
\label{eq:sol_alg_1-st_3}
\end{align}
The right-hand sides of \eqref{eq:sol_alg_1-st_1} and \eqref{eq:sol_alg_1-st_3} correspond to the operation in Line~13 of Algorithm~\ref{alg:main}, and it follows from \clref{cl:spade} that the desired value $\mu_{\langle \bvec{t} \rangle}^{(\infty)}$ is obtained for every $\bvec{0} \le \bvec{t} \le \bvec{r}$ satisfying $0 \le t_{k} \le t_{k}^{(1)}$.
On the other hand, since $k = m$ if and only if $\lambda_{l, m}(t_{l}^{(1)}+1, t_{m}^{(1)}+1) \le \rho_{l, m}(t_{l}^{(1)}+1, t_{m}^{(1)}+1)$ (see the while loop in Lines~6--12 of Algorithm~\ref{alg:main}), it follows from \eqref{def:lambda}--\eqref{def:beta} that \eqref{eq:sol_alg_1-st_3} can be rewritten as
\begin{align}
\mu_{l, m}^{(\infty)}[\beta](t_{l}^{(1)}+1, t_{m}^{(1)}+1)
=
\sum_{\substack{ \bvec{t} : \bvec{0} \le \bvec{t} \le \bvec{r} , \\ 0 \le t_{k} \le t_{k}^{(1)} }} \varepsilon_{\langle \bvec{t} \rangle} .
\label{eq:sol_alg_1-st_4}
\end{align}
By Line~14 of Algorithm~\ref{alg:main}, the right-hand side of \eqref{eq:sol_alg_1-st_4} corresponds to the next value $\xi^{(2)}$;
therefore, we observe from Line~15 of Algorithm~\ref{alg:main} that \clref{cl:club} holds with $h = 1$.
Now, suppose that $2 \le h \le \Omega_{\mathrm{NT}}(q)$ and \clref{cl:club} holds at the previous round before the $h$-th round.
Since the pair $(k, l)$ satisfies that $l = k$ if $k < m$ (see Line~11 of Algorithm~\ref{alg:main}), and $l < k$ if $k = m$ (see Line~9 of Algorithm~\ref{alg:main}), it follows from \eqref{eq:beta_n}, \clref{cl:spade}, and the induction hypothesis that
\begin{align}
\mu_{\langle \bvec{t}^{(h)} \rangle}^{(\infty)}
& =
\beta_{l, m}( t_{l}^{(h)}+1, t_{m}^{(h)}+1 ) + \min\{ \lambda_{l, m}( t_{l}^{(h)}+1, t_{m}^{(h)}+1 ), \rho_{l, m}( t_{l}^{(h)}+1, t_{m}^{(h)}+1 ) \} - \xi^{(h)} ,
\label{eq:sol_alg_h-th_3}
\end{align}
which is indeed the operation in Line~13 of Algorithm~\ref{alg:main}.
Thus, the desired value $\mu_{\langle \bvec{t} \rangle}^{(\infty)}$ is determined as either \eqref{eq:clubsuit_1} or \eqref{eq:clubsuit_2} for every $\bvec{0} \le \bvec{t} \le \bvec{r}$ satisfying $0 \le t_{c} < t_{c}^{(h+1)}$ for some $1 \le c \le m$.
On the other hand, similar to \eqref{eq:sol_alg_1-st_4}, one has \eqref{eq:clubsuit_3}.
Therefore, \clref{cl:club} is proved by induction.

If $\xi^{(h)} < \xi^{(h+1)} = 1$, then it follows from \clref{cl:club} that the desired asymptotic distribution $( \mu_{d}^{(\infty)} )_{d|q}$ has been evaluated at the end of the $h$-th round of the while loop in Lines~4--15 of Algorithm~\ref{alg:main}.
Note from \eqref{eq:clubsuit_3} that $\xi^{(h+1)} = 1$ if there exists a $1 \le c \le m$ satisfying $t_{c}^{(h+1)} > r_{c}$.
Therefore, the total number of rounds of the while loop in Line~4--15 of Algorithm~\ref{alg:main} is at most $\Omega_{\mathrm{NT}}(q)$; and the output of Algorithm~\ref{alg:main} yields the desired asymptotic distribution $( \mu_{d}^{(\infty)} )_{d|q}$ within a finite number of steps.
Finally, we shall verify the computational complexity of Algorithm~\ref{alg:main}.
As said in this proof, the while loop in Lines~4--15 of Algorithm~\ref{alg:main} is repeated at most $\Omega_{\mathrm{NT}}( q )$ times, and
the while loop in Lines~6--12 of Algorithm~\ref{alg:main} is repeated just $m-1$ times.
In the conditional branch of Lines~7 and~10 of Algorithm~\ref{alg:main}, both $\lambda_{i, j}(t_{i} + 1, t_{j} + 1)$ and $\lambda_{i, j}(t_{i} + 1, t_{j} + 1)$ can be calculated by a given initial probability vector $( \varepsilon_{d} )_{d|q}$ with at most $\tau( q )$ additions (see \eqref{def:lambda} and \eqref{def:rho}).
Similarly, in Line~13 of Algorithm~\ref{alg:main}, the three values $\beta_{l, m}(t_{l}+1, t_{m}+1)$, $\lambda_{l, m}(t_{l}+1, t_{m}+1)$, and $\rho_{l, m}(t_{l}+1, t_{m}+1)$ can also be calculated by a given initial probability vector $( \varepsilon_{d} )_{d|q}$ with at most $\tau( q )$ additions (see \eqref{def:lambda}--\eqref{def:beta}).
Therefore, we conclude that Algorithm~\ref{alg:main} runs in time $\mathrm{O}( \omega_{\mathrm{NT}}( q ) \, \Omega_{\mathrm{NT}}( q ) \, \tau( q ) )$.
Note that all calculations in Algorithm~\ref{alg:main} are addition and subtraction, i.e., there is neither multiplication nor division.
This completes the proof of \thref{th:mu_d}.
\hfill\IEEEQEDhere

\section{Proof of \thref{th:polarization}}
\label{app:polarization}

\thref{th:polarization} can be simply proven by using a similar argument to \appref{app:multilevel} together with Nasser and Telatar's result \cite[Section~VI]{nasser_telatar_it2016} summarized in \eqref{eq:multilevel}.
In the following, we provide an alternative proof of \thref{th:polarization} to make this paper self-contained.

To prove \thref{th:polarization}, we use the following technical lemma.

\begin{lemma}
\label{lem:additive}
For each $n \in \mathbb{N}$, let a nonempty collection $\mathcal{F}_{n}$ of subsets of a set be a field,%
\footnote{Note that this field $\mathcal{F}_{n}$ is a measure theoretic notion satisfying $A^{\complement} \in \mathcal{F}_{n}$ if $A \in \mathcal{F}_{n}$, and $A \cup B \in \mathcal{F}_{n}$ if $A, B \in \mathcal{F}_{n}$, where $A^{\complement}$ denotes the complement of a set $A$.}
and let $f_{n} : \mathcal{F}_{n} \to [0, 1]$ be an additive set function.
For each $i \in \mathbb{N}$, let $( S_{i, n} )_{n}$ be a sequence of sets such that $S_{i, n} \in \mathcal{F}_{n}$ for every $n \in \mathbb{N}$ and $f_{n}( S_{i, n} ) \to 1$ as $n \to \infty$.
Then, it holds that
\begin{align}
\lim_{n \to \infty} f_{n} \left( \bigcap_{i=1}^{k} S_{i, n} \right)
=
1
\quad
\mathrm{for} \ k \in \mathbb{N} .
\end{align}
\end{lemma}

\begin{IEEEproof}[Proof of \lemref{lem:additive}]
See \appref{app:additive}.
\end{IEEEproof}

The proof of \thref{th:polarization} is inspired by Alsan and Telatar's simple proof of polarization \cite[Theorem~1]{alsan_telatar_it2016}.
Let $1 \le i < j \le m$ and $a, b \ge 1$ be given.
Define
\begin{align}
\nu_{i, j}^{(n)}[\theta](a, b)
\coloneqq
\frac{ 1 }{ 2^{n} } \sum_{\bvec{s} \in \{ -, + \}^{n}} \theta_{i, j}^{\bvec{s}}(a, b)^{2}
\end{align}
for each $n \in \mathbb{N}$.
Then, we have that for a fixed $\delta \in (0, 1)$,
\begin{align}
\nu_{i, j}^{(n+1)}[\theta](a, b)
& =
\frac{ 1 }{ 2^{n} } \sum_{\bvec{s} \in \{ -, + \}^{n}} \frac{ 1 }{ 2 } \Big[ \theta_{i, j}^{\bvec{s}-}(a, b)^{2} + \theta_{i, j}^{\bvec{s}+}(a, b)^{2} \Big]
\notag \\
& \overset{\mathclap{\text{(a)}}}{=}
\frac{ 1 }{ 2^{n} } \sum_{\bvec{s} \in \{ -, + \}^{n}} \left[ \left( \frac{ 1 }{ 2 } \Big( \theta_{i, j}^{\bvec{s}-}(a, b) + \theta_{i, j}^{\bvec{s}+}(a, b) \Big) \right)^{2} + \left( \frac{ 1 }{ 2 } \Big(\theta_{i, j}^{\bvec{s}-}(a, b) - \theta_{i, j}^{\bvec{s}+}(a, b) \Big) \right)^{2} \right]
\notag \\
& \overset{\mathclap{\text{(b)}}}{=}
\frac{ 1 }{ 2^{n} } \sum_{\bvec{s} \in \{ -, + \}^{n}} \left[ \Big( \theta_{i, j}^{\bvec{s}}(a, b) + \lambda_{i, j}^{\bvec{s}}(a, b) \, \rho_{i, j}^{\bvec{s}}(a, b) \Big)^{2} + \Big( \theta_{i, j}^{\bvec{s}}(a, b) \, \big[ 1 - \theta_{i, j}^{\bvec{s}}(a, b) \big] + \lambda_{i, j}^{\bvec{s}}(a, b) \, \rho_{i, j}^{\bvec{s}}(a, b) \Big)^{2} \right]
\notag \\
& \ge
\frac{ 1 }{ 2^{n} } \sum_{\bvec{s} \in \{ -, + \}^{n}} \Big[ \theta_{i, j}^{\bvec{s}}(a, b)^{2} + \theta_{i, j}^{\bvec{s}}(a, b)^{2} \big[1 - \theta_{i, j}^{\bvec{s}}(a, b) \big]^{2} \Big]
\notag \\
& \ge
\nu_{i, j}^{(n)}[\theta](a, b) + \frac{ 1 }{ 2^{n} } \sum_{\substack{ \bvec{s} \in \{ -, + \}^{n} : \\ \delta \le \theta_{i, j}^{\bvec{s}}(a, b) \le 1 - \delta }} \theta_{i, j}^{\bvec{s}}(a, b)^{2} \big[1 - \theta_{i, j}^{\bvec{s}}(a, b) \big]^{2}
\notag \\
& \ge
\nu_{i, j}^{(n)}[\theta](a, b) + \frac{ 1 }{ 2^{n} } \sum_{\substack{ \bvec{s} \in \{ -, + \}^{n} : \\ \delta \le \theta_{i, j}^{\bvec{s}}(a, b) \le 1 - \delta }} \delta^{2} (1 - \delta)^{2} ,
\label{eq:nu_theta}
\end{align}
where
\begin{itemize}
\item
(a) follows from the identity
\begin{align}
\frac{ x^{2} + y^{2} }{ 2 } = \Big( \frac{ x + y }{ 2 } \Big)^{2} + \Big( \frac{ x - y }{ 2 } \Big)^{2} ,
\end{align}
and
\item
(b) follows by \lemref{lem:formulas}.
\end{itemize}
This implies that the sequence $\big( \nu_{i, j}^{(n)}[\theta](a, b) \big)_{n=1}^{\infty}$ is nondecreasing.
As $\nu_{i, j}^{(n)}[\theta](a, b) \le 1$ for every $n \in \mathbb{N}$, the sequence $\big( \nu_{i, j}^{(n)}[\theta](a, b) \big)_{n=1}^{\infty}$ is convergent; thus, it holds that $\nu_{i, j}^{(n+1)}[\theta](a, b) - \nu_{i, j}^{(n)}[\theta](a, b) \to 0$ as $n \to \infty$.
We get from \eqref{eq:nu_theta} that
\begin{align}
0
& \le
\frac{ 1 }{ 2^{n} } \Big| \Big\{ \bvec{s} \in \{ -, + \}^{n} \ \Big| \ \delta \le \theta_{i, j}^{\bvec{s}}(a, b) \le 1 - \delta \Big\} \Big|
\le
\frac{ \nu_{i, j}^{(n+1)}[\theta](a, b) - \nu_{i, j}^{(n)}[\theta](a, b) }{ \delta^{2} (1 - \delta)^{2} } .
\end{align}
As $\delta \in (0, 1)$ is a fixed number that does not depend on $n \in \mathbb{N}$, this implies that
\begin{align}
\lim_{n \to \infty} \frac{ 1 }{ 2^{n} } \Big| \Big\{ \bvec{s} \in \{ -, + \}^{n} \ \Big| \ \delta \le \theta_{i, j}^{\bvec{s}}(a, b) \le 1 - \delta \Big\} \Big|
=
0 .
\label{eq:theta_limit}
\end{align}
Therefore, we observe that
\begin{align}
\lim_{n \to \infty} \frac{ 1 }{ 2^{n} } \Big| \Big\{ \bvec{s} \in \{ -, + \}^{n} \ \Big| \ \varepsilon_{d}^{\bvec{s}} < \delta \ \mathrm{or} \ \varepsilon_{d}^{\bvec{s}} > 1 - \delta \Big\} \Big|
=
1
\end{align}
for every fixed $\delta \in (0, 1)$ and $d|q$.
\begin{align}
\limsup_{n \to \infty}\left( \frac{ 1 }{ 2^{n} } \Big| \Big\{ \bvec{s} \in \{ -, + \}^{n} \ \Big| \ \varepsilon_{d}^{\bvec{s}} > 1 - \delta \Big\} \Big| - \mu_{d}^{(n)} \right)
\le
\delta
\end{align}

We now prove \eqref{eq:proportion0}.
It follows from \corref{cor:mu_d} that there exist an integer $0 \le \tilde{m} \le m$ and a sequence $( \bvec{t}^{(h)} )_{h = 0}^{\tilde{m}}$ such that
(i) $\bvec{0} = \bvec{t}^{(0)} \le \bvec{t}^{(1)} \le \cdots \le \bvec{t}^{(\tilde{m})} = \bvec{r}$,
(ii) $\bvec{t}^{(h)} \neq \bvec{t}^{(h^{\prime})}$ whenever $h \neq h^{\prime}$, and
(iii) $\mu_{\langle \bvec{t} \rangle} > 0$ if and only if $\bvec{t} = \bvec{t}^{(h)}$ for some $0 \le h \le \tilde{m}$.
If $\mu_{d}^{(\infty)} = 0$, then we observe that for a fixed $\delta \in (0, 1)$,
\begin{align}
0
& =
\mu_{d}^{(\infty)}
\notag \\
& \overset{\mathclap{\text{(a)}}}{=}
\lim_{n \to \infty} \frac{ 1 }{ 2^{n} } \sum_{\bvec{s} \in \{-, +\}^{n}} \varepsilon_{d}^{\bvec{s}}
\notag \\
& \overset{\mathclap{\text{(b)}}}{\ge}
\limsup_{n \to \infty} \frac{ 1 }{ 2^{n} } \sum_{\bvec{s} \in \{-, +\}^{n} : \varepsilon_{d} \ge \delta} \delta
\notag \\
& =
\delta \limsup_{n \to \infty} \frac{ 1 }{ 2^{n} } \Big| \Big\{ \bvec{s} \in \{ -, + \}^{n} \ \Big| \ \varepsilon_{d}^{\bvec{s}} \ge \delta \Big\} \Big| ,
\end{align}
where
\begin{itemize}
\item
(a) follows by the definition of $\mu_{d}^{(\infty)}$ and the hypothesis that $\mu_{d}^{(\infty)} = 0$, and
\item
(b) follows from the fact that
\begin{align}
\frac{ 1 }{ 2^{n} } \sum_{\bvec{s} \in \{-, +\}^{n}} \varepsilon_{d}^{\bvec{s}}
& \ge
\frac{ 1 }{ 2^{n} } \sum_{\bvec{s} \in \{-, +\}^{n} : \varepsilon_{d} \ge \delta} \varepsilon_{d}
\notag \\
& \ge
\frac{ 1 }{ 2^{n} } \sum_{\bvec{s} \in \{-, +\}^{n} : \varepsilon_{d} \ge \delta} \delta .
\end{align}
\end{itemize}
This implies that
\begin{align}
\lim_{n \to \infty} \frac{ 1 }{ 2^{n} } \Big| \Big\{ \bvec{s} \in \{ -, + \}^{n} \ \Big| \ \varepsilon_{d}^{\bvec{s}} < \delta \Big\} \Big|
=
1 ,
\label{eq:eps_vanish}
\end{align}
provided that $\mu_{d}^{(\infty)} = 0$.
Therefore, it suffices to verify that
\begin{align}
\lim_{n \to \infty} \frac{ 1 }{ 2^{n} } \Big| \Big\{ \bvec{s} \in \{ -, + \}^{n} \ \Big| \ \delta \le \varepsilon_{\langle \bvec{t}^{(h)} \rangle}^{\bvec{s}} \le 1 - \delta \Big\} \Big|
=
0
\label{eq:intermediate_eps_sequence}
\end{align}
for every $h = 0, 1, \dots, \tilde{m}$.
We prove \eqref{eq:intermediate_eps_sequence} by induction.
Firstly, consider the case where $h = \tilde{m}$, where note that $\bvec{t}^{(\tilde{m})} = \bvec{r}$ and $\langle \bvec{t}^{(\tilde{m})} \rangle = \langle \bvec{r} \rangle = q$.
Since $\bvec{t}^{(\tilde{m}-1)} \le \bvec{t}^{(\tilde{m})}$ and $\bvec{t}^{(\tilde{m}-1)} \neq \bvec{t}^{(\tilde{m})}$, there exists an index $1 \le i \le m$ satisfying $t_{i}^{(\tilde{m}-1)} < t_{i}^{(\tilde{m})}$, which implies that $\mu_{\langle \bvec{t} \rangle}^{(\infty)} = 0$ for every $\bvec{0} \le \bvec{t} \le \bvec{r}$ satisfying $\bvec{t} \neq \bvec{r}$ and $(t_{i}, t_{j}) = (r_{i}, r_{j})$ for some $j \neq i$.
For such an appropriate choice of $(i, j)$, we have that
\begin{align}
0
& \overset{\mathclap{\text{(a)}}}{=}
\lim_{n \to \infty} \frac{ 1 }{ 2^{n} } \Big| \Big\{ \bvec{s} \in \{-, +\}^{n} \ \Big| \ \delta \le \theta_{i, j}^{\bvec{s}}(r_{i}, r_{j}) \le 1 - \frac{ \delta }{ \tau( q ) } \Big\} \Big|
\notag \\
& \ge
\limsup_{n \to \infty} \frac{ 1 }{ 2^{n} } \left| \Big\{ \bvec{s} \in \{-, +\}^{n} \ \Big| \ \delta \le \theta_{i, j}^{\bvec{s}}(r_{i}, r_{j}) \le 1 - \frac{ \delta }{ \tau( q ) } \Big\} \cap \left( \bigcap_{\substack{ \bvec{t} : \bvec{0} \le \bvec{t} \le \bvec{r}, \bvec{t} \neq \bvec{r}, \\ (t_{i}, t_{j}) = (r_{i}, r_{j}) }} \Big\{ \bvec{s} \in \{ -, + \}^{n} \ \Big| \ \varepsilon_{\langle \bvec{t} \rangle}^{\bvec{s}} < \frac{ \delta }{ \tau( q ) } \Big\} \right) \right|
\notag \\
& \overset{\mathclap{\text{(b)}}}{\ge}
\limsup_{n \to \infty} \frac{ 1 }{ 2^{n} } \left| \Big\{ \bvec{s} \in \{-, +\}^{n} \ \Big| \ \delta \le \varepsilon_{q}^{\bvec{s}} \le 1 - \delta \Big\} \cap \left( \bigcap_{\substack{ \bvec{t} : \bvec{0} \le \bvec{t} \le \bvec{r}, \bvec{t} \neq \bvec{r}, \\ (t_{i}, t_{j}) = (r_{i}, r_{j}) }} \Big\{ \bvec{s} \in \{ -, + \}^{n} \ \Big| \ \varepsilon_{\langle \bvec{t} \rangle}^{\bvec{s}} < \frac{ \delta }{ \tau( q ) } \Big\} \right) \right|
\notag \\
& \overset{\mathclap{\text{(c)}}}{=}
\limsup_{n \to \infty} \frac{ 1 }{ 2^{n} } \left( \Big| \Big\{ \bvec{s} \in \{-, +\}^{n} \ \Big| \ \delta \le \varepsilon_{q}^{\bvec{s}} \le 1 - \delta \Big\} \Big| + \left| \bigcap_{\substack{ \bvec{t} : \bvec{0} \le \bvec{t} \le \bvec{r}, \bvec{t} \neq \bvec{r}, \\ (t_{i}, t_{j}) = (r_{i}, r_{j}) }} \Big\{ \bvec{s} \in \{ -, + \}^{n} \ \Big| \ \varepsilon_{\langle \bvec{t} \rangle}^{\bvec{s}} < \frac{ \delta }{ \tau( q ) } \Big\} \right| \right.
\notag \\
& \qquad \qquad \qquad
\left. {} - \left| \Big\{ \bvec{s} \in \{-, +\}^{n} \ \Big| \ \delta \le \varepsilon_{q}^{\bvec{s}} \le 1 - \delta \Big\} \cup \left( \bigcap_{\substack{ \bvec{t} : \bvec{0} \le \bvec{t} \le \bvec{r}, \bvec{t} \neq \bvec{r}, \\ (t_{i}, t_{j}) = (r_{i}, r_{j}) }} \Big\{ \bvec{s} \in \{ -, + \}^{n} \ \Big| \ \varepsilon_{\langle \bvec{t} \rangle}^{\bvec{s}} < \frac{ \delta }{ \tau( q ) } \Big\} \right) \right| \right)
\notag \\
& \overset{\mathclap{\text{(d)}}}{\ge}
\limsup_{n \to \infty} \frac{ 1 }{ 2^{n} } \left( \Big| \Big\{ \bvec{s} \in \{-, +\}^{n} \ \Big| \ \delta \le \varepsilon_{q}^{\bvec{s}} \le 1 - \delta \Big\} \Big| + \left| \bigcap_{\substack{ \bvec{t} : \bvec{0} \le \bvec{t} \le \bvec{r}, \bvec{t} \neq \bvec{r}, \\ (t_{i}, t_{j}) = (r_{i}, r_{j}) }} \Big\{ \bvec{s} \in \{ -, + \}^{n} \ \Big| \ \varepsilon_{\langle \bvec{t} \rangle}^{\bvec{s}} < \frac{ \delta }{ \tau( q ) } \Big\} \right| - 2^{n} \right)
\notag \\
& \overset{\mathclap{\text{(e)}}}{\ge}
\limsup_{n \to \infty} \frac{ 1 }{ 2^{n} } \Big| \Big\{ \bvec{s} \in \{-, +\}^{n} \ \Big| \ \delta \le \varepsilon_{q}^{\bvec{s}} \le 1 - \delta \Big\} \Big| + \liminf_{n \to \infty} \frac{ 1 }{ 2^{n} } \left| \bigcap_{\substack{ \bvec{t} : \bvec{0} \le \bvec{t} \le \bvec{r}, \bvec{t} \neq \bvec{r}, \\ (t_{i}, t_{j}) = (r_{i}, r_{j}) }} \Big\{ \bvec{s} \in \{ -, + \}^{n} \ \Big| \ \varepsilon_{\langle \bvec{t} \rangle}^{\bvec{s}} < \frac{ \delta }{ \tau( q ) } \Big\} \right| - 1
\notag \\
& \overset{\mathclap{\text{(f)}}}{=}
\limsup_{n \to \infty} \frac{ 1 }{ 2^{n} } \Big| \Big\{ \bvec{s} \in \{-, +\}^{n} \ \Big| \ \delta \le \varepsilon_{q}^{\bvec{s}} \le 1 - \delta \Big\} \Big| ,
\label{eq:eps_q_vanish}
\end{align}
where
\begin{itemize}
\item
(a) follows from \eqref{eq:theta_limit}, i.e.,
\begin{align}
0
& =
\lim_{n \to \infty} \frac{ 1 }{ 2^{n} } \Big| \Big\{ \bvec{s} \in \{-, +\}^{n} \ \Big| \ \delta \le \theta_{i, j}^{\bvec{s}}(r_{i}, r_{j}) \le 1 - \delta \Big\} \Big|
\notag \\
& \le
\liminf_{n \to \infty} \frac{ 1 }{ 2^{n} } \Big| \Big\{ \bvec{s} \in \{-, +\}^{n} \ \Big| \ \delta \le \theta_{i, j}^{\bvec{s}}(r_{i}, r_{j}) \le 1 - \frac{ \delta }{ \tau( q ) } \Big\} \Big|
\notag \\
& \le
0 
\end{align}
with $\tau( q ) \coloneqq \prod_{i = 1}^{m}(r_{i} + 1)$,
\item
(b) follows from the identities
\begin{align}
\theta_{i, j}^{\bvec{s}}(r_{i}, r_{j})
\overset{\eqref{def:theta}}{=}
\sum_{\substack{ \bvec{t} : \bvec{0} \le \bvec{t} \le \bvec{r} , \\ (t_{i}, t_{j}) = (r_{i}, r_{j}) }} \varepsilon_{\langle \bvec{t} \rangle}^{\bvec{s}}
=
\varepsilon_{q}^{\bvec{s}} + \sum_{\substack{ \bvec{t} : \bvec{0} \le \bvec{t} \le \bvec{r} , \bvec{t} \neq \bvec{r} , \\ (t_{i}, t_{j}) = (r_{i}, r_{j}) }} \varepsilon_{\langle \bvec{t} \rangle}^{\bvec{s}}
\end{align}
and the inclusions
\begin{align}
&
\Big\{ \bvec{s} \in \{-, +\}^{n} \ \Big| \ \delta \le \theta_{i, j}^{\bvec{s}}(r_{i}, r_{j}) \le 1 - \frac{ \delta }{ \tau( q ) } \Big\} \cap \left( \bigcap_{\substack{ \bvec{t} : \bvec{0} \le \bvec{t} \le \bvec{r}, \bvec{t} \neq \bvec{r}, \\ (t_{i}, t_{j}) = (r_{i}, r_{j}) }} \Big\{ \bvec{s} \in \{ -, + \}^{n} \ \Big| \ \varepsilon_{\langle \bvec{t} \rangle}^{\bvec{s}} < \frac{ \delta }{ \tau( q ) } \Big\} \right)
\notag \\
& \qquad \supset
\left\{ \bvec{s} \in \{-, +\}^{n} \ \middle| \ \delta \le \varepsilon_{q}^{\bvec{s}} \le 1 - \frac{ \delta }{ \tau( q ) } - \sum_{\substack{ \bvec{t} : \bvec{0} \le \bvec{t} \le \bvec{r} , \bvec{t} \neq \bvec{r} \\ (t_{i}, t_{j}) = (r_{i}, r_{j}) }} \varepsilon_{\langle \bvec{t} \rangle}^{\bvec{s}} \right\} \cap \left( \bigcap_{\substack{ \bvec{t} : \bvec{0} \le \bvec{t} \le \bvec{r}, \bvec{t} \neq \bvec{r}, \\ (t_{i}, t_{j}) = (r_{i}, r_{j}) }} \Big\{ \bvec{s} \in \{ -, + \}^{n} \ \Big| \ \varepsilon_{\langle \bvec{t} \rangle}^{\bvec{s}} < \frac{ \delta }{ \tau( q ) } \Big\} \right)
\notag \\
& \qquad \supset
\left\{ \bvec{s} \in \{-, +\}^{n} \ \middle| \ \delta \le \varepsilon_{q}^{\bvec{s}} \le 1 - \frac{ \delta }{ \tau( q ) } - \sum_{\substack{ \bvec{t} : \bvec{0} \le \bvec{t} \le \bvec{r} , \bvec{t} \neq \bvec{r} \\ (t_{i}, t_{j}) = (r_{i}, r_{j}) }} \frac{ \delta }{ \tau( q ) } \right\} \cap \left( \bigcap_{\substack{ \bvec{t} : \bvec{0} \le \bvec{t} \le \bvec{r}, \bvec{t} \neq \bvec{r}, \\ (t_{i}, t_{j}) = (r_{i}, r_{j}) }} \Big\{ \bvec{s} \in \{ -, + \}^{n} \ \Big| \ \varepsilon_{\langle \bvec{t} \rangle}^{\bvec{s}} < \frac{ \delta }{ \tau( q ) } \Big\} \right)
\notag \\
& \qquad \supset
\Big\{ \bvec{s} \in \{-, +\}^{n} \ \Big| \ \delta \le \varepsilon_{q}^{\bvec{s}} \le 1 - \frac{ \delta }{ \tau( q ) } - (\tau(q) - 1) \frac{ \delta }{ \tau( q ) } \Big\} \cap \left( \bigcap_{\substack{ \bvec{t} : \bvec{0} \le \bvec{t} \le \bvec{r}, \bvec{t} \neq \bvec{r}, \\ (t_{i}, t_{j}) = (r_{i}, r_{j}) }} \Big\{ \bvec{s} \in \{ -, + \}^{n} \ \Big| \ \varepsilon_{\langle \bvec{t} \rangle}^{\bvec{s}} < \frac{ \delta }{ \tau( q ) } \Big\} \right)
\notag \\
& \qquad =
\Big\{ \bvec{s} \in \{-, +\}^{n} \ \Big| \ \delta \le \varepsilon_{q}^{\bvec{s}} \le 1 - \delta \Big\} \cap \left( \bigcap_{\substack{ \bvec{t} : \bvec{0} \le \bvec{t} \le \bvec{r}, \bvec{t} \neq \bvec{r}, \\ (t_{i}, t_{j}) = (r_{i}, r_{j}) }} \Big\{ \bvec{s} \in \{ -, + \}^{n} \ \Big| \ \varepsilon_{\langle \bvec{t} \rangle}^{\bvec{s}} < \frac{ \delta }{ \tau( q ) } \Big\} \right) ,
\label{eq:inclusions}
\end{align}
\item
(c) follows by the inclusion-exclusion principle,
\item
(d) follows from the fact that
\begin{align}
\left| \Big\{ \bvec{s} \in \{-, +\}^{n} \ \Big| \ \delta \le \varepsilon_{q}^{\bvec{s}} \le 1 - \delta \Big\} \cup \left( \bigcap_{\substack{ \bvec{t} : \bvec{0} \le \bvec{t} \le \bvec{r}, \bvec{t} \neq \bvec{r}, \\ (t_{i}, t_{j}) = (r_{i}, r_{j}) }} \Big\{ \bvec{s} \in \{ -, + \}^{n} \ \Big| \ \varepsilon_{\langle \bvec{t} \rangle}^{\bvec{s}} < \frac{ \delta }{ \tau( q ) } \Big\} \right) \right|
\le
2^{n}
\label{eq:card_le_pow-of-2}
\end{align}
\item
(e) follows from the fact that
\begin{align}
\limsup_{n \to \infty} (a_{n} + b_{n})
\ge
\limsup_{n \to \infty} a_{n} + \liminf_{n \to \infty} b_{n}
\label{eq:limsup_liminf}
\end{align}
for two sequences $(a_{n})_{n}$ and $(b_{n})_{n}$, and
\item
(f) follows from \lemref{lem:additive} and \eqref{eq:eps_vanish}.
\end{itemize}
Thus, it follows from \eqref{eq:eps_q_vanish} that
\begin{align}
\lim_{n \to \infty} \frac{ 1 }{ 2^{n} } \Big| \Big\{ \bvec{s} \in \{-, +\}^{n} \ \Big| \ \delta \le \varepsilon_{\langle \bvec{t}^{(\tilde{m})} \rangle}^{\bvec{s}} \le 1 - \delta \Big\} \Big|
=
0 .
\end{align}
We now suppose that for some integer $0 \le h < \tilde{m}$, it holds that
\begin{align}
\lim_{n \to \infty} \frac{ 1 }{ 2^{n} } \Big| \Big\{ \bvec{s} \in \{-, +\}^{n} \ \Big| \ \delta \le \varepsilon_{\langle \bvec{t}^{(h^{\prime})} \rangle}^{\bvec{s}} \le 1 - \delta \Big\} \Big|
=
0
\label{eq:hypo_eps_h}
\end{align}
for every $h < h^{\prime} \le \tilde{m}$.
Note that $\mu_{\langle \bvec{t}^{(h^{\prime})} \rangle}^{(\infty)} > 0$ for every $h \le h^{\prime} \le \tilde{m}$, and $\mu_{\langle \bvec{t} \rangle}^{(\infty)} = 0$ for every $\bvec{t}^{(h)} \le \bvec{t} \le \bvec{r}$ satisfying $\bvec{t} \neq \bvec{t}^{(h^{\prime})}$ for all $h \le h^{\prime} \le \tilde{m}$.
If $h > 0$, then since $\bvec{t}^{(h-1)} \le \bvec{t}^{(h)}$ and $\bvec{t}^{(h-1)} \neq \bvec{t}^{(h)}$, there exists an index $1 \le i \le m$ satisfying $t_{i}^{(h-1)} < t_{i}^{(h)}$, which implies that $\mu_{\langle \bvec{t} \rangle}^{(\infty)} = 0$ for every $\bvec{0} \le \bvec{t} \le \bvec{r}$ satisfying $\bvec{t} \neq \bvec{t}^{(h^{\prime})}$ for all $h \le h^{\prime} \le \tilde{m}$ and $(t_{i}, t_{j}) \ge (t_{i}^{(h)}, t_{j}^{(h)})$ for some $j \neq i$.
If $h = 0$, then it is obvious that $\mu_{\langle \bvec{t} \rangle}^{(\infty)} = 0$ for every $\bvec{0} \le \bvec{t} \le \bvec{r}$ satisfying $\bvec{t} \neq \bvec{t}^{(h^{\prime})}$ for all $0 \le h^{\prime} \le \tilde{m}$.
For such an appropriate choice of $(i, j)$, similar to \eqref{eq:eps_q_vanish}, we have that 
\begin{align}
0
& \overset{\mathclap{\text{(a)}}}{=}
\lim_{n \to \infty} \frac{ 1 }{ 2^{n} } \Big| \Big\{ \bvec{s} \in \{-, +\}^{n} \ \Big| \ \delta \le \theta_{i, j}^{\bvec{s}}(t_{i}^{(h)}, t_{j}^{(h)}) \le 1 - \frac{ \delta }{ \tau( q ) } \Big\} \Big|
\notag \\
& \ge
\limsup_{n \to \infty} \frac{ 1 }{ 2^{n} } \left| \Big\{ \bvec{s} \in \{-, +\}^{n} \ \Big| \ \delta \le \theta_{i, j}^{\bvec{s}}(t_{i}^{(h)}, t_{j}^{(h)}) \le 1 - \frac{ \delta }{ \tau( q ) } \Big\} \cap \left( \bigcap_{\substack{ \bvec{t} : \bvec{0} \le \bvec{t} \le \bvec{r}, \\ \bvec{t} \neq \bvec{t}^{(h^{\prime})} \, \forall h^{\prime} \ge h , \\ (t_{i}, t_{j}) \ge (t_{i}^{(h)}, t_{j}^{(h)}) }} \Big\{ \bvec{s} \in \{ -, + \}^{n} \ \Big| \ \varepsilon_{\langle \bvec{t} \rangle}^{\bvec{s}} < \frac{ \delta }{ \tau( q ) } \Big\} \right) \right|
\notag \\
& \overset{\mathclap{\text{(b)}}}{\ge}
\limsup_{n \to \infty} \frac{ 1 }{ 2^{n} } \left| \Bigg\{ \bvec{s} \in \{-, +\}^{n} \ \Bigg| \ \delta \le \sum_{h^{\prime} = h}^{\tilde{m}} \varepsilon_{\langle \bvec{t}^{(h^{\prime})} \rangle}^{\bvec{s}} \le 1 - \delta \Bigg\} \cap \left( \bigcap_{\substack{ \bvec{t} : \bvec{0} \le \bvec{t} \le \bvec{r}, \\ \bvec{t} \neq \bvec{t}^{(h^{\prime})} \, \forall h^{\prime} \ge h , \\ (t_{i}, t_{j}) \ge (t_{i}^{(h)}, t_{j}^{(h)}) }} \Big\{ \bvec{s} \in \{ -, + \}^{n} \ \Big| \ \varepsilon_{\langle \bvec{t} \rangle}^{\bvec{s}} < \frac{ \delta }{ \tau( q ) } \Big\} \right) \right|
\notag \\
& \overset{\mathclap{\text{(c)}}}{\ge}
\limsup_{n \to \infty} \frac{ 1 }{ 2^{n} } \left( \Bigg| \Bigg\{ \bvec{s} \in \{-, +\}^{n} \ \Bigg| \ \delta \le \sum_{h^{\prime} = h}^{\tilde{m}} \varepsilon_{\langle \bvec{t}^{(h^{\prime})} \rangle}^{\bvec{s}} \le 1 - \delta \Bigg\} \Bigg| + \left| \bigcap_{\substack{ \bvec{t} : \bvec{0} \le \bvec{t} \le \bvec{r}, \\ \bvec{t} \neq \bvec{t}^{(h^{\prime})} \, \forall h^{\prime} \ge h , \\ (t_{i}, t_{j}) \ge (t_{i}^{(h)}, t_{j}^{(h)}) }} \Big\{ \bvec{s} \in \{ -, + \}^{n} \ \Big| \ \varepsilon_{\langle \bvec{t} \rangle}^{\bvec{s}} < \frac{ \delta }{ \tau( q ) } \Big\} \right| - 2^{n} \right)
\notag \\
& \overset{\mathclap{\text{(d)}}}{\ge}
\limsup_{n \to \infty} \frac{ 1 }{ 2^{n} } \Bigg| \Bigg\{ \bvec{s} \in \{-, +\}^{n} \ \Bigg| \ \delta \le \sum_{h^{\prime} = h}^{\tilde{m}} \varepsilon_{\langle \bvec{t}^{(h^{\prime})} \rangle}^{\bvec{s}} \le 1 - \delta \Bigg\} \Bigg| + \liminf_{n \to \infty} \frac{ 1 }{ 2^{n} } \left| \bigcap_{\substack{ \bvec{t} : \bvec{0} \le \bvec{t} \le \bvec{r}, \\ \bvec{t} \neq \bvec{t}^{(h^{\prime})} \, \forall h^{\prime} \ge h , \\ (t_{i}, t_{j}) \ge (t_{i}^{(h)}, t_{j}^{(h)}) }} \Big\{ \bvec{s} \in \{ -, + \}^{n} \ \Big| \ \varepsilon_{\langle \bvec{t} \rangle}^{\bvec{s}} < \frac{ \delta }{ \tau( q ) } \Big\} \right| - 1
\notag \\
& \overset{\mathclap{\text{(e)}}}{=}
\limsup_{n \to \infty} \frac{ 1 }{ 2^{n} } \Bigg| \Bigg\{ \bvec{s} \in \{-, +\}^{n} \ \Bigg| \ \delta \le \sum_{h^{\prime} = h}^{\tilde{m}} \varepsilon_{\langle \bvec{t}^{(h^{\prime})} \rangle}^{\bvec{s}} \le 1 - \delta \Bigg\} \Bigg| ,
\label{eq:eps_sum_vanish}
\end{align}
where
\begin{itemize}
\item
(a) follows from \eqref{eq:theta_limit},
\item
(b) follows from the the identities
\begin{align}
\theta_{i, j}^{\bvec{s}}(t_{i}^{(h)}, t_{j}^{(h)})
& =
\sum_{\substack{ \bvec{t} : \bvec{0} \le \bvec{t} \le \bvec{r}, \\ (t_{i}, t_{j}) \ge (t_{i}^{(h)}, t_{j}^{(h)}) }} \varepsilon_{\langle \bvec{t} \rangle}^{\bvec{s}}
\notag \\
& =
\left( \sum_{h^{\prime} = h}^{\tilde{m}} \varepsilon_{\langle \bvec{t}^{(h^{\prime})} \rangle}^{\bvec{s}} \right) + \left( \sum_{\substack{ \bvec{t} : \bvec{0} \le \bvec{t} \le \bvec{r}, \\ \bvec{t} \neq \bvec{t}^{(h^{\prime})} \, \forall h^{\prime} \ge h , \\ (t_{i}, t_{j}) \ge (t_{i}^{(h)}, t_{j}^{(h)}) }} \varepsilon_{\langle \bvec{t} \rangle}^{\bvec{s}} \right) ,
\label{eq:inclusions_2}
\end{align}
which arises from \eqref{def:theta}, and the inclusions as in \eqref{eq:inclusions},
\item
(c) follows by the inclusion-exclusion principle and the fact that
\begin{align}
\left| \Bigg\{ \bvec{s} \in \{-, +\}^{n} \ \Bigg| \ \delta \le \sum_{h^{\prime} = h}^{\tilde{m}} \varepsilon_{\langle \bvec{t}^{(h^{\prime})} \rangle}^{\bvec{s}} \le 1 - \delta \Bigg\} \cup \left( \bigcap_{\substack{ \bvec{t} : \bvec{0} \le \bvec{t} \le \bvec{r}, \\ \bvec{t} \neq \bvec{t}^{(h^{\prime})} \, \forall h^{\prime} \ge h , \\ (t_{i}, t_{j}) \ge (t_{i}^{(h)}, t_{j}^{(h)}) }} \Big\{ \bvec{s} \in \{ -, + \}^{n} \ \Big| \ \varepsilon_{\langle \bvec{t} \rangle}^{\bvec{s}} < \frac{ \delta }{ \tau( q ) } \Big\} \right) \right|
\le
2^{n}
\label{eq:card_le_pow-of-2_part2}
\end{align}
\item
(d) follows from \eqref{eq:limsup_liminf}, and
\item
(e) follows from \lemref{lem:additive} and \eqref{eq:eps_vanish}.
\end{itemize}
Hence, it follows from \eqref{eq:eps_sum_vanish} that
\begin{align}
\lim_{n \to \infty} \frac{ 1 }{ 2^{n} } \Bigg| \Bigg\{ \bvec{s} \in \{-, +\}^{n} \ \Bigg| \ \delta \le \sum_{h^{\prime} = h}^{\tilde{m}} \varepsilon_{\langle \bvec{t}^{(h^{\prime})} \rangle}^{\bvec{s}} \le 1 - \delta \Bigg\} \Bigg|
& =
0 .
\label{eq:eps_sum_vanish2}
\end{align}
Furthermore, we observe that
\begin{align}
0
& \overset{\mathclap{\text{(a)}}}{=}
\lim_{n \to \infty} \frac{ 1 }{ 2^{n} } \Bigg| \Bigg\{ \bvec{s} \in \{-, +\}^{n} \ \Bigg| \ \delta \le \sum_{h^{\prime} = h}^{\tilde{m}} \varepsilon_{\langle \bvec{t}^{(h^{\prime})} \rangle}^{\bvec{s}} \le 1 - \frac{ \delta }{ \tilde{m} } \Bigg\} \Bigg|
\notag \\
& \ge
\limsup_{n \to \infty} \frac{ 1 }{ 2^{n} } \Bigg| \Bigg\{ \bvec{s} \in \{-, +\}^{n} \ \Bigg| \ \delta \le \sum_{h^{\prime} = h}^{\tilde{m}} \varepsilon_{\langle \bvec{t}^{(h^{\prime})} \rangle}^{\bvec{s}} \le 1 - \frac{ \delta }{ \tilde{m} } \Bigg\} \cap \left( \bigcap_{h^{\prime} = h + 1}^{\tilde{m}} \Big\{ \bvec{s} \in \{-, +\}^{n} \ \Big| \ \frac{ \delta }{ \tilde{m} } \le \varepsilon_{\langle \bvec{t}^{(h)} \rangle}^{\bvec{s}} \le 1 - \frac{ \delta }{ \tilde{m} } \Big\}^{\complement} \right) \Bigg|
\notag \\
& \overset{\mathclap{\text{(b)}}}{\ge}
\limsup_{n \to \infty} \frac{ 1 }{ 2^{n} } \Bigg| \Big\{ \bvec{s} \in \{-, +\}^{n} \ \Big| \ \delta \le \varepsilon_{\langle \bvec{t}^{(h)} \rangle}^{\bvec{s}} \le 1 - \delta \Big\} \cap \left( \bigcap_{h^{\prime} = h + 1}^{\tilde{m}} \Big\{ \bvec{s} \in \{-, +\}^{n} \ \Big| \ \frac{ \delta }{ \tilde{m} } \le \varepsilon_{\langle \bvec{t}^{(h)} \rangle}^{\bvec{s}} \le 1 - \frac{ \delta }{ \tilde{m} } \Big\}^{\complement} \right) \Bigg|
\notag \\
& \overset{\mathclap{\text{(c)}}}{\ge}
\limsup_{n \to \infty} \frac{ 1 }{ 2^{n} } \left( \Big| \Big\{ \bvec{s} \in \{-, +\}^{n} \ \Big| \ \delta \le \varepsilon_{\langle \bvec{t}^{(h)} \rangle}^{\bvec{s}} \le 1 - \delta \Big\} \Big| + \Bigg| \bigcap_{h^{\prime} = h + 1}^{\tilde{m}} \Big\{ \bvec{s} \in \{-, +\}^{n} \ \Big| \ \frac{ \delta }{ \tilde{m} } \le \varepsilon_{\langle \bvec{t}^{(h)} \rangle}^{\bvec{s}} \le 1 - \frac{ \delta }{ \tilde{m} } \Big\}^{\complement} \Bigg| - 2^{n} \right)
\notag \\
& \overset{\mathclap{\text{(d)}}}{\ge}
\limsup_{n \to \infty} \frac{ 1 }{ 2^{n} } \Big| \Big\{ \bvec{s} \in \{-, +\}^{n} \ \Big| \ \delta \le \varepsilon_{\langle \bvec{t}^{(h)} \rangle}^{\bvec{s}} \le 1 - \delta \Big\} \Big| + \liminf_{n \to \infty} \frac{ 1 }{ 2^{n} } \Bigg| \bigcap_{h^{\prime} = h + 1}^{\tilde{m}} \Big\{ \bvec{s} \in \{-, +\}^{n} \ \Big| \ \frac{ \delta }{ \tilde{m} } \le \varepsilon_{\langle \bvec{t}^{(h)} \rangle}^{\bvec{s}} \le 1 - \frac{ \delta }{ \tilde{m} } \Big\}^{\complement} \Bigg| - 1
\notag \\
& \overset{\mathclap{\text{(e)}}}{=}
\limsup_{n \to \infty} \frac{ 1 }{ 2^{n} } \Big| \Big\{ \bvec{s} \in \{-, +\}^{n} \ \Big| \ \delta \le \varepsilon_{\langle \bvec{t}^{(h)} \rangle}^{\bvec{s}} \le 1 - \delta \Big\} \Big| ,
\label{eq:eps_h_vanish}
\end{align}
where
\begin{itemize}
\item
(a) follows from \eqref{eq:eps_sum_vanish2},
\item
(b) follows by the inclusions as in \eqref{eq:inclusions} and \eqref{eq:inclusions_2},
\item
(c) follows by the inclusion-exclusion principle and the fact that
\begin{align}
\Bigg| \Big\{ \bvec{s} \in \{-, +\}^{n} \ \Big| \ \delta \le \varepsilon_{\langle \bvec{t}^{(h)} \rangle}^{\bvec{s}} \le 1 - \delta \Big\} \cup \left( \bigcap_{h^{\prime} = h + 1}^{\tilde{m}} \Big\{ \bvec{s} \in \{-, +\}^{n} \ \Big| \ \frac{ \delta }{ \tilde{m} } \le \varepsilon_{\langle \bvec{t}^{(h)} \rangle}^{\bvec{s}} \le 1 - \frac{ \delta }{ \tilde{m} } \Big\}^{\complement} \right) \Bigg|
\le
2^{n}
\label{eq:card_le_pow-of-2_part3}
\end{align}
\item
(d) follows from \eqref{eq:limsup_liminf}, and
\item
(e) follows from \lemref{lem:additive} and the hypothesis \eqref{eq:hypo_eps_h}.
\end{itemize}
Therefore, it follows from \eqref{eq:eps_h_vanish} that
\begin{align}
\lim_{n \to \infty} \frac{ 1 }{ 2^{n} } \Big| \Big\{ \bvec{s} \in \{-, +\}^{n} \ \Big| \ \delta \le \varepsilon_{\langle \bvec{t}^{(h)} \rangle}^{\bvec{s}} \le 1 - \delta \Big\} \Big|
& =
0 ,
\label{eq:eps_h_vanish2}
\end{align}
which implies by induction of \eqref{eq:intermediate_eps_sequence} together with \eqref{eq:eps_vanish} that \eqref{eq:proportion0} of \thref{th:polarization} holds, i.e.,
\begin{align}
\lim_{n \to \infty} \frac{ 1 }{ 2^{n} } \Big| \Big\{ \bvec{s} \in \{-, +\}^{n} \ \Big| \ \delta \le \varepsilon_{d}^{\bvec{s}} \le 1 - \delta \Big\} \Big|
& =
0
\end{align}
for every fixed $\delta \in (0, 1)$ and every $d|q$.

Finally, we prove \eqref{eq:proportion1} of \thref{th:polarization}.
It follows by the definition \eqref{def:mu_d} that
\begin{align}
\mu_{d}^{(n)}
& =
\frac{ 1 }{ 2^{n} } \sum_{\bvec{s} \in \{ -, + \}^{n}} \varepsilon_{d}^{\bvec{s}}
\notag \\
& \le
\frac{ 1 }{ 2^{n} } \sum_{\substack{ \bvec{s} \in \{ -, + \}^{n} : \\ \varepsilon_{d} < \delta }} \delta + \frac{ 1 }{ 2^{n} } \sum_{\substack{ \bvec{s} \in \{ -, + \}^{n} : \\ \delta \le \varepsilon_{d} \le 1 - \delta }} (1 - \delta) + \frac{ 1 }{ 2^{n} } \sum_{\substack{ \bvec{s} \in \{ -, + \}^{n} : \\ \varepsilon_{d} > 1 - \delta }} 1
\notag \\
& =
\delta + \frac{ 1 }{ 2^{n} } \sum_{\substack{ \bvec{s} \in \{ -, + \}^{n} : \\ \delta \le \varepsilon_{d} \le 1 - \delta }} (1 - 2 \delta) + \frac{ 1 }{ 2^{n} } \sum_{\substack{ \bvec{s} \in \{ -, + \}^{n} : \\ \varepsilon_{d} > 1 - \delta }} (1 - \delta) ,
\notag
\end{align}
which implies together with \eqref{eq:proportion0} that
\begin{align}
\mu_{d}^{(\infty)}
\le
\delta + (1 - \delta) \liminf_{n \to \infty} \frac{ 1 }{ 2^{n} } \Big| \Big\{ \bvec{s} \in \{ -, + \}^{n} \ \Big| \ \varepsilon_{d}^{(n)} > 1 - \delta \Big\} \Big|
\label{eq:proportion1_1}
\end{align}
In addition, we also get
\begin{align}
\mu_{d}^{(n)}
& =
\frac{ 1 }{ 2^{n} } \sum_{\bvec{s} \in \{ -, + \}^{n}} \varepsilon_{d}^{\bvec{s}}
\notag \\
& \ge
\frac{ 1 }{ 2^{n} } \sum_{\substack{ \bvec{s} \in \{ -, + \}^{n} : \\ \delta \le \varepsilon_{d} \le 1 - \delta }} \delta + \frac{ 1 }{ 2^{n} } \sum_{\substack{ \bvec{s} \in \{ -, + \}^{n} : \\ \varepsilon_{d} > 1 - \delta }} (1-\delta) ,
\end{align}
which also implies together with \eqref{eq:proportion0} that
\begin{align}
(1-\delta) \limsup_{n \to \infty} \frac{ 1 }{ 2^{n} } \Big| \Big\{ \bvec{s} \in \{ -, + \}^{n} \ \Big| \ \varepsilon_{d}^{(n)} > 1 - \delta \Big\} \Big|
\le
\mu_{d}^{(\infty)}
\label{eq:proportion1_2}
\end{align}
Since $\delta > 0$ can be chosen arbitrarily small, as in Alsan and Telatar's proof of \cite[Theorem~1]{alsan_telatar_it2016}, it follows from \eqref{eq:proportion1_1} and \eqref{eq:proportion1_2} that \eqref{eq:proportion1} holds.
This completes the proof of \thref{th:polarization}.
\hfill\IEEEQEDhere

\section{Proof of \lemref{lem:additive}}
\label{app:additive}

We prove \lemref{lem:additive} by induction.
Define
\begin{align}
S_{n}^{(k)}
\coloneqq
\bigcap_{i = 1}^{k} S_{i, n}
\end{align}
for each $k, n \in \mathbb{N}$.
By hypothesis, it is clear that
\begin{align}
\lim_{n \to \infty} f_{n}\big( S_{n}^{(1)} \big)
=
\lim_{n \to \infty} f_{n}\big( S_{1, n} \big)
=
1 .
\end{align}
Suppose that
\begin{align}
\lim_{n \to \infty} f_{n}\big( S_{n}^{(k-1)} \big)
=
1 .
\end{align}
for a fixed integer $k \in \mathbb{N}$.
Then, we have
\begin{align}
1
& =
\lim_{n \to \infty} f_{n}\big( S_{n}^{(k-1)} \big)
\notag \\
& \ge
\liminf_{n \to \infty} f_{n}\big( S_{n}^{(k)} \big)
\notag \\
& =
\liminf_{n \to \infty} \Big( f_{n}\big( S_{n}^{(k-1)} \big) + f_{n}\big( S_{k, n} \big) - f_{n}\big( S_{n}^{(k-1)} \cup S_{k, n} \big) \Big)
\notag \\
& \ge
\liminf_{n \to \infty} f_{n}\big( S_{n}^{(k-1)} \big) + \liminf_{n \to \infty} f_{n}\big( S_{k, n} \big) - \limsup_{n \to \infty} f_{n}\big( S_{n}^{(k-1)} \cup S_{k, n} \big)
\notag \\
& \ge
1 + 1 - 1
\notag \\
& =
1 ,
\end{align}
which implies that
\begin{align}
\lim_{n \to \infty} f_{n}\big( S_{n}^{(k)} \big)
=
1 .
\end{align}
This completes the proof of \lemref{lem:additive}.
\hfill\IEEEQEDhere

\bibliographystyle{IEEEtran}
\bibliography{IEEEabrv,mybib}

\begin{IEEEbiographynophoto}{Yuta Sakai}(Member, IEEE)
was born in Japan in 1992.
He is currently a Research Fellow in the Department of Electrical and Computer Engineering at the National University of Singapore (NUS).
He received the B.E.\ and M.E.\ degrees in the Department of Information Science from the University of Fukui in 2014 and 2016, respectively, and the Ph.D.\ degree in the Advanced Interdisciplinary Science and Technology from the University of Fukui in 2018.
His research interests include information theory and coding theory.
\end{IEEEbiographynophoto}

\begin{IEEEbiographynophoto}{Ken-ichi Iwata}(Member, IEEE)
received the B.Ed. degree from Wakayama University in 1993, the M.Sc. degree from Information Science from Japan Advanced Institute of Science and Technology in 1995, and the D.E. degree from the University of Electro-Communications in 2006.
Since 2008 he has been with University of Fukui, where he is an Associate Professor.
\end{IEEEbiographynophoto}

\begin{IEEEbiographynophoto}{Hiroshi Fujisaki}(Member, IEEE)
is an Associate Professor of Kanazawa University from 2011.
He received the B.E.\ and M.E.\ degrees in Electronic Engineering from Kyushu University, Fukuoka, Japan, in 1989 and 1991 respectively.
He received the D.E.\ degree in Communication Engineering from the Department of Computer Science and Communication Engineering, Kyushu University, Japan in 2001.
From 1991 to 1996, he worked as a Research Staff member in Hitachi, Ltd., Ibaraki, Japan.
From 1998 to 2001, he worked as a Research Associate in the Department of Computer Science and Communication engineering, Kyushu University.
From 2001 to 2010, he worked as a Lecturer in Graduate School of Natural Science and Technology, Kanazawa University, Japan.
His research interests are in random number generations based on
one-dimensional ergodic transformations
and their applications to digital communication systems.
\end{IEEEbiographynophoto}

\end{document}